\documentclass[12pt]{article}
\usepackage{amssymb,epsf,graphicx}
\usepackage{amsmath,amssymb,bm,epsfig,epsf,graphicx,color,float,url} 
\usepackage{dsfont}
\usepackage{caption}
\usepackage{subcaption}
\usepackage{colortbl}
\usepackage[all]{xy}

\input epsf.sty
\newcommand{\Tr}{\mathop{\rm Tr}\nolimits}

\def\bra#1{\langle #1 |}
\def\bbraa#1{\langle\!\langle \hspace{0.5pt}  #1 \hspace{0.5pt}|\hspace{-1pt}|}
\def\ket#1{|#1 \rangle}
\def\kett#1{|#1 \rangle\!\rangle}
\def\kkett#1{|\hspace{-1pt}|\hspace{0.5pt} #1 \hspace{0.5pt}\rangle\!\rangle}
\def\aver#1{\left\langle\, #1 \,\right\rangle}

\def\half{\frac{1}{2}}

\let\eps = \varepsilon
\def \be {\begin{equation}}
\def \ee {\end{equation}}
\def \bea {\begin{eqnarray}}
\def \eea {\end{eqnarray}}
\def \bdm {\begin{displaymath}}
\def \edm {\end{displaymath}}

\def \dd {{\cal D}}

\def \hh {{\cal H}}

\def\hsp{\hspace{-8.5pt}}
\def\hspF{\hspace{-6pt}}
\def\Rac#1#2#3#4#5#6{\mbox{\scriptsize${\protect\begin{Bmatrix} #1,&\hsp #2,&\hsp #3\\ #4,&\hsp #5,&\hsp #6 \end{Bmatrix}}$}}
\def\TET#1#2#3#4#5#6{\mbox{\scriptsize${\protect\begin{bmatrix} #1,&\hsp #2,&\hsp #3\\ #4,&\hsp #5,&\hsp #6 \end{bmatrix}}$}^{\mathrm{TET}}}
\def\W6J#1#2#3#4#5#6{\mbox{\scriptsize${\protect\begin{bmatrix} #1,&\hsp #2,&\hsp #3\\ #4,&\hsp #5,&\hsp #6 \end{bmatrix}}$}}
\def\Fmat#1#2#3#4#5#6{F_{#1 #2}{\mbox{\scriptsize${\protect\begin{bmatrix} #3  &\hspF #4\\ #5  &\hspF #6\end{bmatrix}}$}}}
\def\Fblock#1#2#3#4#5#6{F_{#1 #2}^{\mathrm{blocks}}{\mbox{\scriptsize${\protect\begin{bmatrix} #3 &\hspF #4\\ #5 &\hspF #6\end{bmatrix}}$}}}
\def\FRac#1#2#3#4#5#6{F_{#1 #2}^{\mathrm{Rac}}{\mbox{\scriptsize${\protect\begin{bmatrix} #3 &\hspF #4\\ #5 &\hspF #6\end{bmatrix}}$}}}
\def\Umat#1#2#3#4#5#6{U_{#1 #2}{\mbox{\scriptsize${\protect\begin{bmatrix} #3 &\hspF #4\\ #5 &\hspF #6\end{bmatrix}}$}}}

\def \Vir {\mathrm{Vir}}

\def\1{ \mathds{1}}

\begin{document}
{}~ \hfill\vbox{\hbox{}\hbox{MISC-2016-09} }\break
\vskip 2.1cm

\centerline{\large \bf Topological defects in open string field theory}
\vspace*{8.0ex}

\centerline{\large \rm Toshiko Kojita$^{(a,b)}$\footnote{Email: {\tt kojita at cc.kyoto-su.ac.jp }}, Carlo Maccaferri$^{(c)}$\footnote{Email: {\tt maccafer at gmail.com}},
Toru Masuda$^{(a,d)}$\footnote{Email: {\tt masudatoru at cc.nara-wu.ac.jp }},
Martin Schnabl$^{(a)}$\footnote{Email: {\tt schnabl.martin at gmail.com}}}

\vspace*{4.0ex}
\begin{center}

$^{(a)}${\it Institute of Physics of the ASCR, v.v.i. \\
Na Slovance 2, 182 21 Prague 8, Czech Republic}
\vskip .4cm

$^{(b)}${\it Maskawa Institute for Science and Culture, Kyoto Sangyo Univ.,\\
Motoyama, Kamigamo, Kita-Ku, Kyoto-City, Kyoto, Japan}
\vskip .4cm

$^{(c)}${\it Dipartimento di Fisica, Universit\'a di Torino and  INFN    Sezione di Torino\\
Via Pietro Giuria 1, I-10125 Torino, Italy}
\vskip .4cm

$^{(d)}${\it Department of Physics, Nara Women's University \\ Kita-Uoya-Nishimachi,
Nara, Nara, Japan}

\end{center}

\vspace*{6.0ex}

\centerline{\bf Abstract}
\bigskip

We show how conformal field theory topological defects can relate solutions of open string field theory for different boundary conditions.  To this end we generalize the results of Graham and Watts to include the action of defects on boundary condition changing fields. Special care is devoted to the general case when nontrivial multiplicities arise upon defect action. Surprisingly the fusion algebra of defects is realized on open string fields only up to a (star algebra) isomorphism.

 \vfill \eject

\baselineskip=16pt

\tableofcontents

\setcounter{footnote}{0}

\section{Introduction and summary}
\setcounter{equation}{0}
\label{sec:Introduction}

In the past 16  years there has been quite a lot of progress in charting out the space of possible solutions of the classical equations of motion of open string field theory (OSFT) \cite{Witten} by both numerical \cite{SZ, BSZ, MSZ, SZmarg, Michishita,  Bagchi, Karcz, CZJPmarg, KMS, KRS, KM} as well as analytic tools
 \cite{RZ,wedge, Schnabl,Okawa, ElS, Erler-Split,KKT, Ellwood, KOZ, KMS, Murata1,Baba:2012cs} by which new exact solutions have been found or analyzed
 \cite{Schnabl, ES, marg, KORZ, FKP, KO, BMT, KOS, Murata2, Hata:2011ke, Hata:2012cy, Masuda:2012kt,
Erler_analytic, Takahashi:2002ez, simple-marg, EM, MaccSch, KMTT, IKT}. See \cite{Thorn, TZ, SenRev, FK, lightning, OkawaRev} for reviews.

The OSFT action
\be
S_{OSFT} = -\frac{1}{g_o^2} \left[ \frac12 \aver{\Psi * Q_B \Psi} + \frac13 \aver{\Psi*\Psi*\Psi} \right],\label{action}
\ee
can be formulated for an arbitrary system of ``D-branes", coincident or not, and described by a generic Boundary Conformal Field Theory (BCFT, see \cite{CardyReview, PZ2, RunkelPhD, GaberdielReview, RS} for reviews) for composite or fundamental boundary conditions.

Obviously, to describe all solutions for the bewildering space of theories based on arbitrary BCFT is a difficult task. But beside the intrinsic importance of the classification of OSFT solutions,
this program can potentially help in the discovery of new D-brane systems, by encoding new world-sheet boundary conditions into the gauge invariant content of OSFT solutions \cite{KMS, Ellwood, KOZ}. Numerical approaches are useful on a case-by-case basis, especially when one does not know what to expect, i.e. when the problem of classifying all conformal boundary conditions for a given bulk CFT is unsolved. Analytic solutions are scarce and until recently they essentially described only the universal tachyon vacuum or marginal deformations. A notable progress has been achieved with the solution \cite{EM}, by Erler and one of the authors, which can be written down explicitly for any given pair of  reference and target BCFT's. The existence of this solution gives evidence that OSFT can describe the whole landscape of D-branes that are consistent with a given closed string background. However since the solution requires the knowledge of the OPE between the boundary condition changing operators between the two BCFT's, it does not directly help  in the problem of discovering new BCFT's.

It would be  nice to have an organizing principle by which we could simply relate solutions in the same or possibly different theories. Solution generating techniques are scarce and problematic \cite{martin1, ell-sing, EM-sing}. It is well known however that symmetries can be used to generate new solutions. Given a star algebra automorphism $S$
\be
S (\psi*\chi) = S(\psi) * S (\chi),
\ee
commuting with the BRST operator $Q_B$ one can see that if $\Psi$ is a solution of the equation of motion, then so is $S \Psi$. The operator $S$ can correspond to a discrete symmetry, or  a continuous symmetry. In the latter case one has a family of such operators $S_\alpha$ which arise by exponentiation of the infinitesimal generator, a star algebra derivative $P$
\be
P(\psi*\chi) = (P\psi)*\chi + \psi*(P\chi).
\ee
Indeed, assuming $\left[Q_B, P\right]=0$ and setting  $S_\alpha = e^{\alpha P}$, one finds that $S_\alpha$ maps solutions to solutions. The symmetry generator $P$ is often given by a contour integral of a spin one current and upon exponentiation it can be interpreted as a topological defect operator. Even in the case of discrete symmetries, the operator $S$ can be viewed as a so called group-like topological defect operator \cite{FFRS, TFT}.\footnote{Topological defects have played a prominent role in the recent development of two-dimensional CFT, see also \cite{PZ,BBDO,GW,FGRS,Drukker:2010jp}.}

The main goal of this paper is to extend this analogy further. For every topological defect in a given BCFT we construct an operator ${\cal D}$ which maps the state space of one BCFT into another, in such a way that
\be\label{distributivity}
{\cal D} (\psi*\chi) = {\cal D} (\psi) * {\cal D}(\chi).
\ee
What makes a defect topological is that the defect operator commutes with the energy momentum tensor, and hence  also commutes with the BRST charge $Q_B$. Then it immediately follows that if $\Psi$ is a classical solution of OSFT for a given BCFT, then ${\cal D}\Psi$ is a solution of OSFT built upon another BCFT.

The explicit action of the defect operator ${\cal D}$ on the open string fields turns out to be quite tricky. In general the string field algebra is not given by a single BCFT Hilbert space with a single boundary condition but is given rather by a direct sum of $A$-$B$ bimodules $\bigoplus_{a,b} \hh^{(ab)}$, where $a$ and $b$ label the boundary conditions for the endpoints of open string stretched by two D-branes. The algebras $A$ and $B$ represent a set of boundary fields on a D-brane for the $a$  and $b$ boundary condition respectively, and are themselves  bimodules with the left and right multiplication provided by the operator product. As the defect operator must commute with the Virasoro generators (single surviving copy on the upper half-plane) it must act as
\be
{\cal D}^d \phi_i^{ab} = \sum_{a',b'} X_{ia'b'}^{dab} \phi_i^{a'b'}.
\ee
An important feature is that this maps boundary operators intertwining between two given boundary conditions into a {\em sum} of operators intertwining between different boundary conditions allowed by the defect fusion rules of the theory. It thus maps, in general, the original star algebra into a bigger star algebra. This contrasts with the action of the defect operators on the closed string Hilbert space where it maps the whole space into itself \cite{PZ}.

For the sake of simplicity and concreteness we limit our discussion in this paper to topological defects of minimal models with diagonal partition function. These have been fully classified \cite{PZ} and they are in one-to-one correspondence with primary operators. For these defects we find from the
distributivity requirement (\ref{distributivity}), in a canonical normalization for boundary fields (\ref{eq:C-canonical}), that the $X$ coefficients are given in terms of the normalized $g$-functions $g'$ (\ref{norm-g}) and the normalized 6J-symbols (\ref{Normalized6J})
\be
X_{ia'b'}^{dab} =\left(g'_ag'_bg'_{a'}g'_{b'}\right)^{\frac14}\W6J{a}{a'}{d}{b'}{b}{i}.
\ee
For the special case of $a=b$ this reduces, up to a normalization, to the result of Graham and Watts \cite{GW}.

\begin{figure}
\centering
\includegraphics[]{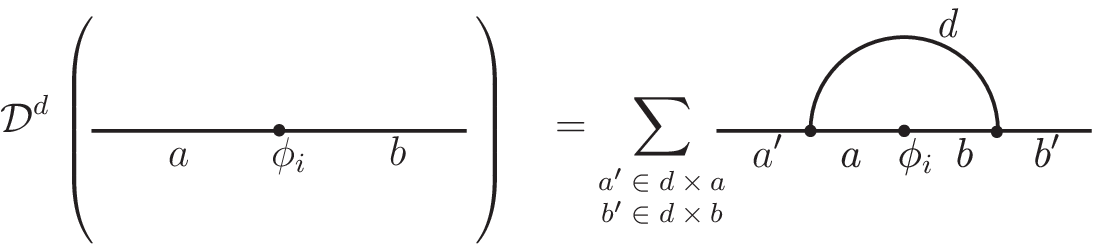}
\caption{Action of the topological defect on a boundary field $\phi_i^{ab}$ can be described by enclosing it with an open defect attached to the boundary. The result is a collection (direct sum) of boundary fields $\phi_i^{a'b'}$ in all possible new boundary conditions allowed by fusion. The dots at the junctions represent simple normalization factors determined in subsection \ref{subsec:geom_constr}.}
\label{Fig:defect_action}
\end{figure}

The action of the defect on boundary fields can be conveniently understood in terms of defects attached onto the boundary as in Figure \ref{Fig:defect_action}. The defect endpoints can be freely moved along the boundary without changing correlators as long as the defect does not cross any operator insertion. The junction point can be viewed as an insertion of the identity operator (and not a traditional boundary condition changing operator) up to a normalization factor which we determine in subsection \ref{subsec:geom_constr}.

Quite surprisingly however, the fusion rules of the open defect operators are twisted by an orthogonal similarity transformation when multiple boundary condition are generated by the defect. So instead of
\be
D^d D^c = \sum_e N_{dc}^{\;\;\; e} D_e
\ee
which holds for the defect action on bulk states, the action on boundary operators obeys
\be
\dd^d \dd^c = U_{dc} \left( \bigoplus_e N_{dc}^{\;\;\; e} \dd_e \right) U_{dc}^{-1}.
\ee
Here $U$ is a matrix, which for fixed $c$ and $d$ describes a discrete transformation in the space of multiplicity labels, which for fixed initial and final boundary conditions has its rows labeled by the intermediate boundary condition created by $\dd^c$, and columns labeled by $e$ the summation parameter of the direct sum on the right hand side.
When we include also the initial and the final boundary conditions as part of the multiindices, i.e.  $\lbrace a,a',a'' \rbrace$ and $[e; a, a'']$ respectively, the $U$ matrix becomes orthogonal and it is surprisingly given by the Racah symbol (\ref{Racah})
\be
(U_{dc})^{\{a\,a'\,a''\}[e;\,b,\,b'']}
=\delta_{ab}\,\delta_{a''\,b''}\,\Rac{c}{a}{a'}{a''}{d}{e}.
\ee
%

From the explicit formula \cite{KMS} for the boundary state in terms of the OSFT classical solution it follows that applying the defect operator to the open string field  results in a boundary state encircled by the defect operator which gives a new consistent boundary state of the kind considered by Graham and Watts \cite{GW}. In formulas
\be
\kkett{B_{\dd\Psi}}=D\kkett{B_\Psi}.
\ee
Therefore, assuming  that a given solution $\Psi_{X \to Y}$ describes BCFT$_Y$ in terms of BCFT$_X$, upon action of the defect operator ${\cal D}$, it will describe BCFT$_{{\cal D} Y}$ in terms of BCFT$_{{\cal D} X}$, i.e.
\be
{\cal D} \Psi_{X \to Y} = \Psi_{{\cal D} X \to {\cal D} Y}.
\ee
As a byproduct of our analysis we discover the interesting relation
\be
D^d_j\,\,^{(a)}\!B_j^i\,g_{a}=\sum_{a'\in d\times a}\,X_{ia'a'}^{daa}\,^{(a')}\!B_j^i\,g_{a'},
\ee
relating bulk to boundary structure constants for different boundary conditions, where $D$ is the coefficient of the defect operator (\ref{eq:Y^a_i=}) acting on a spinless bulk field (labeled by $j$).

In order to be as self-contained as possible the paper includes some review material and it is organized as follows. In section \ref{sec:CFT-Defects} we review the basic definitions and properties of defects and defect networks in two dimensional CFT. We also introduce the duality matrices as generic solutions to the pentagon identity (which is a consequence of the topological structure of the networks and the basic fusion rules of the defects) and the related 6J and Racah symbols. In section \ref{sec:Boundaries} we review the basic construction of boundary states in diagonal minimal models and how topological defects act on them. In addition we  review Runkel's derivation of the boundary OPE coefficients and identify a particularly useful normalization for boundary fields. Section \ref{sec:Defects} is devoted to the main results of our work, namely the construction of open topological defect operators as
maps between two boundary operator algebras which is compatible with the OPE. We present two independent derivations of our results, one which is algebraic and which builds on the initial analysis by Graham and Watts \cite{GW} and one which is geometric and uses the properties of defect networks in presence of boundaries. Both our constructions clearly show that the composition of open topological defect operators follows the fusion rules only up to a similarity transformation whose precise structure is encoded in the Racah symbols.  In section \ref{Sec:OSFT} we consider open topological defects as new operators in OSFT which map solutions to solutions. We show that the way OSFT observables are affected by the action of defects is consistent with the BCFT description and  the interpretation of OSFT solutions as describing a new BCFT using the degrees of freedom of a reference BCFT.  In section \ref{sec:Ising} we concretely present our constructions in the explicit example of the Ising model BCFT. Few appendices contain further results which are used in the main text.

\section{Defects in conformal field theory}
\setcounter{equation}{0}
\label{sec:CFT-Defects}


Consider two, generally distinct, 2d CFT's
glued along a one-dimensional interface.
We assume that energy is conserved across
this interface.
Let $(x^0,x^1)$ be the coordinates of the system and
the interface is placed at $x^1=0$.
From the conservation law $\partial_0T^{00}+\partial_{1}T^{10}=0$, we
see that if $T^{10}$ is continuous then the total energy is conserved
\be
\frac{\partial}{\partial x^{0}} \int dx^{1}T^{00} =\int dx^{1}\frac{\partial}{\partial x^{1}} T^{10} =0.
\ee
Then we require that the momentum density $T^{01}=T^{10}$ is continuous across the interface
\be
T^{01}(x^0, x^1)\big|_{x^1\to 0+}
=
T^{01}(x^0, x^1)\big|_{x^1\to 0-} .
\ee
Introducing the complex coordinates {$z=x^0 +ix^1 $ and $\bar z=x^0-ix^1$},  the above condition is written as
\be
\lim_{x^1\to 0}\left(T(z)-\bar T(\bar z)\right)\big|_{z=x^0+ix^1}=\lim_{x^1\to 0}
\left(T(z)-\bar T(\bar z)\right)\big|_{z=x^0-ix^1}.
\label{eq:conformal_condition}
\ee
This condition also means that the system has invariance under  conformal transformations
which leave the shape of the defect line untouched, and the interface
enjoying \eqref{eq:conformal_condition}
 is called a conformal interface or a conformal defect.
The gluing condition
\eqref{eq:conformal_condition}
is usually implemented by giving a rule for how fields of
these two CFTs are related at the interface, and
conformal defects give a mapping from a field configuration of one theory to that of the other theory.
%
The concept of conformal defects comes from the study of one dimensional impurity system \cite{Wong:1994np, Oshikawa:1996dj}. For a recent discussion of conformal defects see
\cite{Bachas:2004sy,Quella:2006de,Bachas:2007td,Bajnok:2013waa}.

There are two special classes of conformal defects: the factorized defects and the topological defects.
The factorized defects are purely reflective with respect to the energy flow,
and the two CFTs do not communicate at all. This condition is
given by requiring the energy current $T^{01}$ to be zero at the defect,
or equivalently
\be
\lim_{x^1\to 0}T(z)\big|_{z=x^0+ix^1}=\lim_{x^1\to 0}\bar T(\bar z)\big|_{z=x^0+ix^1},
 \qquad
\lim_{x^1\to 0}{T}({z})\big|_{\bar{z}=x^0-ix^1}=
\lim_{x^1\to 0}\bar{T}(\bar{z})\big|_{\bar{z}=x^0-ix^1}.
\label{eq:condition_factorized_defects}
\ee
From \eqref{eq:condition_factorized_defects}
we see that
the system is reduced to two separated BCFTs sharing the defect line as their common boundary.
Also note that a conformal boundary can be viewed
as an example of a factorized defect between a given bulk CFT and an empty $c=0$ theory.

On the other hand,
topological defects
are purely transmissive with respect to the energy, and
this condition is expressed by the momentum conservation across the defect.
From the conservation law $\partial_0 T^{01}+\partial_1 T^{11}=0$, we see that
if $T^{11}$ is continuous across the defect, the momentum is conserved.
 In  complex coordinates,
this condition is given by
\be
\lim_{x^1\to 0}T(z)\big|_{z=x^0+ix^1}=\lim_{x^1\to 0}T(z)\big|_{z=x^0-ix^1},
 \qquad
\lim_{x^1\to 0}\bar{T}(\bar{z})\big|_{\bar{z}=x^0+ix^1}=
\lim_{x^1\to 0}\bar{T}(\bar{z})\big|_{\bar{z}=x^0-ix^1}.
\ee
The energy momentum tensor
does not see the defect,
since its components are continuous
across the defect line.
Therefore, continuous deformations of  topological
defects do not change the value of correlation functions.

A familiar example of a topological defect appears in the Ising model CFT,
which is equivalent to  the minimal model $\mathcal M(3,4)$ with three primary fields $\{\mathds 1,\  \eps,\  \sigma\}$.
In addition, there exists the disorder field $\mu$ which however is not mutually local with the spin field $\sigma$.
The correlation functions containing both spin fields and disorder fields have branch cuts
on their Riemann surface, and they are represented by disorder lines which connect a pair of disorder fields.
Clearly the value of such correlation functions does not change by continuous deformations of disorder lines.

In the above example, topological defects are curve segments with both end-points at disorder fields
but  we can consider more generic configurations in which defects join or end on a boundary, as we will review and discuss later.

%
%

\subsection{Closed topological defects}\label{sect:closed-defects}
A particularly class of defect operators  are closed topological defects which
are associated to homotopy classes of cycles on a punctured surface.
The closed topological defects give rise naturally to closed string operators $D$ which act on bulk fields by encircling them with the defect.
Since both the holomorphic and antiholomorphic component of the energy momentum are continuous across the defect, the encircling defect loop can be arbitrarily smoothly deformed and
the defect action should also commute with the Virasoro generators
\be
\left[ L_n, D \right] =  [\tilde L_n, D ] =0, \qquad \forall n.
\ee
By Schur's lemma the action of the defect operator on the bulk states must  be constant on every Verma module.
Then we can concentrate on
the action of $D$ on the bulk primary operator $\phi_{(i,\bar i,\alpha,\bar \alpha)}(z,\bar z)$, where $i$ and $\bar i$ are labels for the Virasoro representation of the holomorphic and
antiholomorphic part and $\alpha$ and $\bar \alpha$ are corresponding
multiplicity labels.
Let $a$ be a label classifying the topological defects then, following Petkova and Zuber \cite{PZ}, we can write
\be
D^a \phi_{(i,\bar i,\alpha,\bar \alpha)}(z,\bar z)=D^a_{(i,\bar i,\alpha,\bar \alpha)}\phi_{(i,\bar i,\alpha,\bar \alpha)}(z,\bar z),
\ee
with constant coefficients $D^a_{(i,\bar i,\alpha,\bar \alpha)}$.
This can be written by
\be
D^a = \sum_{(i,\bar i,\alpha,\bar \alpha)} D^a_{(i,\bar i,\alpha,\bar \alpha)} P^{(i,\bar i,\alpha,\bar \alpha)},
\label{eq:defect_projector}
\ee
where $P^{(i,\bar i,\alpha,\bar \alpha)}$ is the projector on the Verma module labeled by ${(i,\bar i,\alpha,\bar \alpha)}$.

To determine the coefficients $D^a_{(i,\bar i,\alpha,\bar \alpha)}$, let us consider
the modular transformation of the torus partition function with a
pair of closed topological defect lines. There are two ways for evaluating this.
One way is to consider  time slices parallel to the defect line, obtaining the
following expression
\be
\begin{split}
Z_{a|b}
=&
{\rm Tr}\left((D^a)^\dagger D^b\tilde q^{L_0-\frac{c}{24}}\bar{\tilde q}^{\bar L_0-\frac{c}{24}}\right)\\
=&
\sum_{{(j,\bar j,\alpha,\bar \alpha)}} (D^a_{(j,\bar j,\alpha,\bar \alpha)})^*
D^b_{(j,\bar j,\alpha,\bar \alpha)}\chi_j(\tilde q)\chi_{\bar j}(\bar{\tilde q}).
\end{split}
\label{eq:partiton_defect1}
\ee
The other way is to consider  time evolution
along the defect line.
Let  ${V_{i\bar i;x}}^y$ denotes the multiplicity of
the Virasoro representation $(i\bar i)$
appearing in the spectrum
in this time slicing
\be
\mathcal H_{a|b}={V_{i\bar i;a}}^b\mathcal R_i\otimes \bar{\mathcal R}_{\bar i},
\ee
then  the partition function is written as
\be
Z_{a|b}=\sum_{{(j,\bar j,\alpha,\bar \alpha)}} {V_{j\bar j;a}}^b\chi_j(q)\chi_{\bar j}(\bar q).
\label{eq:partiton_defect2}
\ee
From modular invariance of the Virasoro characters, we can connect \eqref{eq:partiton_defect1}
and \eqref{eq:partiton_defect2}, and obtain a bootstrap equation
\be
{V_{i\bar i;a}}^b=\sum_{j\bar j}{S_{ji}}{S_{\bar j\bar i}}D^a_{(i,\bar i,\alpha,\bar \alpha)}\left(D^b_{(i,\bar i,\alpha,\bar \alpha)}\right)^*,
\label{eq:bootstrap_closed_defct}
\ee
which in paticular states that the rhs should be a positive integer.

For the diagonal minimal models there is a simple solution to
\eqref{eq:bootstrap_closed_defct}
\be
D^a_i=\frac{S_{ai}}{S_{\mathds{1}i}},\qquad {V_{i\bar i;a}}^b=\sum_k {N_{ai}}^k{N_{k\bar i}}^b.
\label{eq:Y^a_i=}
\ee
We can check this result with the help of the Verlinde formula \cite{Verlinde}
\be
N_{ij}^k=\sum_l \frac{S_{il}S_{jl}S^*_{kl}}{S_{\mathds{1}l}}.\label{eq:Verlinde}
\ee
On the right hand side of the first equation of  \eqref{eq:Y^a_i=}, index $a$ runs over the irreducible representations of
the Virasoro algebra, and
from this expression, we see that
there are as many distinct topological defects as primary operators.
Plugging back to \eqref{eq:defect_projector}, we obtain that
\be
D_a = \sum_i \frac{S_{ai}}{S_{\mathds{1}i}} P^i.
\ee
It then follows that these operators obey the fusion algebra
\be
D_a D_b = \sum N_{ab}^{\;\;\; c} D_c,
\ee
just like the conformal families of the primary fields
\be
[\phi_a]\times [\phi_b] = \sum_c N_{ab}^{\;\;\; c} [\phi_c].
\ee
This follows by a simple computation
\be
\left(\sum_i \frac{S_{ai}}{S_{\mathds{1}i}} P^i \right) \left(\sum_j \frac{S_{bj}}{S_{\mathds{1}j}} P^j \right)
= \sum_i \frac{S_{ai}}{S_{\mathds{1}i}} \frac{S_{bi}}{S_{\mathds{1}i}} P^i
= \sum_i \sum_c N_{ab}^{\;\;\; c} \frac{S_{ci}}{S_{\mathds{1}i}} P^i,
\ee
where in the last equality we used  Verlinde formula (\ref{eq:Verlinde}).
%

\subsection{Defect networks}


To get a more conceptual understanding of topological defects, it is useful to consider networks of topological defects, where the defects are allowed to join in trivalent vertices. Let us state now some minimalistic assumptions: defects can be decomposed into elementary ones. The labels of the elementary defects define naturally an associative algebra. For two elementary defects $a$ and $b$ we define the product of labels $a \times b$ as a free sum of labels of the elementary defects which arise upon fusing defect $a$ with defect $b$. The associativity follows from the topological nature of defects.

Another important assumption is that there is always an identity defect, labeled as 1 or $\mathds{1}$, which can be freely drawn or attached anywhere without changing anything.\footnote{At this point our conventions differ from some of the literature on the subject e.g. \cite{Runkel:T-systems}.}

\begin{figure}
\centering
\includegraphics[]{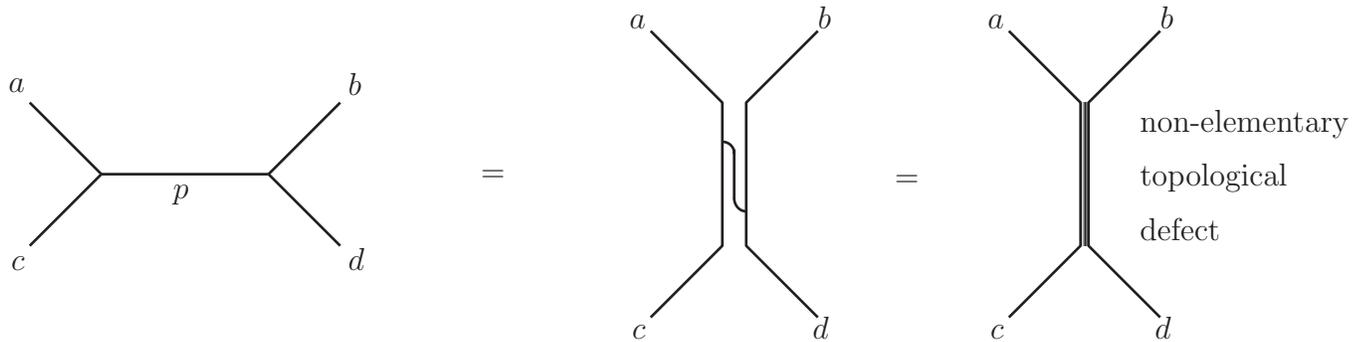}
\caption{The topological nature of  defects implies that an``s-channel" configuration with an elementary defect should be equivalent to a ``t-channel" configuration with a composite, but still topological defect.}
\label{Fig:defect_motiv}
\end{figure}

\begin{figure}
\centering
\includegraphics[]{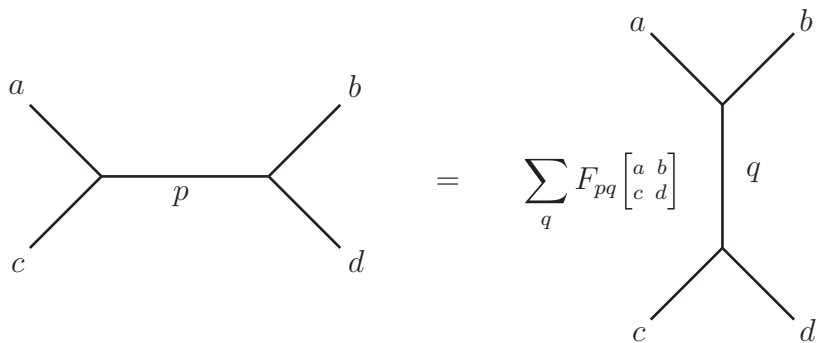}
\caption{Elementary defect network move. }
\label{Fig:defect_fusion}
\end{figure}

The most powerful property of topological defects, is that they can be freely deformed, without changing the value of any correlator, as long as they do not cross the position of any operator insertion or another defect. Therefore, a small piece of defect network in the``s-channel"-like configuration, shown on the left hand side of Figure \ref{Fig:defect_motiv}, composed of four defects joined by an intermediate one, can be deformed into an alternate ``t-channel"-like configuration where the new defect will in general no longer be elementary one.
Decomposing it into a linear combination of elementary defects, defines a set of a priori unconstrained coefficients $\Fmat{p}{q}{a}{b}{c}{d}$, see Figure \ref{Fig:defect_fusion}.
In many circumstances these coefficients are known, but in order to be self-contained and perhaps more general, let us ignore this knowledge and proceed by following the consequences of consistency.

\begin{figure}[!htbp]
\centering
\includegraphics[]{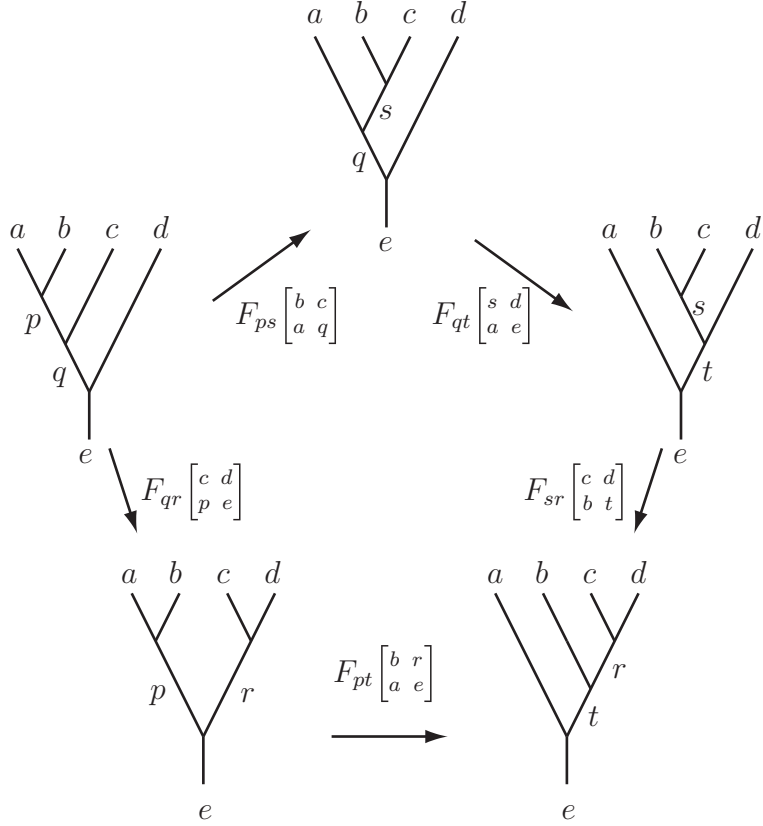}
\caption{Pentagon identity as a consistency condition for fusion of defects.}
\label{Fig:defect_pentagon}
\end{figure}

The most important consistency condition comes from considering the defect network shown in Figure~\ref{Fig:defect_pentagon}.
By following the fusion rule in Figure \ref{Fig:defect_fusion} from one starting to one final configuration along two different paths depicted by arrows in Figure \ref{Fig:defect_pentagon}, one finds the celebrated pentagon identity\footnote{This identity differs from the one given in \cite{MS} by transposition of columns in every $F$. For unoriented defects this difference is immaterial as $\Fmat{p}{q}{a}{b}{c}{d}=\Fmat{p}{q}{b}{a}{d}{c}$. For oriented defects one has to specify carefully an orientation. Our implicit orientation (always downwards in
Figure~\ref{Fig:defect_pentagon}) is the same as in \cite{KR} and others \cite{AGGS}, but differs from the one required to match the formulas of Moore and Seiberg \cite{MS, Moore-Seiberg}. \label{Footnote:orientation}}
\be
\label{pentagon}
\sum_s
\Fmat{p}{s}{b}{c}{a}{q} \Fmat{q}{t}{s}{d}{a}{e} \Fmat{s}{r}{c}{d}{b}{t}
=\Fmat{q}{r}{c}{d}{p}{e} \Fmat{p}{t}{b}{r}{a}{e}.
\ee
By the MacLane coherence theorem, this equation is enough to guarantee the consistency of the fusion rule in Figure \ref{Fig:defect_fusion} for any possible defect network. This is not to say that other identities are not of interest, but that those required for consistency are implied by the pentagon identity.

A simple property of the $F$ following from our definition and the natural normalization for the trivial identity defect is that
\be
\Fmat{p}{q}{a}{b}{c}{d} = 1 \qquad \mbox{whenever } 1 \in \lbrace a,b,c,d \rbrace.
\ee
Using this  in the pentagon identity by setting
$e=1$ (which requires also $q=d$, $a=t$ and $p=r$)  one finds the orthogonality relation
\be
\label{orthogonality}
\sum_s
\Fmat{p}{s}{b}{c}{a}{d} \Fmat{s}{r}{c}{d}{b}{a}
= \delta_{pr}.
\ee
For $p \ne r$ the left hand side of (\ref{pentagon}) still makes sense, but the right hand side would contain fusion matrix elements with non-admissible label, so the only consistent value for the left hand side is zero. Alternatively, one can of course apply the elementary move again to the right hand side of Figure \ref{Fig:defect_fusion} viewed sideways.

\begin{figure}[!htbp]
\centering
\includegraphics[]{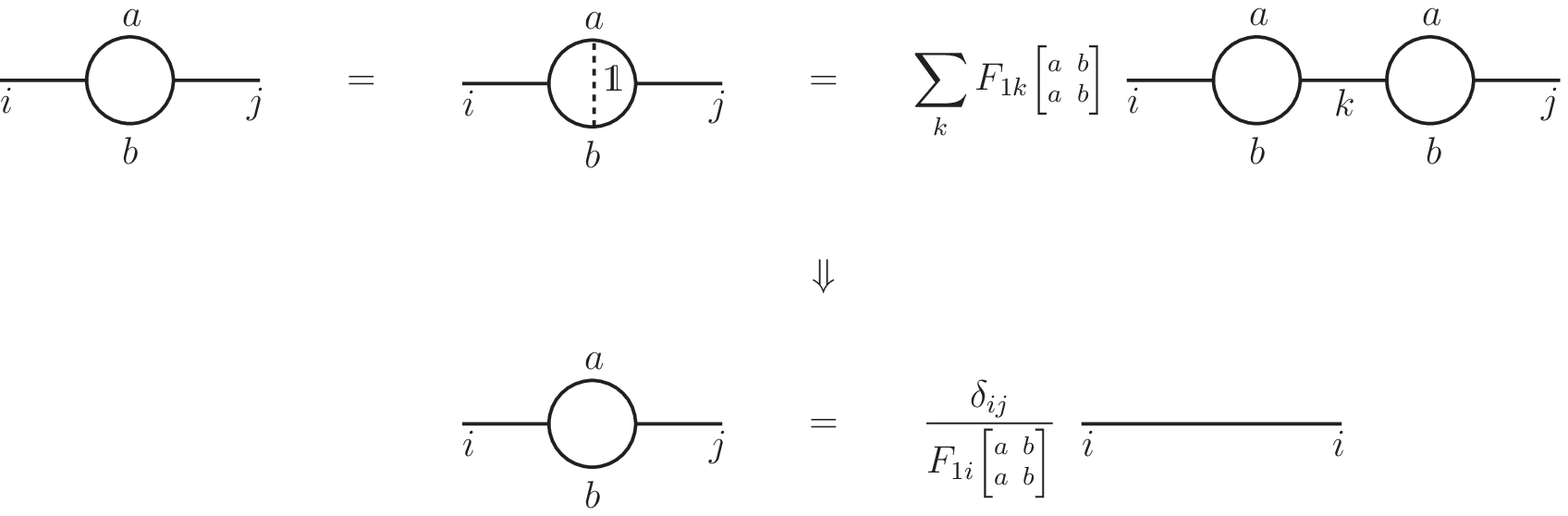}
\caption{Defect bubble network. When no operators are present inside, the topological bubble can be shrank to zero size, yielding a numerical factor symmetric under the exchange of $a$ and $b$ labels.}
\label{Fig:defect_bubble}
\end{figure}

Let us now consider defect networks containing closed loops. An arbitrary such network can be reduced via the elementary move in Figure \ref{Fig:defect_fusion} to a network without any closed loops.
The simplest defect network with a loop is of course a single loop, but let us start with a slightly more general configuration shown in Figure \ref{Fig:defect_bubble} of a bubble with two external defect lines attached.
As long as there are no operators put inside the bubble, it can shrink leaving behind a pure number. This already tells, that the two external lines must carry the same label otherwise, upon shrinking, one would expect a defect changing operator insertion. These operators carry nontrivial conformal weight and thus the correlator scaling properties would contradict those expected for a network of topological defects. Attaching an auxiliary line of identity defect (see Figure \ref{Fig:defect_bubble}) allows us to find a simple expression for the bubble in terms of the $F$-matrix. In rotation invariant theories for unoriented defects the numerical factor $\Fmat{1}{i}{a}{b}{a}{b}$ must be symmetric in $a$ and $b$, as follows by considering the 180$^\circ$ degree rotation.

As a corollary we find that a bubble with no defects attached, which is the same thing as if two identity defects were attached, is equivalent to an overall factor
\be\label{norm-g}
g_a' = \frac{1}{\Fmat{1}{1}{a}{a}{a}{a}},
\ee
which we identify with the {\em normalized} $g$-function of the defect.\footnote{In the context of the minimal models this is equal to $\frac{S_{1a}}{S_{11}}$, where $S$ is the modular $S$-matrix. It is thus the value of the $g$-function of the boundary condition associated to the defect via the folding trick, normalized by the $g$-function of the trivial defect. In non-unitary theories (e.g. for the Lee-Yang model) or theories with oriented defects this quantity may coincide with the usual normalized $g$-function only up to a sign. }

\begin{figure}[!htbp]
\centering
\includegraphics[]{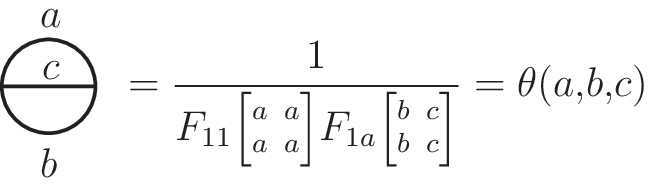}
\caption{Defect sunset network gives rise to an $S_3$ symmetric factor $\theta(a,b,c)$.}
\label{Fig:defect_sunset}
\end{figure}

Another corollary is an expression for the sunset diagram in Figure~\ref{Fig:defect_sunset}.
Since it can be viewed as a symmetric bubble on a loop in two possible ways, it follows that
\be
\theta(a,b,c) = \theta(a,c,b) = \theta(c,b,a),
\ee
and hence it enjoys the full $S_3$ permutation symmetry. Analogously, we can introduce
\be
\tilde\theta(a,b,c) = \frac{1}{\Fmat{1}{1}{a}{a}{a}{a} \Fmat{a}{1}{b}{b}{c}{c}},
\ee
which, as a consequence of pentagon identity (setting $t=q=1$ together with  $s=e=d=a$, $r=b$ and $p=c$ in (\ref{pentagon})) satisfies
\be\label{DDt}
g_a' g_b' g_c' = \theta(a,b,c)  \tilde\theta(a,b,c),
\ee
and hence $\tilde\theta(a,b,c)$ also possess the $S_3$ permutation symmetry.

\begin{figure}[!htbp]
\centering
\includegraphics[]{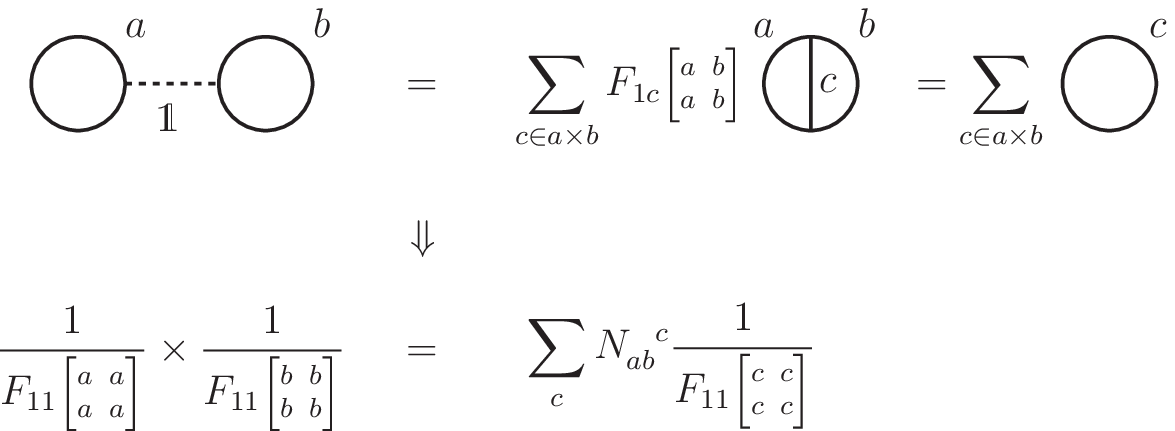}
\caption{Derivation of the Verlinde-like formula. In the second step we use the $S_3$ symmetry of the bubble factor $\theta(a,b,c)$.}
\label{Fig:Verlinde}
\end{figure}

Another peculiar identity which can be obtained by the sequence of moves in Figure~\ref{Fig:Verlinde}
is
\be\label{VerlindeF}
\frac{1}{\Fmat{1}{1}{a}{a}{a}{a}} \frac{1}{\Fmat{1}{1}{b}{b}{b}{b}} = \sum_{c} N_{ab}^{\;\;\;c} \frac{1}{\Fmat{1}{1}{c}{c}{c}{c}}.
\ee
It might come as a surprise that this equation is a consequence of the polynomial pentagon identity (\ref{pentagon}). To see that it is indeed the case, one may proceed in two steps. Starting with the orthogonality relation (\ref{orthogonality}), setting $p=r=1$,  adjusting accordingly the other indices, and using the relation (\ref{DDt}) one derives easily (\ref{VerlindeF}).

For the minimal models, the relation (\ref{VerlindeF}) is in fact nothing but the Verlinde formula for the first row (or column) of the $S$-matrix
\be
\frac{S_{\mathds{1}a} S_{\mathds{1}b}}{S_{\mathds{1}\mathds{1}}} =
\sum_c N_{ab}^{\;\;\;c} S_{\mathds{1}c}.
\label{eq:Verlinde1column}
\ee
thanks to the relation between the modular $S$-matrix and the $F$-matrices (which follows from the formulas in \cite{Moore-Seiberg}, see also e.g. (E.9) in \cite{BPPZ})
\be
\frac{S_{ij}}{S_{11}} = \sum_{ k \in i \times j} e^{2\pi i (h_i+h_j-h_k)} \frac{1}{\Fmat{1}{1}{k}{k}{k}{k}},
\ee
which for $i=1$ simplifies to $g_j'=\frac{S_{1j}}{S_{11}} = \left(\Fmat{1}{1}{j}{j}{j}{j}\right)^{-1}$.

\begin{figure}[!htbp]
\centering
\includegraphics[]{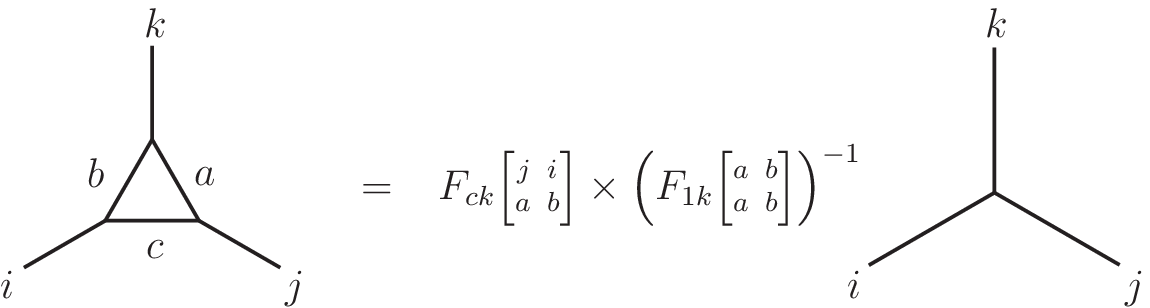}
\caption{Triangular defect network. When no operators are present inside, the topological triangle can be shrank to zero size, yielding a numerical factor with $S_3$ permutation symmetry under the exchange of $(i,a)$, $(j,b)$ and $(k,c)$ labels.}
\label{Fig:defect_triangle}
\end{figure}

The next natural step is to consider a defect loop with three external defect lines attached, as in Figure~\ref{Fig:defect_triangle}. Applying the elementary defect network move to any pair of vertices connected by an internal defect line, we find the elementary vertex with a bubble on one of the external lines, and as before we can replace the bubble by the corresponding factor.
In general this would yield a $Z_3$ symmetric expression which follows directly from the pentagon identity (\ref{pentagon}) by setting $t=1$ which requires also $s=d$, $r=b$ and $e=a$ with the help of (\ref{DDt}).  For unoriented parity-invariant defects in parity invariant CFT's the symmetry is enhanced to $S_3$.

\begin{figure}[!htbp]
\centering
\includegraphics[]{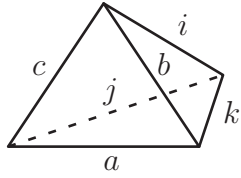}
\caption{Tetrahedral defect network. When no operators are present inside the faces, the topological tetrahedron can be shrank to zero size, yielding a numerical factor with $S_4$ permutation symmetry under the exchange of the faces with $(a,b,c)$, $(a,j,k)$, $(b,k,i)$ and $(c,i,j)$ labels.}
\label{Fig:defect_tetrahedron}
\end{figure}

The symmetries of the $F$ matrices are actually much larger and can be nicely manifested by considering a defect network in the shape of tetrahedron, see Figure~\ref{Fig:defect_tetrahedron}. It can be drawn as such on a Riemann sphere, but the resulting identities should have universal validity.
Let us now choose any triangular face, e.g. $(abc)$, and shrink it to a point picking up the triangle factor as in Figure~\ref{Fig:defect_triangle}. This results in the sunset diagram, see Figure~\ref{Fig:defect_sunset}, which we have already evaluated. Combining the factors, one quickly arrives at
\bea
\TET{i}{j}{k}{a}{b}{c} &\equiv& \frac{\Fmat{c}{k}{j}{i}{a}{b}}{\Fmat{1}{k}{a}{b}{a}{b} \Fmat{1}{k}{i}{j}{i}{j} \Fmat{1}{1}{k}{k}{k}{k}} \nonumber\\
&=& \frac{1}{g_k'} \theta(a,b,k) \theta(i,j,k) \Fmat{c}{k}{j}{i}{a}{b}.
\eea
The notation is such that the labels in the upper row always form an admissible triplet, i.e. $i \in j  \times k$.

We could have chosen an arbitrary face of the tetrahedron for reducing the triangle and due to the $Z_3$ cyclicity of the defect triangle in Figure~\ref{Fig:defect_triangle}, we would have obtained one out of three possible expressions. Altogether we get 12 different expressions which must be equal to each other, and which correspond to the orientation preserving subgroup $A_4$ of the tetrahedral group $S_4$. The $Z_3$ subgroup cyclically permutes the columns
\be\label{WidZ3}
\TET{i}{j}{k}{a}{b}{c} = \TET{j}{k}{i}{b}{c}{a} = \TET{k}{i}{j}{c}{a}{b},
\ee
while the $Z_2\times Z_2$ subgroup is switching upper and lower labels simultaneously in two different columns
\be\label{WidZ2}
\TET{i}{j}{k}{a}{b}{c} = \TET{i}{b}{c}{a}{j}{k} = \TET{a}{b}{k}{i}{j}{c} = \TET{a}{j}{c}{i}{b}{k}.
\ee
The equality of the three expressions (\ref{WidZ3}) follows already from the pentagon identity, but (\ref{WidZ2}) does not. The reason is that in deriving (\ref{WidZ2}) we have assumed that all defects were unoriented. For oriented defects some labels in the identity (\ref{WidZ2}) must be replaced by the conjugate labels to account for the change of orientation.

In the special case of parity invariant defects in parity invariant theories the tetrahedral defect network is invariant under the full tetrahedral group $S_4$ which in addition to the generators of $A_4$ contains also 12 transformations combining rotations with a single reflection. The additional identities can be generated with the help of
\be\label{WidOdd}
\TET{i}{j}{k}{a}{b}{c} = \TET{j}{i}{k}{b}{a}{c}.
\ee
These are the symmetries of the classical or quantum Wigner's 6J symbol. This object however differs from the 6J symbol by a tetrahedral invariant normalization factor (see discussion in section \ref{6J&Racah}), and resembles thus an object often called $TET$ in the literature, see e.g. \cite{Coquereaux:2005hc} or \cite{6J-book}. Another difference would arise in the case of oriented defect, where the tetrahedral invariance of the defect network would be broken.\footnote{Also because of the fact, that up to the normalization factor, this object obeys the pentagon identity without signs, it is more reminiscent of the classical Racah W-coefficient.}

\subsubsection{Specular symmetries}

As follows from the definition of the $F$-matrix, see Figure~\ref{Fig:defect_fusion}, invariance under 180$^\circ$ rotation for unoriented defects implies
\be\label{Fid180}
\Fmat{p}{q}{a}{b}{c}{d} = \Fmat{p}{q}{d}{c}{b}{a}.
\ee
Similarly parity invariance of both the theory and the defect (with respect to any axis) implies a stronger condition
\be\label{FidSpec}
\Fmat{p}{q}{a}{b}{c}{d} = \Fmat{p}{q}{c}{d}{a}{b} = \Fmat{p}{q}{b}{a}{d}{c}.
\ee
Of course, the two identities (\ref{FidSpec}) together imply (\ref{Fid180}). All these identities are true in the Virasoro minimal models.

Mathematically, the symmetry of the $F$ matrix (\ref{Fid180}) is equivalent to the condition (\ref{WidZ2}) under the assumption of the pentagon identity, or in particular (\ref{WidZ3}).
Analogously, under the same assumption, the symmetries (\ref{FidSpec}) are equivalent to
$\TET{i}{j}{k}{a}{b}{c} = \TET{b}{a}{k}{j}{i}{c}$ and $\TET{i}{j}{k}{a}{b}{c} = \TET{j}{i}{k}{b}{a}{c}$ respectively.


\subsubsection{6J symbols, Racah symbols and their identities}\label{6J&Racah}

From the defect network manipulations we have seen that $\TET{i}{j}{k}{a}{b}{c}$ obeys the same tetrahedral symmetries as classical or quantum Wigner symbol. Such an object is by no means unique, since a product over the four tetrahedron vertices of an $S_3$ invariant function of the three corresponding edges will always have the tetrahedral symmetry. A particularly useful combination is what we call the normalized 6J symbol
\bea
\W6J{i}{j}{k}{a}{b}{c} &=& \frac{1}{\sqrt{\theta(i,j,k)\theta(i,b,c)\theta(a,j,c)\theta(a,b,k)}} \TET{i}{j}{k}{a}{b}{c}
\\\nonumber
\\
&=& \frac{1}{g_k'} \sqrt{\frac{\theta(a,b,k) \theta(i,j,k)}{\theta(i,b,c)\theta(a,j,c)}} \Fmat{c}{k}{j}{i}{a}{b},
\label{Normalized6J}
\eea
which enjoys also the full tetrahedral symmetry for unoriented defects. If it were not for the prefactor $1/g_k'$ it would have obeyed the pentagon identity, since
\be\label{Fgt6J}
\Fmat{p}{q}{a}{b}{c}{d} \to \frac{\Lambda(a,b,q)\Lambda(c,q,d)}{\Lambda(c,a,p)\Lambda(p,b,d)} \Fmat{p}{q}{a}{b}{c}{d},
\ee
is an exact symmetry of the pentagon identity (\ref{pentagon}) for an arbitrary function $\Lambda(i,j,k)$ of an admissible triplet\footnote{Further discussion of  this ``gauge symmetry" observed by Moore and Seiberg \cite{Moore-Seiberg} is relegated to appendix~\ref{app:Gauge_Symmetry}. }. Due to cyclicity of $\theta$, one can arrange their arguments, so that the full prefactor under the square root in (\ref{Normalized6J}) can be viewed as a gauge transformation.

Hence the 6J symbol obeys the following pentagon-like identity
\be\label{6Jpentagon}
\sum_s g_s' \W6J{c}{b}{s}{a}{q}{p} \W6J{d}{s}{t}{a}{e}{q} \W6J{d}{c}{r}{b}{t}{s} = \W6J{d}{c}{r}{p}{e}{q} \W6J{r}{b}{t}{a}{e}{p}.
\ee
When one of the entries equals 1, the 6J symbol simplifies
\be
\W6J{i}{c}{b}{1}{b}{c} = \frac{1}{\sqrt{g_b' g_c'}}.
\ee
This offers a very promising procedure to solve the polynomial equations in the general case. Given the fusion rules, we first solve the much simpler system $g_a' g_b'=\sum_c N_{ab}^{\;\;\;c} g_c'$. This system of equations in rational CFT's has as many solutions as we have labels. Every solution corresponds to a single column of the modular $S$-matrix normalized by its first element. Then we can solve the pentagon identities (\ref{6Jpentagon}) by imposing the symmetries on the 6J symbol. We have checked that for Lee-Yang model, Ising model and tricritical Ising model these over-determined polynomial systems have a unique solution. \bigskip
%
%

A fundamental property of the 6J symbol is gauge invariance under the symmetry (\ref{Fgt6J}) which holds for any $\Lambda$ which is $S_3$-invariant and subject to the additional condition $\Lambda(1,a,a)=1$.

%

Another very useful object which will play an important role in the following sections of this paper is what we call Racah symbol following the terminology of Coquearaux\footnote{What Coquearaux \cite{Coquereaux:2005hc} calls geometrical Racah symbols, or Carter et al \cite{6J-book} call the 6j symbol are our $F$ matrices.}
\be
\Rac{i}{j}{k}{a}{b}{c} \equiv \sqrt{g_k' g_c'} \W6J{i}{j}{k}{a}{b}{c}.
\label{Racah}
\ee
It obeys the full pentagon identity,
\be\label{RacPentagon}
\sum_s \Rac{c}{b}{s}{a}{q}{p} \Rac{d}{s}{t}{a}{e}{q} \Rac{d}{c}{r}{b}{t}{s} = \Rac{d}{c}{r}{p}{e}{q} \Rac{r}{b}{t}{a}{e}{p},
\ee
and can be therefore identified with the $F$ matrix in a given special gauge
\be
\Rac{c}{b}{s}{a}{q}{p}=\FRac{p}{s}{b}{c}{a}{q}.
\ee
A peculiarity of this gauge is that
\be
\FRac{1}{i}{a}{b}{a}{b}=\FRac{i}{1}{a}{a}{b}{b}=\sqrt{\frac{g'_i}{g'_a g'_b}}.
\ee
When one of the entries of the first two columns equal 1, the Racah symbol simplifies
\be
\Rac{i}{c}{b}{1}{b}{c} = 1.
\ee

This object, like the 6J symbol, is also invariant under the gauge transformation (\ref{Fgt6J}).
Just like a generic solution to the pentagon identity, the Racah symbols obey the orthogonality condition
\be
\sum_q \Rac{b}{a}{q}{c}{d}{p}\Rac{c}{a}{s}{b}{d}{q}=\delta_{ps}.\label{Racah-ortho}
\ee
When we manipulate defect networks we don't have necessarily to fix a gauge for the involved $F$ matrices. This has to be contrasted with the $F$ matrices arising from the transformations of the
conformal blocks which are uniquely determined once the conformal blocks are normalized, by giving the coefficient of their leading term. It would be interesting to know whether there is a specific normalization choice which also fixes the gauge for the defect networks. This may
involve a careful study of defect changing fields, which goes beyond the scope of this paper.

\section{Boundaries in conformal field theory}
\setcounter{equation}{0}
\label{sec:Boundaries}


Conformal boundary conditions in 2D CFT's have to satisfy a number of consistency conditions spelled out explicitly in \cite{Lewellen, Sagnotti}. This section is a review of some of these consistency  conditions in diagonal minimal models and of the action of
topological defects on the fundamental boundary states.

\subsection{Boundary conditions in minimal models}

From the bulk perspective, conformal boundary conditions in 2D CFT's are encoded in the conformal boundary states which are required to obey a number of necessary conditions. The most elementary requirement of preserving the conformal symmetry forbids the two-dimensional energy and momentum to flow through the boundary, which leads to the gluing condition
\be
\left(L_n - \bar L_{-n} \right) \kkett{B} = 0.
\label{gluing}
\ee
The set of linearly independent solutions was written down by Ishibashi \cite{Ishibashi}. The Ishibashi states are in one-to-one correspondence with spinless bulk primaries $V^\alpha$
\bea
\kett{V_\alpha} &=& \sum_{IJ} M^{IJ}(h_\alpha) L_{-I}\bar L_{-J} \ket{V_\alpha} \\
&=& \sum_n \ket{n,\alpha} \otimes \overline{\ket{n,\alpha}} \\
&=& \Big[ 1+\frac1{2h_\alpha}L_{-1}\bar L_{-1} + \cdots \Big] \ket{V_\alpha}.
\eea
The multi-indices $I,J$, with $I=\{i_1,...,i_n\}$ appearing in the first line label the non-degenerate descendants in the conformal family of $V_\alpha$,  and  $M^{IJ}(h_\alpha)$ is  the inverse of the Gram matrix $\bra{V^\alpha}L_{I}L_{-J}\ket{V_\alpha}$. We denoted   $L_{I} = L_{i_1} L_{i_2} \ldots L_{i_n}$ and $L_{-I} = L_{-i_n} L_{-i_{n-1}} \ldots L_{-i_1}$. The Ishibashi state (see the second line) is  as a sum over a  basis of states (which are orthonormal wrt the Gram matrix) in the Verma module over the chiral part of the primary $V_\alpha$.

A highly non-trivial consistency requirement is  given by Cardy's condition. Consider the partition function on a finite cylinder with two boundary conditions $a$ and $b$. Viewed in the ``closed string" channel, the diagram can be interpreted
as a matrix element between two boundary states $\kkett{a}$ and $\kkett{b}$. In the ``open string'' channel  it becomes a trace over the Hilbert space of the CFT with the two boundary conditions $a$ and $b$
\be
\bbraa{a} {\tilde q}^{\frac{1}{2}(L_0+\bar{L}_0 -\frac{c}{12})} \kkett{b} = \Tr_{{\cal H}_{ab}^{\mathrm{open}}} \left( q^{L_0-\frac{c}{24}} \right),
\label{cardy's condition}
\ee
where
\be
q= e^{2\pi i \tau}, \qquad\qquad \tilde q =e^{-2\pi i /\tau},
\ee
and $\tau=R/L$ is given by the radius and length of the cylinder.

It is well known that for minimal models with diagonal partition function  Cardy's condition is solved by a set of fundamental boundary states, explicitly given by \cite{Cardy}
\be
\kkett{B_i} = \sum_j \frac{S_{i}^{\; j}}{\sqrt{S_{\1}^{\; j}}} \, \kett{j},
\label{Cardy_sol}
\ee
where $S_i^{\;j}$ are the entries of the modular matrix and $i$ and $j$ denote the Virasoro representation which are present in the minimal model. Therefore in this case there is a one-to-one correspondence between chiral primaries and fundamental boundary conditions. The most general boundary condition consistent with Cardy's condition is obtained by taking positive integer linear combinations of the above fundamental boundary states, in the case of diagonal minimal models.

\subsection{Runkel's solution for boundary structure constants}\label{Runkel}
Boundary operators generally change the boundary conditions. The multiplicity of a boundary operator in the Virasoro representation $k$, changing the boundary conditions from $i$ to $j$ is the integer coefficient $N_{ij}^{\;\;k}$ appearing in  the fusion rules of the theory
\be
\phi_i \times \phi_j = \sum_k N_{ij}^{\;\;\,k} \, \phi_k.
\label{fusion}
\ee

While one could extract the spectrum of boundary operators from the cylinder amplitude between two boundary states, to compute their OPE structure constants one has to resort to the 4-pt conformal bootstrap.
To this end, let us consider a 4pt boundary function
\be
{\cal G}^{(abcd)}_{ijkl}(\xi)\equiv\aver{I \circ \phi_i^{ab}(0)\phi_j^{bc}(1)\phi_k^{cd}(\xi)\phi_l^{da}(0)}_{\rm UHP},
\ee
where $I(z)=-\frac1z$. We can compute it in two ways using different OPE channels
\bea
{\cal G}^{(abcd)}_{ijkl}(\xi)&=&\aver{I \circ \phi_i^{ab}(0)\phi_j^{bc}(1)\phi_k^{cd}(\xi)\phi_l^{da}(0)} \nonumber\\
&=&\aver{h\circ\phi_l^{da}(0) h\circ I \circ \phi_i^{ab}(0)h\circ \phi_j^{bc}(1)h\circ\phi_k^{cd}(\xi)}
\nonumber\\
&=&\xi^{h_i+h_j-h_l-h_k}\aver{I \circ \phi_l^{da}(0) \,\phi_i^{ab}(1) \,\phi_j^{bc}(1-\xi)\, \phi_k^{cd}(0)}\nonumber\\
&=&\xi^{h_i+h_j-h_l-h_k}{\cal G}^{(dabc)}_{lijk}(1-\xi),\label{eq:four-1}
\eea
where $h(z) = \frac{z-\xi}{z}$.
We now express the four point functions in terms of the structure constants and the  four point  conformal blocks
\be
{\cal G}_{ijkl}^{(abcd)}(\xi)= \sum_p C_{ij}^{(abc)\,p}C_{kl}^{(cda)\,p} G_{pp}^{(aca)} {\cal F}(i,j,k,l;p)(\xi),\label{eq:four-2}
\ee
where the conformal blocks are given by the formula
\be
{\cal F}(i,j,k,l;p)(\xi) = \sum_{I,J} \beta_I(h_i,h_j,h_p) \beta_J(h_k,h_l,h_p) G^{IJ}(h_p) \xi^{h_p+|J|-h_k-h_l},
\ee
where $I$ is a Virasoro multiindex, $G^{IJ}$ is the matrix of inner products in a highest weight representation of Virasoro algebra, and the $\beta$ coefficients are defined via
\be
\phi_i^{ab}(x)\phi_j^{bc}(y) = \sum_{p,I}  C_{ij}^{(abc)\,p} \frac{\beta_{I}(h_i,h_j,h_k)}{(x-y)^{h_i + h_j - h_k - |I|}} L_{-I} \phi_p^{ac}(y).
\ee
Notice that ${\cal F}$ do {\em not} depend on any normalization of boundary operators.\\
The conformal blocks in $\xi$ can be linearly related to the the conformal blocks in $1-\xi$ via {\it precisely chosen} $F$ matrices
\be
{\cal F}(k,l,i,j;p)(\xi)=\sum_q \Fblock{p}{q}{l}{i}{k}{j}\, {\cal F}(i,l,k,j;q)(1-\xi).\label{eq:Fblocks-trans}
\ee
Then, using (\ref{eq:four-2}) and (\ref{eq:Fblocks-trans}),  we find from (\ref{eq:four-1})
\be\label{bootstrap}
C_{li}^{(dab)\,p}C_{jk}^{(bcd)\,p} G_{pp}^{(dbd)} = \sum_q \Fblock{q}{p}{l}{i}{k}{j} C_{ij}^{(abc)\,q} C_{kl}^{(cda)\,q} G_{qq}^{(aca)},
\ee
where the two point functions are given in terms of the three point functions as
\be
G_{pp}^{(aca)} = C_{pp}^{(aca)\,1} g_a.
\ee
This can be further simplified to\footnote{Use $C_{kl}^{(cda)\,q}C_{qq}^{(aca)\,1}=C_{qk}^{(acd)\,l}C_{ll}^{(ada)\,1}$ and $C_{ii}^{(aba)\,1}g_a=C_{ii}^{(bab)\,1}g_b$.}
\be
C_{ip}^{(abd)\,l}C_{jk}^{(bcd)\,p}  = \sum_q \Fblock{q}{p}{l}{i}{k}{j} C_{ij}^{(abc)\,q} C_{qk}^{(acd)\,l}.
\ee
Runkel \cite{Runkel:1998he} has observed\footnote{Generalizations have been studied in \cite{BPPZ, FFFS1, FFFS2}.} that this equation can be exactly solved by setting
\be\label{RunkelSol}
C_{ij}^{(abc)\,k} = \Fblock{b}{k}{a}{c}{i}{j},
\ee
thanks to the pentagon identity.

We can find a more convenient expression for the structure constants by changing the normalization of the boundary operators. To this end, let us express $\Fblock{p}{q}{a}{b}{c}{d}$ in terms of the normalized 6J symbols
\be
\Fblock{p}{q}{a}{b}{c}{d} = g_q' \sqrt{\frac{\theta(b,d,p) \theta(c,a,p)}{\theta(c,d,q)\theta(b,a,q)}}\W6J{b}{a}{q}{c}{d}{p}.
\ee
Runkel's solution then becomes
\bea
C_{ij}^{(abc)\,k} &=& \Fblock{b}{k}{a}{c}{i}{j}= g_k' \sqrt{\frac{\theta(c,j,b) \theta(i,a,b)}{\theta(i,j,k)\theta(c,a,k)}}\W6J{c}{a}{k}{i}{j}{b}
\\
&=& \sqrt{\frac{g_k'}{g_b'}} \sqrt{\frac{\theta(c,j,b) \theta(i,a,b)}{\theta(i,j,k)\theta(c,a,k)}}\Rac{c}{a}{k}{i}{j}{b}
\\
&=&  \sqrt{\frac{\frac{\theta(c,j,b)}{\sqrt{g_c'g_j' g_b'}} \frac{\theta(i,a,b)}{\sqrt{g_i'g_a' g_b'}}}{\frac{\theta(i,j,k)}{\sqrt{g_i'g_j' g_k'}}\frac{\theta(c,a,k)}{\sqrt{g_c'g_a' g_k'}}}}\Rac{c}{a}{k}{i}{j}{b}.
\eea
From here we see  that  there is a special choice of normalization of the boundary fields, in which\footnote{This expression for the boundary structure constants is particularly interesting since the square of the prefactor coincides with the bulk structure constants, $C_{ijk}^{\rm bulk}=\frac{\sqrt{g_i'g_j' g_k'}}{\theta(i,j,k)}$ in a canonical normalization for bulk operators where the coefficient of the two point functions are set universally to one times the sphere partition function.}
\be\label{eq:C-canonical}
\hat C_{ij}^{(abc)\,k} =  \sqrt{\frac{\sqrt{g_i'g_j' g_k'}}{\theta(i,j,k)}}\Rac{c}{a}{k}{i}{j}{b}.
\ee
The Racah symbol is gauge invariant, but the object $\theta(i,j,k)$ has to be computed from $\Fblock{p}{q}{a}{b}{c}{d}$, i.e.
\be
\theta(i,j,k) = \frac{g_i'}{\Fblock{1}{i}{j}{k}{j}{k}}.
\ee

\subsection{Defect action on boundary states}

As we have  seen in section \ref{sect:closed-defects}, defects act naturally on bulk operators by encircling them. Cardy boundary states are (non-normalizable) states in the space of bulk operators
\be
\kkett{B_a} = \sum_i\frac{S_{ai}}{\sqrt{S_{\mathds 1i}}} \kett{i},
\ee
 so, analogously, the action of the defect operator is given by \cite{GW}
\be
D_a \kkett{B_b} = \sum_c N_{ab}^{\;\;\; c} \kkett{B_c},
\ee
as follows by a computation almost identical to section \ref{sect:closed-defects}, using again Verlinde formula and the fact that the projectors obey $P^i \kett{j} = \delta_j^i \kett{j}$.

Now we can easily offer an alternative proof for the observation of \cite{KRS} that OSFT makes predictions for the coefficients of the boundary states $\kkett{B_X} = \sum_\beta B_X^\beta \kett{\beta}$
\be\label{boundary states relation}
\frac{B_{X}^\beta B_{Y}^\beta}{B_{R}^\beta} = \sum_Z N_{X Y}^{\;\;\;\; Z} B_Z^\beta,
\ee
under the assumption that the reference D-brane $\kkett{R}$ allowed an OSFT solution describing $\kkett{X}$ and that such a solution could have been re-interpreted on a D-brane $\kkett{Y}$ sharing the relevant Verma modules as $\kkett{R}$.

Assuming that the defect operator acts as a multiple of the identity on each Verma module and that as an operator it is selfconjugate (or possibly antiselfconjugate) under BPZ conjugation 
then the coefficients of the boundary states satisfy
\be
\frac{\bra{V^\beta} D  \kkett{R}}{\langle V^\beta \kkett{R}} = \frac{\bra{V^\beta} D  \kkett{X}}{\langle V^\beta \kkett{X}},
\ee
and hence
\be
\frac{B_{DR}^\beta}{B_{R}^\beta} = \frac{B_{DX}^\beta}{B_{X}^\beta},
\ee
where $DR$ and $DX$ stands for D-branes obtained by fusing defect $D$ onto $R$ or $X$ branes. The new $DX$ brane itself is either fundamental or should be an integer linear combination of such
and therefore
\be
\frac{B_{X}^\beta B_{DR}^\beta}{B_{R}^\beta} = B_{DX}^\beta = \sum_Z N_{D X}^{\;\;\;\; Z} B_Z^\beta,
\ee
which matches the formula (\ref{boundary states relation}) derived from OSFT by reinterpreting a solution $\Psi_{R \to X}$ on the $DR$ brane. It is one of the goals of this paper to explain this coincidence.



\section{Attaching defects to boundaries}
\setcounter{equation}{0}
\label{sec:Defects}


In this section we define and study the action of topological defects on boundary fields. This was partially done by Graham and Watts \cite{GW} for boundary operators which do not change the boundary conditions. We will generalize their algebraic approach to the full open string spectrum, including the important case of boundary condition changing operators. In addition we will provide
an independent geometric derivation using defect networks.

We determine how an open string defect acts as an operator mapping the boundary operator algebra of a system of boundary conditions to  a  closed subset of the operator algebra of a new system of boundary conditions. Then we study the composition of such operators and we show how is this related to the fusion rules of the theory. We first proceed in a completely algebraic way, imposing the condition that the OPE must commute with the action of an open topological defect. This is a  non-trivial constraint that can be solved for the coefficients defining the open string defect (see later) and, together with an appropriate twist-invariance condition, allows to uniquely determine such coefficients, thanks to the pentagon identity.  Then we show that the composition of open topological defects is governed by the fusion rules of the theory but, differently from the closed string case, there is a non trivial rotation in the Chan-Paton's labels corresponding to coincident final boundary conditions. This rotation is in fact a similarity transformation. In the second subsection we show that our algebraic results can be independently obtained in a purely geometric way by manipulating the involved defect networks with boundary.



Our goal is to define an action of defects on the open string Hilbert space
\be
\dd:\;{\cal H}_{\rm open}\to{\cal H}_{\rm open}.
\ee
This general action is further specified by decomposing ${\cal H}_{\rm open}$ into fundamental boundary conditions
\be
{\cal H}_{\rm open}=\bigoplus_{a,b} {\cal H}^{(ab)},
\ee
where, when $a\neq b$, the corresponding states are boundary condition changing fields.
Let $d$ be a label for a topological defect, then the open string topological defect is  a linear map
\be
{\cal D}^d: {\cal H}^{(ab)}\to
\bigoplus_{\scriptsize\begin{matrix}a'\in d\times a\\
b'\in d\times b\end{matrix}}{\cal H}^{(a'b')}.
\ee
This map is injective but in general is not surjective (the defect maps from a given open string Hilbert space, onto a ``bigger'' one). This is how open strings feel the fact
that a topological defect, in general,  maps a single D-brane into a system of multiple D-branes, according to the fusion rules of the underlining bulk CFT.
As in the bulk case,  Schur's lemma implies that the operator $\dd$ restricted  to $\hh^{(a,b)} \to \hh^{(a',b')}$ should be a multiple of the identity on every Verma module $\Vir_i$ in $\hh^{(a,b)}$.
Assuming that all states can be obtained by acting Virasoro operators on primary states, which is true in unitary CFT's, the open string defect action is fully specified by
\be
{\cal D}^d  \, \phi_i^{ab} = \sum_{\scriptsize{\begin{matrix} a' \in d \times a \\ b' \in d \times b \end{matrix}}} \! X_{ia'b'}^{dab} \, \phi_i^{a'b'},
\label{Xdef}
\ee
where the $\phi$'s  are the boundary primary fields, allowed by the involved boundary conditions.
\subsection{Algebraic construction}

In this subsection we will determine the above-defined $X$-coefficient in an algebraic way. Then we will inspect how the composition of open topological defects is related to the fusion of defects in the bulk and to the fusion rules of the theory.

\subsubsection{Defect coefficients from OPE}
A simple consistency condition for the action of open-string defects has been introduced by Graham and Watts  \cite{GW}
\be\label{distrib-prim}
\dd^d \left( \phi_i^{ab}(x) \phi_j^{bc}(y)\right) = \left( \dd^d  \phi_i^{ab}(x) \right) \left( \dd^d  \phi_j^{bc}(y)\right).
\ee
Using the general ansatz (\ref{Xdef}) as well as the operator product expansion, the constraint (\ref{distrib-prim}) takes the explicit form
\be\label{distrib-X}
X^{dac}_{ka'c'} C_{ij}^{(abc) k} = \sum_{b'\in d \times b}  C_{ij}^{(a'b'c') k}X^{dab}_{ka'b'} X^{dbc}_{kb'c'}\;,
\ee
where $C_{ij}^{(abc) k}$ are the boundary structure constants.
Restricting ourselves to the $A_n$ series of the minimal models, see section \ref{Runkel}, the boundary structure constants can be written as
\be\label{RunkelRacahSolution}
C_{ij}^{(abc) k}=\frac {n_i^{ab}n_j^{bc}}{n_k^{ac}}\sqrt{\frac{\sqrt{g_i'g_j' g_k'}}{\theta(i,j,k)^{\rm blocks}}}\Rac{c}{a}{k}{i}{j}{b},
\ee
where $n_i^{ab}$ are generic normalizations of boundary fields with the convention that $n_i^{ab}=1$ for the canonical normalization (\ref{eq:C-canonical}).
With this explicit form of the structure constants it immediately follows that (\ref{distrib-X}) admits a general solution
\be
\label{eq: X=(N/N)F}
X_{ia'b'}^{dab}  =\frac {n_i^{ab}}{n_i^{a'b'}}  \Rac{b'}{i}{a}{d}{a'}{b},
\frac{N(d,a,a')}{N(d,b,b')}\ee
thanks to the pentagon identity (\ref{RacPentagon}).
The pentagon identity doesn't fix the constants $N(d, c,c')$, but their are in fact fixed by imposing the parity condition\footnote{In this work we consider only 2D CFT's which are separately invariant under $C$, $P$ and $T$ discrete symmetries. The parity symmetry $P$ is related to the twist symmetry in the corresponding SFT \cite{Yang:2005rx}.
}

\be
X_{ia'b'}^{dab}=X_{ib'a'}^{dba},
\ee
which, using the properties of the Racah symbol in section \ref{6J&Racah}, gives
\be
\left(\frac{N(d,a,a')}{N(d,b,b')}\right)^2=\sqrt{\frac{g'_{a'}g'_b}{g'_ag'_{b'}}},
\ee
so that we can take
\be
N(x,y,z)=\sqrt{\Rac{x}{x}{1}{z}{z}{y}}=\left(\frac{g'_y}{g'_xg'_z}\right)^{\frac14}.
\ee
The $X$ coefficients take the explicit form
\bea\label{X-sym}
X_{ia'b'}^{dab}  &=&
\frac {n_i^{ab}}{n_i^{a'b'}} \FRac{d}{i}{a}{b}{a'}{b'} \frac{\sqrt{\FRac{1}{a'}{a}{d}{a}{d} \FRac{1}{b'}{b}{d}{b}{d}}}{ \FRac{1}{i}{a}{b}{a}{b}} \\
&=&\frac {n_i^{ab}}{n_i^{a'b'}} \,\left(g'_ag'_bg'_{a'}g'_{b'}\right)^{\frac14}\W6J{a}{a'}{d}{b'}{b}{i}.
\eea
Notice that, differently from the boundary structure constants, the defect coefficients  $X$ don't depend on the crossing symmetry properties of the conformal blocks, since
only the Racah symbols are involved in their definitions.


\subsubsection{Fusion of open string defects}
\label{sec:fusion of open string defects}

Let us now consider the fusion of topological defects on general boundary fields.
To this end, we need to calculate the subsequent action of $\dd^c$ and $\dd^d$ on $\phi_i^{ab}$.
As we saw in \eqref{Xdef},
after the first action of $\dd^c$ there are multiple boundary conditions in general,
and it is natural to arrange the r.h.s. of  \eqref{Xdef} into a matrix regarding $a'$ and $b'$ as matrix indices.
That is
\begin{align}
\dd^c\phi_i^{ab}
=
\bordermatrix{%
&{\scriptstyle b'_1}
&{\dots}
&{\scriptstyle b'_n}  \cr
{\scriptstyle a'_1}
& X^{cab}_{ia'_1b'_1}\phi^{a'_1b'_1}_i&\dots&X^{cab}_{ia'_1b'_n}\phi^{a'_1b'_n}_i \cr
{\scriptstyle \vdots}
& \vdots & \dots & \vdots \cr
{\scriptstyle a'_m}
& X^{cab}_{ia'_mb'_1}\phi^{a'_mb'_1}_i&\dots&X^{cab}_{ia'_mb'_n}\phi^{a'_mb'_n}_i
},
\label{eq: Dphi}
\end{align}
where $a_i'\in c\times a$ and $b_j' \in c\times b$.
The number of labels $a_i'$ is given by the number of nonzero ${N_{ca}}^{i}$s,
and the number of labels $b_j'$ is that of ${N_{cb}}^j$s.
The right hand side is a $m\times n$ matrix, and $m$ and $n$ are given by
\be
m=\sum_{i}{N_{ca}}^{a_i'},
\qquad
n=\sum_{j}{N_{cb}}^{b_j'},
\ee
respectively.
We also write this equivalently as
$
(\dd^c \phi_i^{ab})^{a'b'}=X^{cab}_{ia'b'}\phi^{a'b'}_i.
$
Similarly, after the subsequent action of $\dd^d$, we have
\begin{align}
\dd^d\dd^c\phi_i^{ab}
=
\bordermatrix{%
&{\scriptstyle b'_1}
&{\dots}
&{\scriptstyle b'_n}  \cr
{\scriptstyle a'_1}
& M_{a'_1b'_1}&\dots&M_{a'_1b'_n} \cr
{\scriptstyle \vdots}
& \vdots & \dots & \vdots \cr
{\scriptstyle a'_m}
& M_{a'_mb'_1}&\dots&M_{a'_mb'_n}
},
\label{eq:DDphi}
\end{align}
where the submatrix $M_{a'_pb'_q}$ is given by
\begin{align}
M_{a'_pb'_q}
\equiv
\dd^d\left(X^{cab}_{ia_p'b_q'}\phi^{a_p'b_q'}_i\right)
&=
\bordermatrix{%
&{\scriptstyle b''_1}
&{\dots}
&{\scriptstyle b''_t}  \cr
{\scriptstyle a''_1}
& X^{cab}_{ia'_pb'_q}X^{da'_pb'_q}_{ia''_1b''_1}\phi^{a''_1b''_1}_i
          & \dots & X^{cab}_{ia'_pb'_q}X^{da'_pb'_q}_{ia''_1b''_t}\phi^{a''_1b''_t}_i  \cr
{\scriptstyle \vdots}
& \vdots & \dots & \vdots \cr
{\scriptstyle a_s''}
& X^{cab}_{ia'_pb'_q}X^{da'_pb'_q}_{ia''_sb''_1}\phi^{a''_sb''_1}_i  & \dots
          & X^{cab}_{ia'_pb'_q}X^{da'_pb'_q}_{ia''_sb''_t}\phi^{a''_sb''_t}_i
}.
\label{eq:M=XDphi}
\end{align}
This is a $s\times t$ matrix, and $s$ and $t$ are given by
\be
s=\sum_{i}{N_{d{a_p}'}}^{a_i''}, \qquad t=\sum_{j}{N_{db'_q}}^{b''_j},
\ee
and the size of the matrix $\dd^d\dd^c\phi_i^{ab}$ is given by
\be
\left(\sum_{a'_p\in c\times a}s\right)\times\left(\sum_{b'_q\in c\times b}t\right)
=
\left(\sum_{i,\, p} {N_{da_p'}}^{a''_i}{N_{ca}}^{a_p'}\right)\times\left( \sum_{j,\,q}{N_{db_q'}}^{b_j''}{N_{cb}}^{b_q'}\right).
\label{eq:size_of_matrix}
\ee
From \eqref{eq:DDphi} and \eqref{eq:M=XDphi} we see that to identify the position of components in
$(\dd^d\dd^c\phi_i^{ab})$,
we need to refer to both the intermediate boundary condition $(a'b')$ and the final boundary condition $(a''b'')$. We then introduce a composite label $\{a\,a'\,a''\}$ and express the above result as
\be
\left(\dd^d\dd^c \phi_i\right)^{\{a\,a'\,a''\}\{b\,b'\,b''\}} \equiv
\left(\dd^d\left(\dd^c \phi_i^{ab}\right)^{a'b'}\right)^{a''b''} =
 X_{ia'b'}^{cab} X_{ia''b''}^{da'b'} \phi^{a''b''}_i.
 \label{eq:a}
\ee

For the defect action on the bulk space we have $D^d D^c = \sum_e N_{dc}^{\;\;\; e} D^e$. For the action on the boundary operators we have to replace the ordinary sum by a direct sum, since different defects map to different Hilbert spaces
\be
\bigoplus_{e\in d\times c}
\left(\dd^e  \phi_i^{ab}\right)^{a''b''} =
\begin{pmatrix}
\widetilde M_{e_1}&&\\
&\ddots&\\
&&\widetilde M_{e_k}
\end{pmatrix},
\label{eq:direct_sumNDphi}
\ee
where
\be
\widetilde M_{e_j}
\equiv
\dd^{e_j}  \phi_i^{ab}
=
\begin{pmatrix}
X^{e_jab}_{ia''_1b''_1}\phi^{a''_1b''_1}_i&\dots&X^{e_jab}_{ia''_1b''_h}\phi^{a''_1b''_h}_i\\
\vdots&\dots&\vdots\\
X^{e_jab}_{ia''_g b''_1}\phi^{a''_g b''_1}_i&\dots&X^{e_jab}_{ia''_g b''_h}\phi^{a''_g b''_h}_i
\end{pmatrix}.
\ee
Now we introduce the labels $[e;\,a,\,a'']$ to represent
\be
\left(
\bigoplus_{e\in d\times c}
\dd^e  \phi_i\right)^{[e;\,a,\,a''][f;\,b,\,b'']}
\equiv
\left(
\dd^e  \phi_i^{ab}\right)^{a''b''}\delta_{ef}
=
X_{ia''b''}^{eab}\phi_i^{a''b''}\delta_{ef}.
\label{eq:b}
\ee
Clearly the two expressions \eqref{eq:a} and \eqref{eq:b} are different, but notice that they have the same dimensions thanks to the identity
\be
\sum_k {N_{ij}}^{k}{N_{kl}}^m=\sum_k {N_{im}}^k{N_{kl}}^{j}.
\ee
That is, as explained in \eqref{eq:size_of_matrix}, the number of the labels $\{a\,a'\,a''\}$ is given by
$\sum_{a''}\left(\sum_{a'}{N_{ac}}^{a'}{N_{a'd}}^{a''}\right)$,
 while the number of the labels $[e;\,a,\,a'']$ is given by
$\sum_{a''}\left(\sum_{e}{N_{cd}}^e{N_{ae}}^{a''}\right)$.
These two numbers are equal, as explained e.g. in \cite{Moore-Seiberg}.
Similarly, we also conclude that the number of the labels $\{b\,b'\,b''\}$ and that of the labels $[e;\,b,\,b'']$ are the same.

This suggests that there might be a similarity transformation linking the two matrices,
\be
\left(\dd^d\dd^c \phi_i\right)^{\{a\,a'\,a''\}\{b\,b'\,b''\}}
 =
\left[ {U_{dc}
 \left(\bigoplus_{e\in d\times c}\dd^e  \phi_i\right) U_{dc}^{-1} }
 \right]
 ^{\{a\,a'\,a''\}\{b\,b',\,b''\}},
 \label{eq:sim}
 \ee
where $U_{dc}$ is a real invertible matrix with matrix indices $\{a\,a'\,a''\}$ and $[e;\,{\tilde a},\,{\tilde a}'']$.
Substituting \eqref{eq:a} for $(\dd^d\dd^c \phi_i)$ and \eqref{eq:b} for $\left(\bigoplus_{e\in d\times c}\dd^e  \phi_i\right)$, this equation is expressed as
\be
X^{da'b'}_{ia''b''}X^{cab}_{ia'b'}
=
\sum_{[e;\,{\tilde a}\,{\tilde a}'']}\sum_{[f;\,{\tilde b},\,{\tilde b}'']}
U_{dc}^{\{a\,a'\,a''\}[e;\,\tilde a\,{\tilde a}'']}
X^{e\tilde a\tilde b}_{i{\tilde a}''{\tilde b}''}\delta^{ef}
 (U^{-1}_{dc})^{[f;\,{\tilde b},\,{\tilde b}'']\{b\,b'\,b''\}}.
 \label{eq:XX=UXU}
\ee
Since $X_{ia'b'}^{dab}$ is the Racah symbol  with some extra factors \eqref{eq: X=(N/N)F}, the equation \eqref{eq:XX=UXU} is again reminiscent of the pentagon identity \eqref{RacPentagon}.
In fact, we find that the following $U_{dc}$ is a solution
\be
 (U_{dc})^{\{a\,a'\,a''\}[e;\,{\tilde a},\,{\tilde a}'']}
 =
 \begin{cases}

 \displaystyle
 \frac{N(c,a,a')N(d,a',a'')}{M(d,e,c)N(e,a,a'')}\Rac{c}{a}{a'}{a''}{d}{e} \qquad
                                        &\text{$(a=\tilde a)$ and $(a''={\tilde a}'')$},\vspace{3mm}\\
 0\qquad
                                        &\text{$(a\ne \tilde a)$ or $(a''\ne{\tilde a}'')$},\vspace{1.5mm}
 \end{cases}
\label{eq:U_general}
\ee
%
%
 where the factor $N(x,y,z)$ is the same as that appearing in \eqref{eq: X=(N/N)F}, and $M(x,y,z)$ is a nonzero arbitrary real number.
To check \eqref{eq:XX=UXU}, notice that the  inverse matrix $U_{dc}^{-1}$ is given by
\be
(U_{dc}^{-1})^{[e;\,a,\,a'']\{a\,a'\,a''\}}
=
 \frac{M(d,e,c)N(e,a,a'')}{N(c,a,a')N(d,a',a'')}\Rac{a''}{a}{e}{c}{d}{a'},
\label{eq:U_inv_general}
\ee
as can be checked from the orthogonality relation \eqref{Racah-ortho}.
%
A natural choice for $M(x,y,z)$ is
\be
M(x,y,z)=N(x,y,z)=\sqrt{\FRac{y}{1}{z}{z}{x}{x}}=\left(\frac{g'_y}{g'_zg'_x}\right)^{\frac14},\label {M(x,y,z)}
\ee
 which makes $U_{dc}$ an orthogonal matrix
\be
(U_{dc})^{\{a\,a'\,a''\}[e;\,{\tilde a},\,{\tilde a}'']}
=
(U_{dc}^{-1})^{[e;\,{\tilde a},\,{\tilde a}'']\{a\,a'\,a''\}}.
\label{eq:orthogonality}
\ee
This can be easily checked by substituting \eqref{M(x,y,z)} into \eqref{eq:U_general} and \eqref{eq:U_inv_general},  obtaining
\be
(U_{dc})^{\{a\,a'\,a''\}[e;\,a,\,a'']}
=\Rac{c}{a}{a'}{a''}{d}{e},
 \label{eq:U_explicit}
\ee
%
and, using  the tetrahedral symmetry of the Racah symbol
\be
(U_{dc}^{-1})^{[e;\,a,\,a'']\{a\,a'\,a''\}}
=\Rac{a''}{a}{e}{c}{d}{a'}=\Rac{c}{a}{a'}{a''}{d}{e}=(U_{dc})^{\{a\,a'\,a''\}[e;\,a,\,a'']}.
 \label{eq:U_inv_explicit}
\ee


A comment on the appearance of the matrix structures in the above discussion.
We have arranged the elements of the matrices in a particular way as in \eqref{eq:DDphi}, \eqref{eq:M=XDphi} and \eqref{eq:direct_sumNDphi} for illustrative purpose, but  we don't have to necessarily adhere to this ordering of  rows and columns.
Indeed, from \eqref{eq:U_general} we see that the mixing only occurs  when $(a,b)=(\tilde a,\tilde b)$ and $(a'',b'')=({\tilde a}'',{\tilde b}'')$, and with a suitable ordering of the columns and the rows we can bring $U_{dc}$ into a block-diagonal form.
The net mixing  is therefore given by
\begin{equation}
X^{da'b'}_{ia''b''}X^{cab}_{ia'b'}
=
\sum_{e,\,f}
U_{a'e}{\mbox{\scriptsize${\protect\begin{bmatrix} d \;\; c\\ a'' \;\; a\end{bmatrix}}$}}
X^{eab}_{ia''b''}\delta^{ef}
U_{b'f}{\mbox{\scriptsize${\protect\begin{bmatrix} d \;\; c\\ b'' \;\; b\end{bmatrix}}$}},
\end{equation}
where
\begin{equation}
U_{a'e}{\mbox{\scriptsize${\protect\begin{bmatrix} d \;\; c\\ a'' \;\; a\end{bmatrix}}$}} = U_{dc}^{\{a\,a'\,a''\}[e;\,a\,a'']}=\Rac{c}{d}{e}{a''}{a}{a'}.
\end{equation}
In section 4, we will explicitly work out this block-diagonalization in the example of the Ising model CFT.

\subsection{Geometric construction}
\label{subsec:geom_constr}

\begin{figure}[!t]
\centering
\includegraphics[]{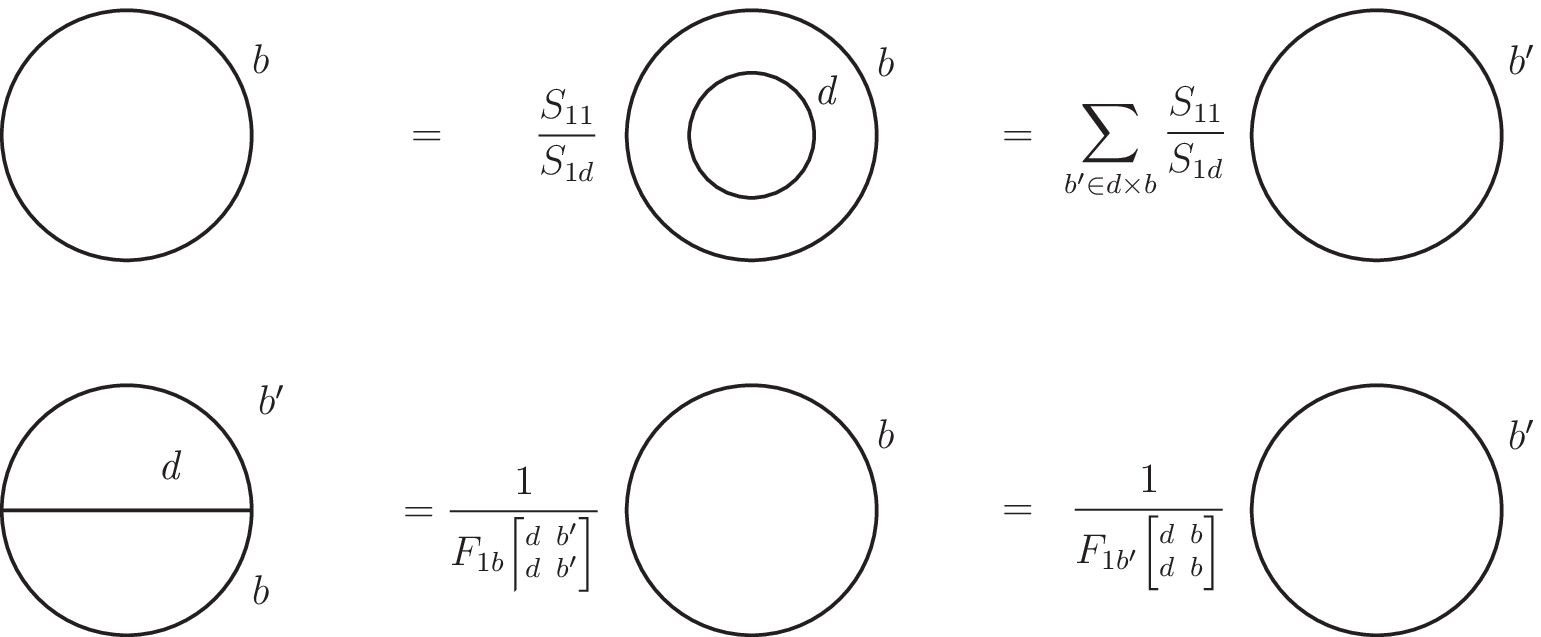}
\caption{Defects can be used to derive a relation between the $g$-functions of different boundary conditions. Defect attached to the boundary can be shrank in two different directions, producing consistent answer thanks to the identities for the fusion matrix $F$.}\label{fig:def-on-disk}
\end{figure}

Imagine a disk correlator with a number of bulk operator insertions. Placing a topological defect parallel to the boundary and sufficiently close to it, so that there are no bulk operators between the defect and the boundary,  one can smoothly deform the defect so that it fuses onto the boundary without affecting any correlator. From the bulk perspective, as we reviewed in Section \ref{sec:CFT-Defects}, these correlators can be viewed as overlaps of the new boundary state $D \kkett{B}$ with the vacuum excited by the vertex operators. Already by considering disk amplitudes without operator insertions, we find a number of interesting relations, illustrated in Figure \ref{fig:def-on-disk}
\be
g_b =  \sum_{b' \in d \times b}
\Fmat{1}{1}{d}{d}{d}{d}\, g_{b'}=\sum_{b' \in d \times b}\frac {g_{b'}}{g'_d},
\ee
from which it follows (assuming the existence of the identity boundary condition) that the normalized $g$ function of the defect is in fact the $g$ function of the corresponding boundary condition, normalized by the $g$-function of the identity boundary condition
\be
g'_d\equiv\frac {1}{\Fmat{1}{1}{d}{d}{d}{d} }=\frac{g_d}{g_1},\label{g-defect-boundary}
\ee
or, considering the ``sunset'' disk diagrams
\be
\frac{g_b}{\Fmat{1}{b}{d}{b'}{d}{b'}} = \frac{g_{b'}}{\Fmat{1}{b'}{d}{b}{d}{b}},
\ee
whose actual numerical value depends on the chosen gauge for the $F$ matrices.
We remind that consistency of defect network manipulations only imply that the involved $F$ matrices obey the pentagon identity and therefore the gauge for the $F$ matrices used for defect manipulations is not fixed. This has to be constrasted with the $F^{\rm blocks}$ matrices entering the boundary structure constants which, as we have reviewed in section \ref{Runkel}, imply a very specific gauge choice once the conformal blocks are canonically normalized.

 Now, what happens by attaching a defect  to a boundary, when there are boundary operators present? The conformal weight of such operators cannot change, so they must become new boundary operators in the same Virasoro representation, but interpolating between the new boundary conditions, and possibly modified by a new normalization constant.

\begin{figure}[!htbp]
\centering
\includegraphics[]{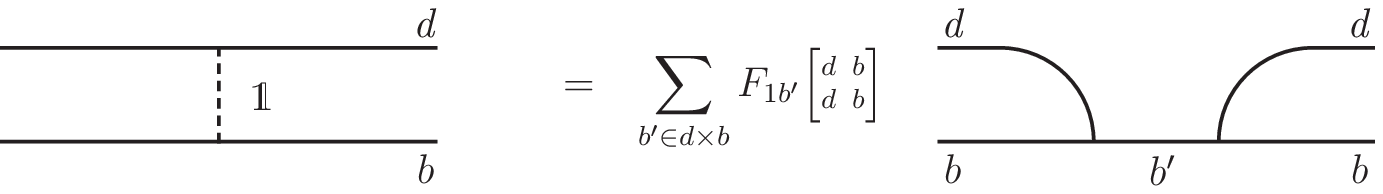}
\caption{Partial fusing of a defect onto a boundary.}
\label{Fig:defect_boundary_rules}
\end{figure}

To understand what happens to the boundary operators it is convenient to proceed in steps. First, imagine to partially fuse a defect $d$ on a boundary segment $b$ . Such a fusion brings in a nontrivial factor $\Fmat{1}{b'}{d}{b}{d}{b}$, see Figure \ref{Fig:defect_boundary_rules}.

\begin{figure}[]
\centering
\includegraphics[]{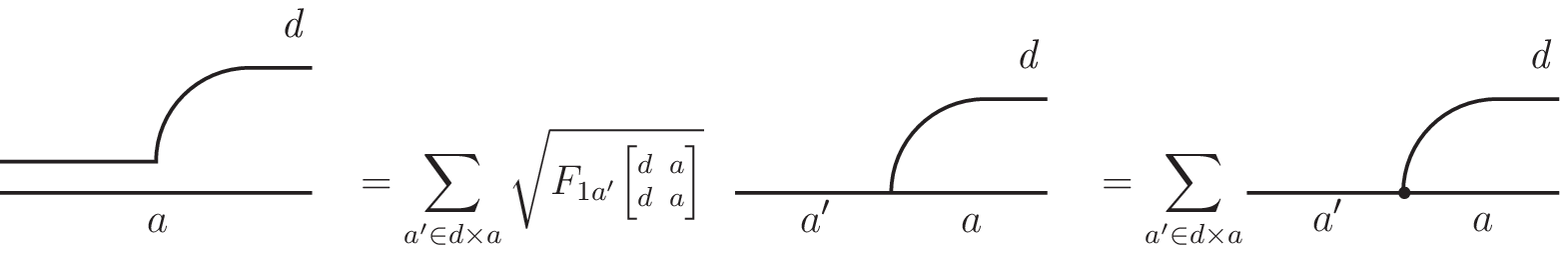}
\caption{A defect fusing onto a boundary. Every produced fundamental boundary condition is accompanied by a junction factor which is graphically represented as a thick dot.}
\label{fig:def-bound-junc}
\end{figure}

\begin{figure}[]
\centering
\includegraphics[]{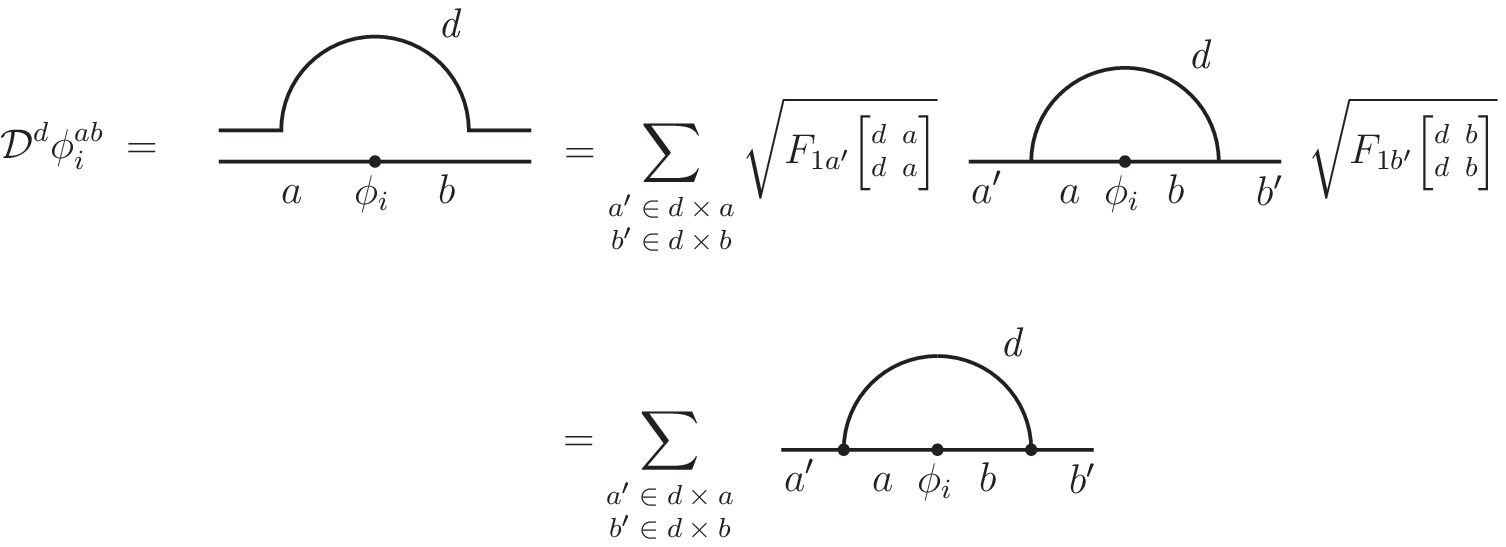}
\caption{The geometric description of a defect action on a boundary field. Notice the $\sqrt F$ factors at the junctions.}
\label{fig:open-defect}
\end{figure}

This factor can be assigned to the left and right junctions between the defect, the original boundary and the new boundary and it is natural to distribute it evenly between the two junctions. This implies that   when a $d$ defect fuses on a boundary $a$  to give a superposition of boundary conditions $a'=d\times a$,  the involved junction must be accompanied by a factor of $\sqrt{\Fmat{1}{a'}{d}{a}{d}{a}}$. This factor will be represented by a boldface dot at the junction, see Figure \ref{fig:def-bound-junc}.
%
The action of a defect on an open string state is explicitly defined in Figure \ref{fig:open-defect}

\begin{figure}[!htbp]
\centering
\includegraphics[]{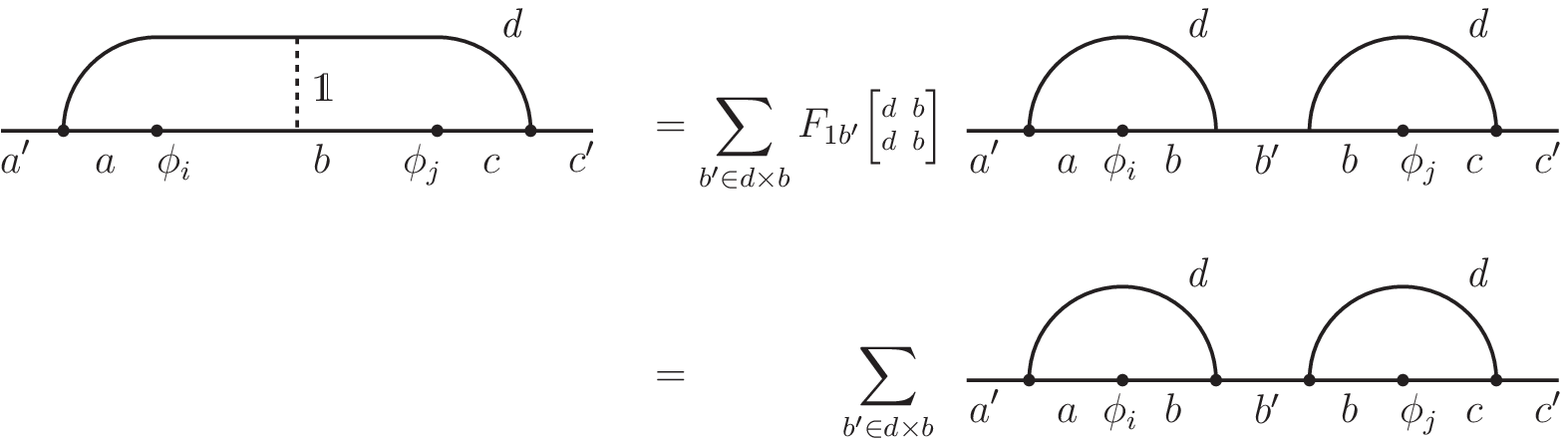}
\caption{Defect distributivity. In the second line the factor $\protect\Fmat{1}{b'}{d}{b}{d}{b}$ has been absorbed into the two $c$-number insertions at the junction points denoted by thick dots.}
\label{Fig:defect_distrib}
\end{figure}

Once the above geometrical defintion is given, defect distributivity
\be
\dd^d \left( \phi_i^{ab}(x) \phi_j^{bc}(y)\right) = \left( \dd^d  \phi_i^{ab}(x) \right) \left( \dd^d  \phi_j^{bc}(y)\right),
\ee
follows rather naturally. Since the defect can be partially fused onto the boundary as in Figure \ref{Fig:defect_distrib} (using the rule in Figure \ref{Fig:defect_boundary_rules}), the only issue one has to take care of is the normalization factor $\Fmat{1}{b'}{d}{b}{d}{b}$ which is accounted by the non-trivial normalization of the junctions.


\subsubsection{Defect action on a boundary field from network manipulations}
\label{sec:geomX}

\begin{figure}[!htbp]
\centering
\includegraphics[]{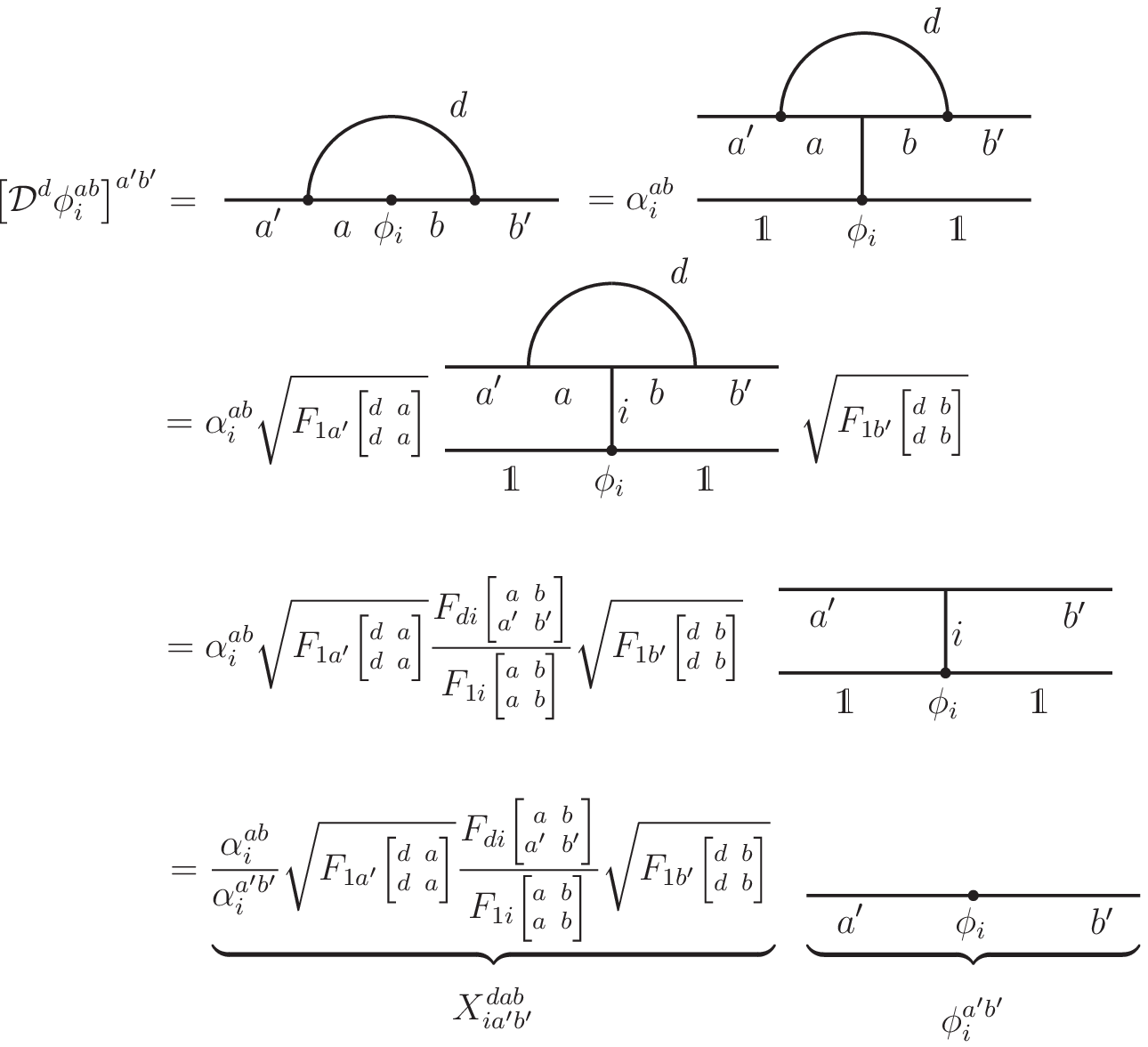}
\caption{Defect network manipulations determining the defect coefficients $X^{dab}_{ia'b'}$. Notice that the boundary insertion $i$ is traded for a defect ending on the boundary. Along such a defect
an $ab$-bubble is collapsed, after $F$-crossing on the original defect line $d$. The two junctions at which the $d$ defect joins the boundary corresponds to square roots of $F$ matrix elements.}
\label{Fig:def-coeff-geom}
\end{figure}

It remains to compute the explicit $X$ coefficients of the defect action
\be
\dd^d \phi_i^{ab}=\sum_{a'\in d\times a}\sum_{b'\in d\times b}X_{ia'b'}^{dab}\,\phi_i^{a'b'}.
\ee
In order to make use of the needed defect network manipulations, we uplift the boundary conditions and the boundary insertions into a defect network with a line
carrying the $i$ representation and ending on a chiral defect-ending  field, placed at the boundary with identity boundary conditions. This is explained in detail in appendix \ref{app:TFT}.  In our setting this move is essentially equivalent to a corresponding three-dimensional manipulation in the topological field theory description of defects in RCFT \cite{TFT},  see in particular  \cite{{Runkel:T-systems}}, but it has its own two-dimensional description given in \ref{app:TFT}. After this topological move, the defect coefficient $X$ can be easily computed as in Figure \ref{Fig:def-coeff-geom}.
The $\alpha$ coefficients explicitly depend on the chosen normalization for boundary fields (with the convention that $n_i^{ab}=$1 for the canonical choice (\ref{eq:C-canonical})  )and on the chosen gauge for defect networks, see (\ref{alphaTFT}), giving in total
\bea
X_{ia'b'}^{dab}&=&\frac{n_i^{ab}}{n_i^{a'b'}} \sqrt{\frac{\gamma(i,a',b')}{\gamma(i,a,b)}}\;\sqrt{\Fmat{1}{a'}{d}{a}{d}{a}}\frac{\Fmat{d}{i}{a}{b}{a'}{b'}}{\Fmat{1}{i}{a}{b}{a}{b}}\sqrt{\Fmat{1}{b'}{d}{b}{d}{b}}\nonumber\\
&=&\frac{n_i^{ab}}{n_i^{a'b'}} \sqrt{\FRac{1}{a'}{d}{a}{d}{a}}\frac{\FRac{d}{i}{a}{b}{a'}{b'}}{\FRac{1}{i}{a}{b}{a}{b}}\sqrt{\FRac{1}{b'}{d}{b}{d}{b}}\nonumber\\
&=&\frac{n_i^{ab}}{n_i^{a'b'}}\left(g'_ag'_bg'_{a'}g'_{b'}\right)^{\frac14}\,\W6J{b'}{a'}{i}{a}{b}{d}.
\eea
Notice in particular that the gauge dependent factors in the $\alpha$'s (\ref{alphaTFT}) conspire together with the gauge dependence of the defect manipulation in Figure \ref{Fig:def-coeff-geom} to give
an overall gauge invariant result which only depends on the normalization choice for the boundary fields, as it should: acting a defect on a boundary field doesn't depend on the defect's gauge.

\subsubsection{Defect fusion from  network manipulations}

\begin{figure}[!htbp]
\centering
\includegraphics[]{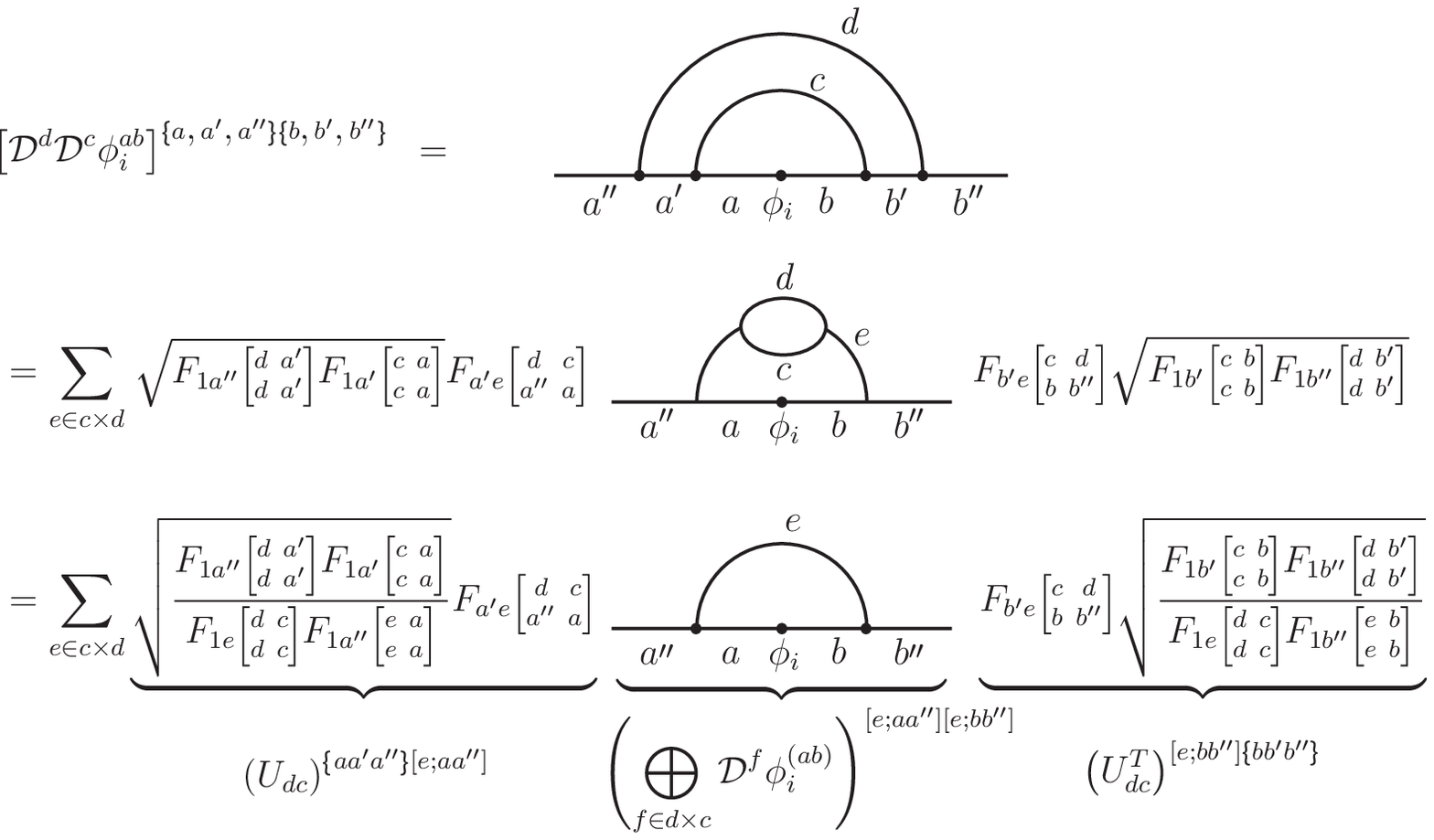}
\caption{Defect network manipulations determining the fusion rules of open string defects.}
\label{Fig:def-fus-geom}
\end{figure}

Let us now see how we can use similar manipulations to reduce the composition of two defects $c$ and, subsequently, $d$   to a direct sum of defects $e$, in the fusion of $c$ and $d$.
We start with a general boundary field $\Phi$ and act the defects on it
\bea
\dd^d\dd^c\Phi&=&\sum_{a,b}\sum_i\sum_{a',b'}\sum_{a'',b''}\left(\dd^d\left(\dd^c\phi_i^{ab}\right)^{a'b'}\right)^{a''b''}\nonumber\\
&=&\sum_{\{a,a',a''\}}\sum_{\{b,b',b''\}}\left(\dd^d\dd^c\Phi\right)^{\{a,a',a''\},\{b,b',b''\}}.
\eea
Then we can perform the manipulations shown in Figure \ref{Fig:def-fus-geom}

\be
\left(U_{dc}\right)^{\{aa'a''\}[e;a,a'']}=\sqrt{\frac{\Fmat{1}{a''}{d}{a'}{d}{a'}\Fmat{1}{a'}{c}{a}{c}{a}}{\Fmat{1}{a''}{e}{a}{e}{a}\Fmat{1}{e}{d}{c}{d}{c}}} \,\Fmat{a'}{e}{d}{c}{a''}{a}=\Rac{c}{d}{e}{a''}{a}{a'},
\ee
which coincides with (\ref{eq:U_explicit})
Notice that in the geometric approach there naturally appears the transpose of the $U$ matrix on the right
\bea\label{DDPhi}
\dd^d\dd^c\Phi=U_{dc}\left(\bigoplus_{e\in d\times c} \dd^e \Phi\right)U_{dc}^T.
\eea

\begin{figure}[!ht]
\centering
\includegraphics[]{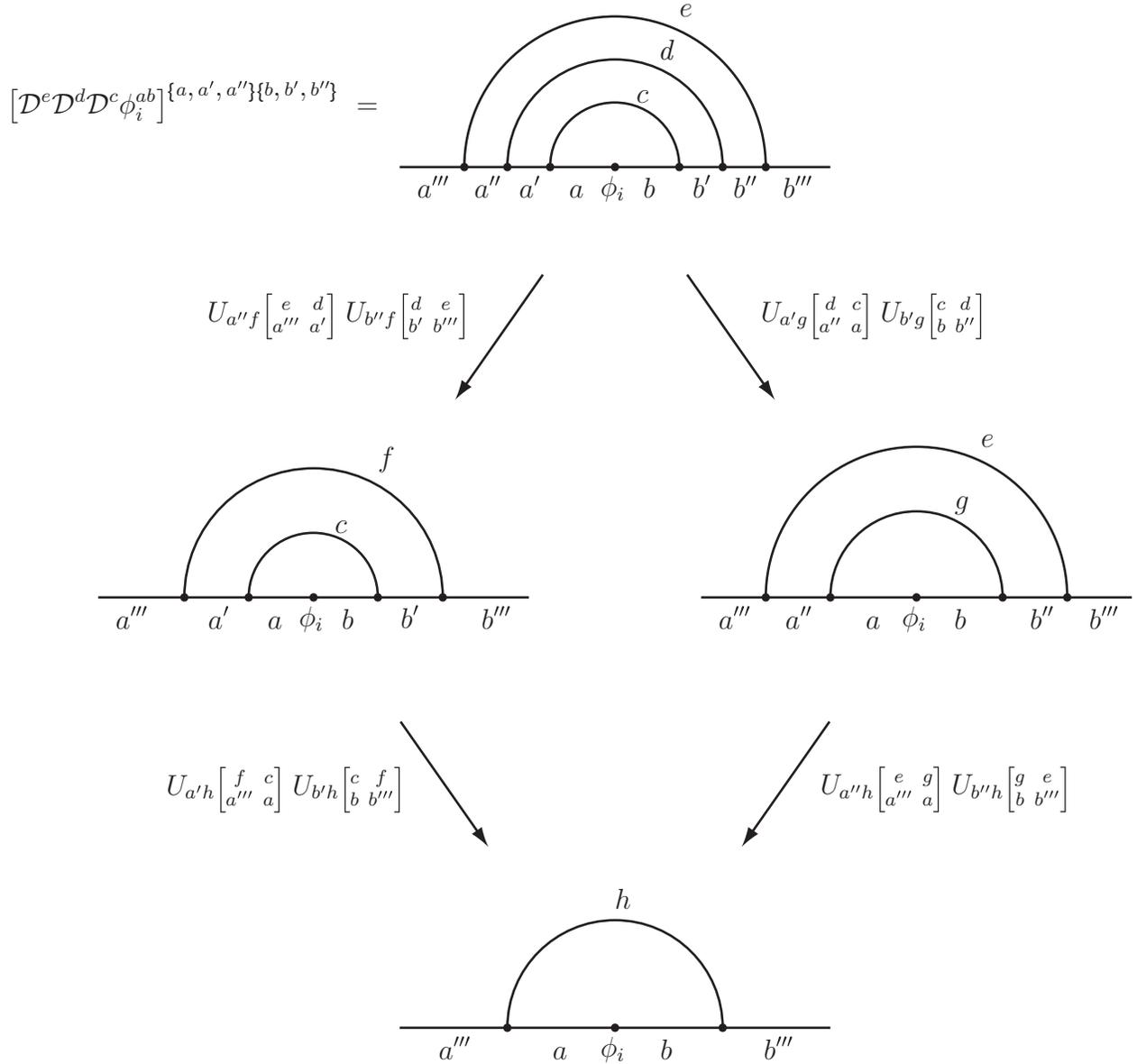}
\caption{Proof of associativity. Fusing together three defects attached to a boundary in two different ways results in a consistency condition (\ref{as-rel-to_prove}) for the $U$ matrix. The resulting condition follows from the pentagon identity for the Racah symbol (\ref{RacPentagon}). }
\label{Fig:assoc}
\end{figure}

Before closing this section, let us comment on one particularly surprising aspect of the relation (\ref{DDPhi}). It is a bit reminiscent of some sort of generalized non-abelian projective representation\footnote{Instead of the usual representation on vectors up to a phase, this behaves as a representation on matrices up to a similarity transformation.} of the closed string defect algebra $D^d D^c = \sum_{e\in d\times c} D^e$ and one may ask whether the corresponding 2-cocycle condition is satisfied. This is equivalent to the condition of associativity of the defect algebra on the open string fields
\be
\left(\dd^e \, \dd^d \right) \dd^c = \dd^e \left(\dd^d  \, \dd^c \right).
\ee
To prove associativity, following the steps in Figure \ref{Fig:assoc}, one has to show that
\be\label{as-rel-to_prove}
\sum_f \Umat{a''}{f}{e}{d}{a'''}{a'} \; \Umat{b''}{f}{d}{e}{b'}{b'''} \Umat{a'}{h}{f}{c}{a'''}{a} \; \Umat{b'}{h}{c}{f}{b}{b'''}  = \sum_g \Umat{a'}{g}{d}{c}{a''}{a} \; \Umat{b'}{g}{c}{d}{b}{b''} \Umat{a''}{h}{e}{g}{a'''}{a} \; \Umat{b''}{h}{g}{e}{b}{b'''}.
\ee
To see that, let us start by writing the pentagon identity for the Racah symbols (\ref{RacPentagon}) in a more convenient form
\be\label{RacPentagon2}
\Rac{p}{e}{r}{d}{c}{q} \Rac{r}{e}{p}{a}{b}{t}  = \sum_s \Rac{c}{b}{s}{a}{q}{p} \Rac{d}{s}{t}{a}{e}{q} \Rac{d}{t}{s}{b}{c}{r}.
\ee
Using this identity we can express the product of the two factors on the left hand side depending on the $a$-type labels (any of the labels $a,a',a''$ and $a'''$)
\bea
\Umat{a''}{f}{e}{d}{a'''}{a'} \; \Umat{a'}{h}{f}{c}{a'''}{a} &=& \Rac{d}{e}{f}{a'''}{a'}{a''} \Rac{c}{f}{h}{a'''}{a}{a'} = \Rac{a'}{a'''}{f}{e}{d}{a''} \Rac{f}{a'''}{a'}{a}{c}{h}
\nonumber\\
&=& \sum_g \Rac{d}{c}{g}{a}{a''}{a'} \Rac{e}{g}{h}{a}{a'''}{a''} \Rac{e}{h}{g}{c}{d}{f}.
\eea
This last expression upon multiplication by the remaining $b$-type terms from the left hand side of (\ref{as-rel-to_prove}) can now easily by summed over label $f$ using the relation (\ref{RacPentagon2}) and one ends up precisely with the right hand side of (\ref{as-rel-to_prove}). This concludes the proof of associativity.


\section{Topological defects in open string field theory}
\label{Sec:OSFT}
\setcounter{equation}{0}



In this section we would like to study how topological defects act on OSFT solutions.
 As we have already stated, OSFT provides a new way to explore the possible conformal boundary conditions of a bulk CFT, by solving the equations of motion.
Let us then briefly review how can we use OSFT to analyze
BCFT's with central charge $c$ different from $26$ \cite{KMS, KRS}.
The open string star algebra is factorized in the Hilbert space of the $c=-26$ ghosts' BCFT, with standard boundary conditions, and in the matter $c=26$ BCFT, whose boundary conditions can be generic. In our application of OSFT to BCFT we will further assume that the matter BCFT is the tensor product of a BCFT$_c$ (whose properties we wish to study) times a compensating ``spectator'' BCFT$_{26-c}$
\be
{\rm BCFT}_{\rm total}={\rm BCFT}_{c}\otimes {\rm BCFT}_{26-c}\otimes{\rm BCFT}_{\rm ghost}.\label{tensor-cft}
\ee
In the total corresponding star algebra we restrict to the subalgebra where only descendants of the identity in the spectator sector are excited. Then we can search for classical solutions with the most general ansatz at ghost number one
\begin{align}
\Psi&=\sum_{a,b}\sum_{i}\sum_{I,J,K}a^{i(ab)}_{IJK}\,
L_{-I}^{c}\ket{\phi_{i}^{ab}}\otimes
L_{-J}^R \ket{0}\otimes
L^{gh}_{-K}\,c_1\ket{0},
\label{eq:Psi_expansion}
\end{align}
where
$L_{-I}^c=L_{-i_n}L_{-i_{n-1}}\cdots L_{-i_1}$ for $0\leq i_1\leq i_2\leq \cdots i_n$,
and $J,K$ are defined in the same way.
The label $i$ runs over all Virasoro representations
 of $\text{BCFT}_c$
that are allowed by the pair of boundary conditions
$a$, $b$  (with, possibly, non-trivial multiplicities).
In the case of diagonal
 minimal models  we have that $i\in a\times b$. If some fundamental boundary
condition appears multiple times (trivial multiplicities), then the coefficients $a_{IJK}^{i(ab)}$ are
matrices in the degeneracy labels (Chan-Paton factors).

A possible approach to search for new boundary conditions in the CFT$_c$ factor (\ref{tensor-cft}) is to solve the OSFT equation of motion with the ansatz  (\ref{eq:Psi_expansion}). In particular the boundary state corresponding to these
new boundary conditions can be computed from OSFT gauge invariant observables \cite{KMS}.

In the previous section we have defined the action of topological defects on generic boundary operators including those which change the boundary conditions. This then naturally defines an action of a topological defect on open string fields
\be
\dd^d\Psi=\sum_{a,b}\sum_{i} \sum_{I,J,K} a^{i(ab)}_{IJK}
L_{-I}^{c}  \left(\dd^d \ket{\phi_{i}^{ab}}\right)\otimes
L_{-J}^R \ket{0}\otimes
L^{gh}_{-K}\,c_1\ket{0},
\ee
because
\be
[L_n^{\mathrm{matter}},\dd]=0.
\label{eq:topological}
\ee
It  follows that, noticing that  $[b_n,\dd]=0$ and $[c_n,\dd]=0$,
\bea
\left[ Q, \dd \right] &=& 0\label{Q-def} .
\eea
It is also not difficult to establish that
\bea
\dd^d (\phi * \chi)  &=& (\dd^d \phi) * (\dd^d\chi) \qquad \forall \phi, \chi.\label{distrib-star}
\eea
Indeed, assuming the BFCT of interest to us is unitary and therefore its total Hilbert space is spanned by the direct sum of the Verma modules over the primaries, this condition
just follows from the compatibility with the OPE  (\ref{distrib-prim}) and the conservation laws of the star product \cite{RZ,wedge, Schnabl}.
Concretely, given two descendants string fields $L_{-I} \phi_i^{ab}(0) \ket{0}$ and $L_{-J}\phi_j^{bc}(0) \ket{0}$, we can express their star product schematically as
\be\label{cons-law}
L_{-I} \phi_i^{ab}(0) \ket{0} * L_{-J}\phi_j^{bc}(0) \ket{0} = \sum_K V_{IJ}^{\;\;\;K}(h_i,h_j) L_{-K} e^{\sum v_k L_{-k}} \phi_i^{ab}(x) \phi_j^{bc}(y) \ket{0},
\ee
where the coefficients $V_{IJ}^{\;\;\;K}$, $v_k$ as well as the insertion points $x$ and $y$ are explicitly known or calculable. Acting with $\dd^d$, one can bring it through all the Virasoro generators, use formula (\ref{distrib-prim}) and use again formula (\ref{cons-law}) to reassemble the left hand side.

Therefore open topological defects map solutions to solutions in OSFT
\be
Q\Psi+\Psi*\Psi=0\quad\rightarrow\quad Q(\dd\Psi)+(\dd\Psi)*(\dd\Psi)=0.
\ee
The main issue is now the physical interpretation of these new solutions.

\subsection{Defect action on string field theory solutions}


In this subsection we will derive the following key result: given a solution $\Psi_{X\to Y}$ which shifts the open string background from BCFT$_X$ to BCFT$_Y$, we will show that the solution $\dd\Psi_{X\to Y}$ shifts
 from BCFT$_{DX}$ to BCFT$_{DY}$, where the subscript denotes the boundary conditions obtained
by fusing the defect $D$ on the $X$ and $Y$ boundary conditions respectively. In formulas
\be\label{def-sol}
{\cal D} \Psi_{X \to Y} = \Psi_{{ D} X \to { D} Y}.
\ee
To do so we will evaluate the OSFT observables of $\dd\Psi_{X\to Y}$ and show that they fully agree with the observables of the r.h.s of (\ref{def-sol}).
In particular we will show that the boundary state of $\dd\Psi_{X\to Y}$ is just the result of the defect action on the boundary state of $\Psi_{X\to Y}$
\be
\kkett{B_{\dd \Psi}}=D\kkett{B_\Psi}.
\ee
Notice that in OSFT there   appears a natural interplay between the open string defect operator $\dd$ and its closed string counterpart $D$.

\subsubsection{Computation of $S\left[\dd \Psi \right]$  }

%


Before studying the full boundary state, it is instructive to look  at
the simpler case of  the OSFT action.
Carrying out the
summation on $I,J$ and $K$ in \eqref{eq:Psi_expansion}  we can write the string field as
\be
 \Psi=\sum_{a,b}\sum_{i}\Psi_i^{ab}
 \label{eq:sum_ab_Psi^ab},
 \ee
where the label $i$ and $ab$
denote the Virasoro representation and boundary conditions
of $\text{BCFT}_c$.
 In order to treat both the quadratic and cubic terms in the action (\ref{action})  at the same time, it is useful to prove the more general statement
 \be
\Tr[(\dd^d\Psi_1)*\cdots*(\dd^d\Psi_n)]=\frac{g_d}{g_1}\Tr[\Psi_1*\cdots*\Psi_n],\label{action1}
\ee
where Tr is the usual Witten integral, combined with trace over Chan-Paton factors.
By defect distributivity
\be
(\dd^d\Psi_1)*\cdots*(\dd^d\Psi_n)=\dd^d(\Psi_1*\cdots*\Psi_n),
\ee
 it is enough to show that
 \be
 \Tr[\dd^d\Psi]=\frac {g_d}{g_1}\Tr[\Psi].\label{action2}
\ee
Using the results of previous sub-sections the l.h.s. can be easily evaluated as
 \bea
\Tr[\dd^d\Psi]&=&\sum_a\;\sum_{i}\;\sum_{a'\in d\times a} X^{daa}_{ia'a'}\Tr[\Psi_{i}^{\{a,a'\},\{a,a'\}}]\nonumber\\
&=&\sum_{a}\,\sum_{i}\delta_{i\mathds{1}}\sum_{a'\in d\times a} \Tr[\Psi_{\mathds{1}}^{\{a,a'\},\{a,a'\}}],
\eea
since only the boundary Virasoro representation $i=1$ can contribute to the Witten integral. Moreover, on general grounds, $X^{daa}_{1a'a'}=1$ since a defect always maps the identity to the identity,  as can be easily checked in the explicit example (\ref{eq: X=(N/N)F}). To continue we simply notice that
\be
\Tr[\Psi_{\mathds{1}}^{\{a,a'\},\{a,a'\}}]=\frac{g_{a'}}{g_a}\Tr[\Psi_{\mathds{1}}^{aa}],
\ee
since the two traces involve the same operator algebra and only differ in the normalization of the vacuum amplitudes $g_a\equiv\aver1_{\rm disk}^{(a)}$, and similarly for $a'$.
We therefore have
\bea
\Tr[\dd^d\Psi]&=&\sum_{a}\Tr[\Psi_{\mathds{1}}^{aa}]\left(\sum_{a'\in d\times a}\frac{g_{a'}}{g_a}\right).
\eea
Using the Pasquier algebra for the $g$-functions, see equation (\ref{VerlindeF}) and (\ref{g-defect-boundary})
\be
\sum_{a'\in d\times a}g_{a'}=\frac{g_d g_a}{g_1},
\ee
concludes the proof of  (\ref{action1}).

In the next section we will see an alternative  geometric approach making use of defect-network manipulations, Figure \ref{Fig:def-Ellwood}.

What we  derived holds at the level of the Witten integral therefore, remembering that the defect trivially commutes with the BRST charge, it is immediate to see
that for any open string field $\Psi$ we have
\be
S_{\rm OSFT}[\dd^d\Psi]=\frac{g_d}{g_1}S_{\rm OSFT}[\Psi].
\ee
Suppose now we have a solution $\Psi_{X\to Y}$ which describes BCFT$_Y$ as a state in BCFT$_X$. This means in particular we have a solution whose action is given by
\be
S_{\rm OSFT}[\Psi_{X\to Y}]=\frac1{2\pi^2}\left(g_{_X}-g_{_Y}\right).
\ee
It follows that the defect-acted solution will have an action given by
\bea
S_{\rm OSFT}[\dd^d\Psi_{X\to Y}]=\frac{g_d}{g_1}S_{\rm OSFT}[\Psi_{X\to Y}]
=\frac1{2\pi^2}\left(\frac{g_d\, g_{_X}}{g_1}-\frac{g_d\, g_{_Y}}{g_1}\right).
\eea
The difference entering into the above equation is nothing but the difference in the identity-coefficients ($g$-functions) of the two boundary states obtained by acting the closed string defect $D^d$ on the source and target boundary states $\kkett{B_X}$, $\kkett{B_Y}$, connected by the solution $\Psi_{X\to Y}$
\bea
D^d\kkett{B_X}&=&D^d\left(g_{_X}\kett1+\cdots\right)=\frac{g_d\, g_{_X}}{g_1}\kett1+\cdots\\
D^d\kkett{B_Y}&=&D^d\left(g_{_Y}\kett1+\cdots\right)=\frac{g_d\, g_{_Y}}{g_1}\kett1+\cdots.
\eea
This is the first non-trivial check that (\ref{def-sol}) indeed holds.


\subsubsection{Computation of the Ellwood invariant}

\begin{figure}[!htbp]
\centering
\includegraphics{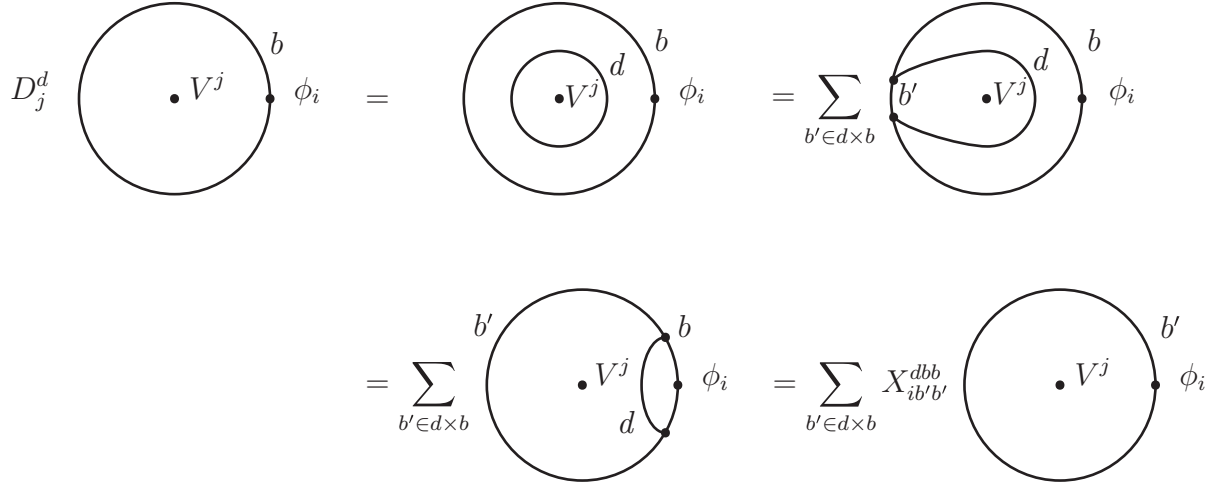}
\caption{Two equivalent ways of computing a bulk-boundary correlator in presence of a closed string defect $d$.}
\label{Fig:def-bulk-bound}
\end{figure}

Let us now see how the Ellwood invariant is affected by the defect action. To start with, it is useful to derive an identity involving the bulk-boundary structure constants $^{(a)}\!B_j^i$,
the $g$-functions and the open and closed defect coefficients. The required identity is obtained by computing a disk amplitude with a spinless bulk field $V^{j}(z,\bar z)$, a boundary field $\phi^{aa}_i(x)$ and a defect
loop $d$ encircling the bulk field. See Figure~\ref{Fig:def-bulk-bound}.

The correlator can be computed by acting the closed string defect on the closed string field $V^j$, or by
partially attaching the defect to the boundary, producing an open string defect acting on the boundary field $\phi_i(x)$. Calling, in generality,  $D^d_i$ the defect
coefficient, i.e. $D=\sum_i D^d_i P_i$, where $P_i$ projects in the $i$-th Verma module in the bulk, we get the identity
\be
D^d_j\,\,^{(a)}\!B_j^i\,g_{a}=\sum_{a'\in d\times a}\,X_{ia'a'}^{daa}\,^{(a')}\!B_j^i\,g_{a'},
\ee
which in the case of diagonal minimal models reads\footnote{Boundary fields are here canonically normalized (\ref{eq:C-canonical}). In the special case where the boundary field carries the identity representation it coincides with (4.42) of \cite{BPPZ}. In this case we have $^{(a)}\!B_j^{1}\,g_{a} =\frac{S_{aj}}{\sqrt{S_{1j}}}$ and analogously
for the new boundary conditions $a'$, and so the relation reduces to the Verlinde formula $$\frac{S_{dj}}{S_{1j}}S_{aj} =\sum_{a'\in d\times a}S_{a'j}.$$}
\be
\frac{S_{dj}}{S_{1j}}\,^{(a)}\!B_j^i\,g_{a}=\sum_{a'\in d\times a}\,\FRac{a}{a'}{a}{d}{i}{a'}\,^{(a')}\!B_j^i\,g_{a'}.
\ee

\begin{figure}
\centering
\includegraphics{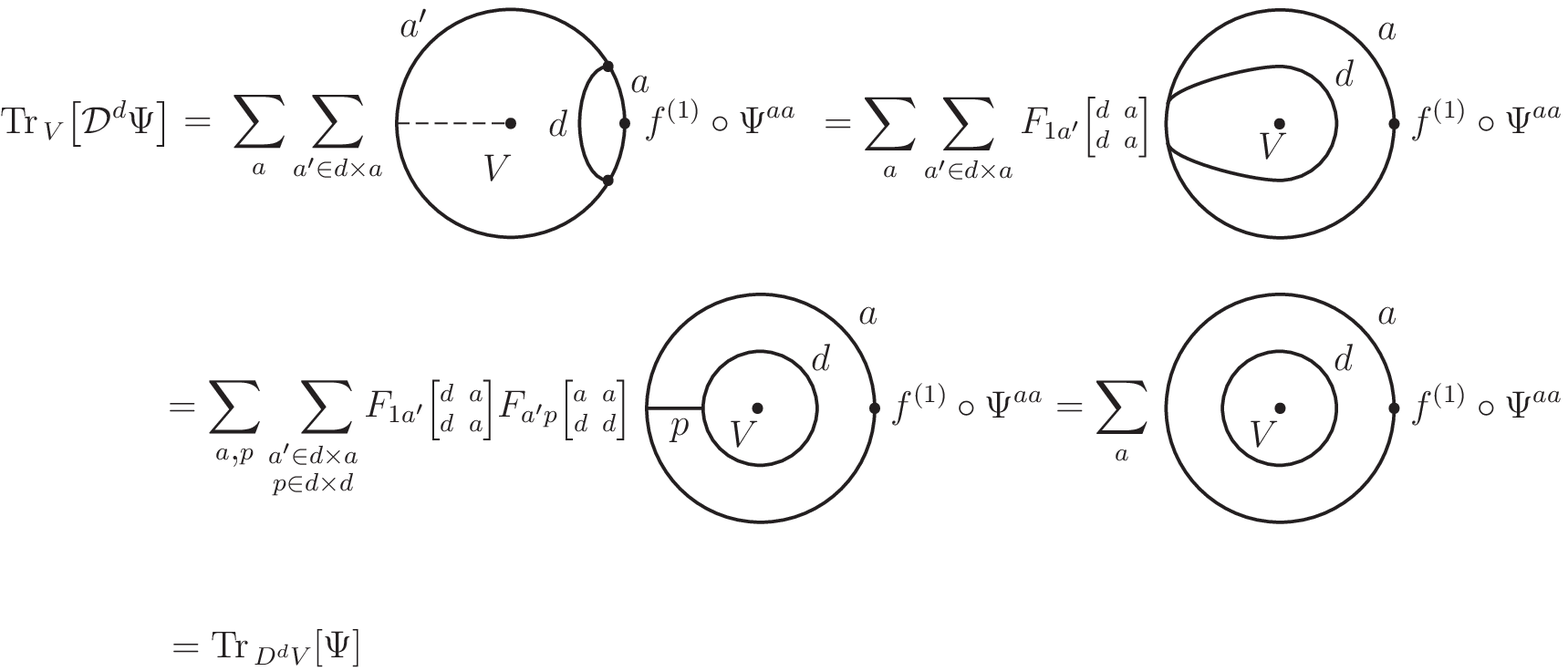}
\caption{Defect networks manipulations for the Ellwood invariant of a defect-acted open string-field. The dashed line corresponds to the identification of
the left and right part of the open string, via the identity conformal map $f^{(1)}(z)=\left(\frac{1+iz}{1-iz}\right)^2$. Notice  that thanks to the junctions normalizations and the consequent $F$-matrix orthogonality relation, only the identity
defect $p=1$ stretches between the boundary and the defect $d$, making $d$ a genuine closed string defect. When $V=1$ this also gives a geometric proof of (\ref{action2}).}
\label{Fig:def-Ellwood}
\end{figure}

Thanks to this relation it is easy to algebraically show that
\bea
\Tr_{V^j}[\dd^d\Psi]&=&\sum_{a}\;\sum_{i}\;\sum_{a'\in d\times a}X^{daa}_{ia'a'}\Tr_{V^j}[\Psi_i^{\{a,a'\},\{a,a'\}}]\nonumber\\
&=&\sum_{a}\;\sum_{i}\;\sum_{a'\in d\times a}\FRac{a}{a'}{a}{d}{i}{a'}\,\frac{^{(a')}\!B_j^{i}\,g_{a'}}{^{(a)}\!B_j^{i}\,g_{a}}\,
\Tr_{V^j}[\Psi_i^{\{a,a'\},\{a,a'\}}]\nonumber\\
&=&\sum_{a}\;\sum_{i}\frac{S_{dj}}{S_{1j}}\Tr_{V^j}[\Psi_i^{aa}]=\sum_{a,\alpha}\;\sum_{i\in a\times a}\Tr_{D^dV^j}[\Psi_i^{aa}]\nonumber\\
&=&\Tr_{D^d V^j}[\Psi],
\eea
where in the second line we have used that the two involved open/closed couplings (carrying the same Virasoro labels but different boundary conditions) only differ by the overall bulk-boundary structure-constant $B$
and the $g$-functions, due to the identical operator algebra involved in their calculation.
Again, this can be shown geometrically manipulating  defect networks, as shown in Figure \ref{Fig:def-Ellwood}, and the result is correctly independent on the gauge used for defect manipulations.

\subsubsection{KMS and KOZ boundary state}

Having obtained in generality how an Ellwood invariant is affected by the action of an open string defect, let us see how the OSFT boundary state constructed in \cite{KMS} (KMS from now on)
behaves under the defect action. The KMS approach gives a simple recipe to directly compute the coefficients of the matter Ishibashi states of the boundary state associated to a given
solution $\Psi$ in terms of a minimal generalization of the Ellwood invariant. In the setting of this paper, where we are assuming that the matter BCFT is the tensor
product of a diagonal rational BCFT of central charge $c$ and a ``spectator'' sector of central charge $(26-c)$, the required generalization of the Ellwood invariant is simply achieved by assuming
that the spectator sector contains a free boson (call it $Y$) with Dirichlet boundary conditions, see \cite{KRS, KMS} for further details. Then the KMS construction is usefully summarized as
\bea
\kkett{B_\Psi}^{\rm KMS}&=&\left(\sum_{j}n_\Psi^j\kett{V^j}\right)^{(c)}\otimes\kkett{B_0}^{(26-c)}\otimes\kkett{B_0}^{\rm ghost}\\
n_\Psi^j&=&(2\pi i)\Tr_{\tilde{\cal V}^j}[\Psi-\Psi_{\rm tv}]\\
{\tilde{\cal V}^j}(z,\bar z)&=&c\bar c\, V^j\, e^{2i\sqrt{1-h_j^2}Y}(z,\bar z).
\eea
Since we have shown that $\Tr_V[\dd \Psi]=\Tr_{DV}[\Psi]$ for any string field, the KMS construction applied to the solution $\dd\Psi$  will obviously give\footnote{A simple consequence of this is that acting with a topological defect on a tachyon vacuum solution, the new solution solution will still be the tachyon vacuum (the boundary state will still vanish), although expressed with the degrees of freedom of the new BCFT $D(X)$.}
\be
\kkett{B_{\dd\Psi}}^{\rm KMS}=D\kkett{B_\Psi}^{\rm KMS}.
\ee
At last let's also consider the other available construction of the boundary state in OSFT, given by Kiermaier, Okawa and Zwiebach (KOZ) \cite{KOZ}.
 KOZ geometrically construct
a BRST-invariant ghost number three closed string state obeying the level matching $b_0^{-}=L_0^{-}=0$. This closed string state is conjectured  to be BRST equivalent to the BCFT boundary
state and in fact the two coincide for many known analytic solutions. The main ingredient of this construction
is a choice of half-propagator strip  in the background of a classical solution $\Psi$, whose left edge and right edge are glued together, to form an annulus-like surface,
which is used to build the closed string state
\be
\kkett{B_\Psi}^{\rm KOZ}=e^{\frac{\pi^2}{s}(L_0+\bar L_0)}\,\oint_s{\rm Pexp}\left[-\int_0^sdt\left[{\cal L}_R(t)+\{ {\cal B}_R(t),\Psi\}\right]\right].
\ee
The various objects entering the above definition are defined in \cite{KOZ}, but for us it is sufficient to recall that the quantity ${\rm Pexp}[...]$ represents a half-propagator strip
of lenght $s$, in the background of the classical solution $\Psi$, and that the symbol $\oint_s$ identifies the initial left edge of the strip with the final right edge.
The internal boundary of this annulus-like surface (corresponding to the propagation of the open string midpoint) defines a closed string state, the KOZ boundary state.

\begin{figure}
\centering
\includegraphics[]{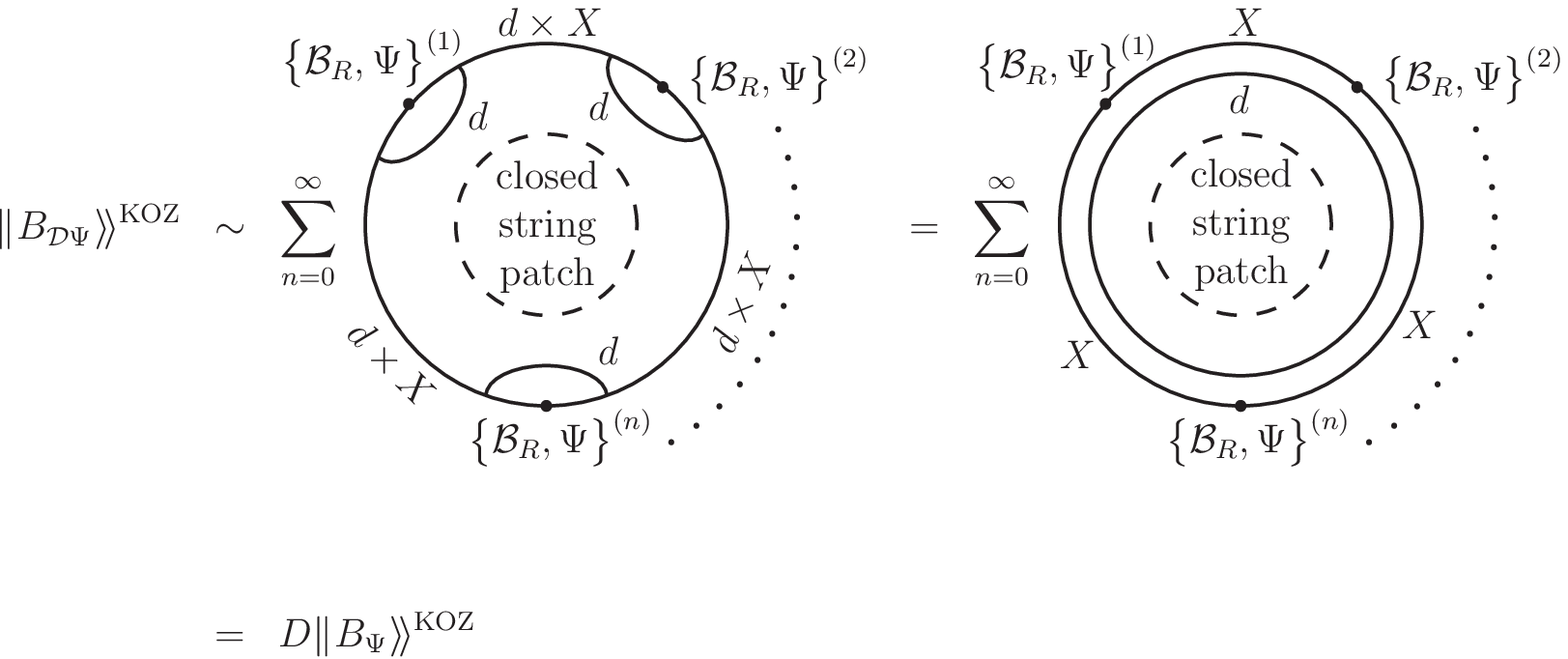}
\caption{Pictorial representation of the KOZ boundary state for a defect-acted solution (the boundary integrals in the ``Pexp'' operation, as well as other details, are not shown, as they are not important for our argument). Thanks to the defect distributivity, the various open string defects result in a single defect-loop encircling the closed string
coordinate patch.}
\label{Fig:KOZ-defect}
\end{figure}

If, instead of the original classical solution $\Psi$ on BCFT$_X$ we replace the defect-acted solution $\dd \Psi$ on BCFT$_{DX}$, then by defect distributivity on the star product (and the
obvious commutativity with the $b$-ghost insertions)
we will recover a closed string defect extending along the midpoint line, which is nothing but a closed-string  defect operator acting on the original KOZ boundary state,
see Figure \ref{Fig:KOZ-defect}
\bea
\kkett{B_{\dd\Psi}}^{\rm KOZ}&=&e^{\frac{\pi^2}{s}(L_0+\bar L_0)}\,\oint_s{\rm Pexp}\left[-\int_0^sdt\left[{\cal L}_R(t)+\{ {\cal B}_R(t),\dd\Psi\}\right]\right]\nonumber\\
&=&D\kkett{B_{\Psi}}^{\rm KOZ}.
\eea

%
%
%
%
%
%
%



\section{Ising OSFT example}
\setcounter{equation}{0}
\label{sec:Ising}

In this section we would like to illustrate our general findings on the concrete example of open string field theory for the Ising model \cite{KRS}.
This is the simplest unitary diagonal minimal model and has  $c=\frac{1}{2}$ . It has three irreducible Virasoro representations, denoted as $\mathds{1}$, $\eps$ and $\sigma$ with well-known fusion rules
\bea
\eps \times \eps &=& \mathds{1}\nonumber \\
\eps \times \sigma &=& \sigma\label{Ising-fusion} \\
\sigma \times \sigma &=& \mathds{1}+\eps.\nonumber
\eea
The bulk fields are all spinless and also labeled by $\mathds{1}$, $\eps$ and $\sigma$.
There are three possible fundamental boundary conditions also denoted as $\mathds{1}$, $\eps$ and $\sigma$ which describe the fixed ($\pm$) and free boundary conditions for the spins in the underlying lattice model.  We will refer to the conformal boundary conditions of the  Ising model as the Ising ``D-branes''.

\subsection{Defect action on Ising boundary fields}

In the Ising model there are three (fundamental) topological defects. They can act on bulk field via closed-string defect operators, with the following composition rules
\bea
D_\eps ^2 &=& D_\mathds{1} \\
D_\sigma D_\eps &=& D_\eps D_\sigma =D_\sigma
\label{eq:D3D2=D3}\\
D_\sigma^2 &=& D_\mathds{1} + D_\eps,
\label{eq:D3^2=D1+D2}
\eea
which realize the fusion rules (\ref{Ising-fusion}).

Now we would like to construct the open string defect operators and study their composition rules according to section \ref{sec:Defects}.

%
The general boundary field on a system of $N_1$ {${\mathds 1}$-branes}, $N_\eps$ $\eps$-branes, and $N_\sigma$ $\sigma$-branes
has the form
\begin{align}
\Psi=
\bordermatrix{%
&{\scriptstyle {\mathds 1}}
&{\scriptstyle \eps}
&{\scriptstyle \sigma}  \cr
{\scriptstyle {\mathds 1}}
& L_{\mathds{1}}^{({\mathds 1}{\mathds 1})}& P_\eps^{({\mathds 1}\eps)}& Q_\sigma^{({\mathds 1}\sigma)} \cr
{\scriptstyle \eps}
& \bar P_\eps^{(\eps {\mathds 1})} & M_{\mathds{1}}^{(\eps\eps)}
& R_\sigma^{(\eps\sigma)} \cr
{\scriptstyle \sigma}
& \bar Q_\sigma^{(\sigma {\mathds 1})} &\bar R_\sigma^{(\sigma\eps)} &
N_{\mathds{1}}^{(\sigma\sigma)}+N_\eps^{(\sigma\sigma)}
},
\label{eq:SF_Ising}
\end{align}
where
the ($a, b$) component is a $N_a\times N_b$ matrix
with respect to the Chan-Paton indices.
We have chosen this presentation since  topological defect operators are blind to  Chan-Paton factors.
The upper indices inside the parenthesis represent
the left and the right boundary conditions,
which are also indicated outside the matrix for later convenience,and
the lower index is the Virasoro label.
Each entry of \eqref{eq:SF_Ising} is a generic matrix-valued state in the Verma module  indicated by the corresponding subscript, and allowed by the boundary conditions. This expression can also represent an open string field of the form (\ref{eq:Psi_expansion}, \ref{eq:sum_ab_Psi^ab}).

As a useful example, let us study the fusion of two $\sigma$ defect operators, $\dd_\sigma^2$, on the boundary field (or open string field)
\eqref{eq:SF_Ising}.
From \eqref{Xdef} and the concrete value of the defect coefficients $X$ for the
Ising model BCFT (see appendix \ref{sec:X_Ising}),
applying the $\dd_\sigma$ on the open string field \eqref{eq:SF_Ising}  results in a matrix with a larger size
\begin{align}
\dd_\sigma\Psi&=\bordermatrix{%
&{\scriptstyle \{{\mathds 1} \sigma\}}
&{\scriptstyle \{\eps \sigma\}}
&{\scriptstyle \{\sigma {\mathds 1}\}}
&{\scriptstyle \{\sigma \eps\}}\cr
{\scriptstyle \{{\mathds 1} \sigma\}}
&L_{\mathds{1}}^{(\sigma\sigma)} & \frac{1}{\sqrt{2}}P_{\eps}^{(\sigma\sigma)}
& \frac{1}{2^{1/4}}Q_{\sigma}^{(\sigma {\mathds 1})}
& \frac{1}{2^{3/4}}Q_{\sigma}^{(\sigma \eps)} \cr
{\scriptstyle \{\eps \sigma\}}
&\frac{1}{\sqrt{2}}\bar P_{\eps}^{(\sigma\sigma)} & M_{\mathds{1}}^{(\sigma\sigma)}
& 2^{1/4}R_{\sigma}^{(\sigma {\mathds 1})}
& -\frac{1}{2^{1/4}}R_{\sigma}^{(\sigma \eps)} \cr
{\scriptstyle \{\sigma {\mathds 1}\}}
&\frac{1}{2^{1/4}}\bar Q_{\sigma}^{({\mathds 1}\sigma)}
& 2^{1/4}\bar R_{\sigma}^{({\mathds 1}\sigma)}
& N_{\mathds{1}}^{({\mathds 1}{\mathds 1})}
& \sqrt{2}N_{\eps}^{({\mathds 1}\eps)}\cr
{\scriptstyle \{\sigma \eps\}}
&\frac{1}{2^{3/4}}\bar Q_{\sigma}^{(\eps\sigma)}
& -\frac{1}{2^{1/4}}\bar R_{\sigma}^{(\eps\sigma)}
& \sqrt{2}N_{\eps}^{(\eps {\mathds 1})}
& N_{\mathds{1}}^{(\eps\eps)}
}.
\label{eq:dd_sigmaPsi}
\end{align}
Here indices $\{ab\}$ $(a,b={\mathds 1},\eps,\sigma)$ outside the matrix
keep track of changes of the left and the right boundary conditions;
for example,  the (1,1) component of the matrix has the indices $\{{\mathds 1}\sigma\}\{{\mathds 1}\sigma\}$, and the boundary condition of the corresponding entry has been changed as
\be
\dd_\sigma : L_{\mathds 1}^{({\mathds 1}{\mathds 1})}\mapsto L_{\mathds 1}^{{(\sigma}{\sigma})}.
\ee
the (1,2) component of the matrix has the indices $\{{\mathds 1}\sigma\}\{\eps\sigma\}$, and correspondingly,
\be
\dd_\sigma : P_\eps^{({\mathds 1}\eps)}
\mapsto X^{\sigma {\mathds 1}\eps}_{\eps\sigma\sigma}
P_\eps^{{(\sigma}{\sigma})},
\ee
where $X^{\sigma {\mathds 1}\eps}_{\eps\sigma\sigma}=\frac{1}{\sqrt{2}}$, and so on.
Note that the action of  $\dd_\sigma$ increases the number of branes in general,
because it changes the $\sigma$-brane into a $\mathds 1$-brane and an $\eps$-brane.
Now the system has $N_{\sigma}$ $\mathds 1$-branes, $N_{\sigma}$ $\eps$-branes and $(N_{\mathds 1}+N_{\eps})$ $\sigma$-branes.

Applying  $\dd_\sigma$ again we obtain
\begin{align}
\dd_\sigma^2\Psi&=
\bordermatrix{%
&{\scriptstyle \{{\mathds 1}\sigma {\mathds 1}\}}
&{\scriptstyle \{{\mathds 1}\sigma \eps\}}
&{\scriptstyle \{\eps \sigma {\mathds 1}\}}
&{\scriptstyle \{\eps\sigma \eps\}}
&{\scriptstyle \{\sigma {\mathds 1} \sigma\}}
&{\scriptstyle \{\sigma\eps \sigma\}} \cr
{\scriptstyle \{{\mathds 1}\sigma {\mathds 1}\}}
&L_{\mathds{1}}^{({\mathds 1}{\mathds 1})} & 0 & 0 & P_{\eps}^{({\mathds 1}\eps)}
& \frac{1}{\sqrt{2}}Q_{\sigma}^{({\mathds 1}\sigma)} &
\frac{1}{\sqrt{2}}Q_{\sigma}^{({\mathds 1}\sigma)}\cr
{\scriptstyle \{{\mathds 1}\sigma \eps\}}
&0 &  L_{\mathds{1}}^{(\eps\eps)} &P_{\eps}^{(\eps {\mathds 1})}& 0 &
\frac{1}{2}Q_{\sigma}^{(\eps\sigma)}
& -\frac{1}{2}Q_{\sigma}^{(\eps\sigma)}\cr
{\scriptstyle \{\eps \sigma {\mathds 1}\}}
&0 & \bar P_{\eps}^{({\mathds 1}\eps)}&M_{\mathds{1}}^{({\mathds 1}{\mathds 1})} & 0 &
R_{\sigma}^{({\mathds 1}\sigma)}
&-R_{\sigma}^{({\mathds 1}\sigma)} \cr
{\scriptstyle \{\eps \sigma \eps\}}
&\bar P_{\eps}^{(\eps {\mathds 1})}& 0 & 0 & M_{\mathds{1}}^{(\eps\eps)}&
\frac{1}{\sqrt{2}}R_{\sigma}^{(\eps \sigma)}
&\frac{1}{\sqrt{2}}R_{\sigma}^{(\eps \sigma)}\cr
{\scriptstyle \{\sigma {\mathds 1}\sigma\}}
&\frac{1}{\sqrt{2}}\bar Q_{\sigma}^{(\sigma {\mathds 1})}
& \frac{1}{2}\bar Q_{\sigma}^{(\sigma\eps)}&
\bar R_{\sigma}^{(\sigma {\mathds 1})}
& \frac{1}{\sqrt{2}}\bar R_{\sigma}^{(\sigma\eps)}&
N_{\mathds{1}}^{(\sigma\sigma)} & N_{\eps}^{(\sigma\sigma)}\cr
{\scriptstyle \{\sigma \eps \sigma\}}
&\frac{1}{\sqrt{2}}\bar Q_{\sigma}^{(\sigma {\mathds 1})}
&-\frac{1}{2}\bar Q_{\sigma}^{(\sigma\eps)}&
-\bar R_{\sigma}^{(\sigma {\mathds 1})}
&\frac{1}{\sqrt{2}}\bar R_{\sigma}^{(\sigma\eps)}&
N_{\eps}^{(\sigma\sigma)} & N_{\mathds{1}}^{(\sigma\sigma)}
},
\label{eq:D3^2Psi}
\end{align}
which
has an equal  number of rows and columns
 given by $2(N_{\mathds 1}+N_\eps+N_\sigma)$.

On the other hand, $\left(\dd_{\mathds 1}\oplus \dd_\eps\right)\Psi$ is given by
\begin{align}
\bordermatrix{%
& {\scriptstyle [{\mathds{1}};{\mathds{1}}{\mathds{1}}]}
& {\scriptstyle [{\mathds{1}};\eps\eps]}
& {\scriptstyle [{\mathds{1}};\sigma\sigma]}
& {\scriptstyle [\eps; {\mathds{1}}\eps ]}
& {\scriptstyle [\eps; \eps {\mathds{1}}]}
& {\scriptstyle [\eps;\sigma\sigma]} \cr
{\scriptstyle [{\mathds{1}};{\mathds{1}}{\mathds{1}}]}&
L_{\mathds{1}}^{({\mathds{1}}{\mathds{1}})} & P_{\eps}^{({\mathds{1}}\eps)} & Q_{\sigma}^{({\mathds{1}}\sigma)} & 0 & 0 & 0\cr
{\scriptstyle [{\mathds{1}};\eps\eps]}&
\bar P_{\eps}^{(\eps {\mathds{1}})} & M_{\mathds{1}}^{(\eps\eps)}
& R_{\sigma}^{(\eps\sigma)} & 0 & 0 & 0\cr
{\scriptstyle [{\mathds{1}};\sigma\sigma]}&
\bar Q_{\sigma}^{(\sigma {\mathds{1}})}&\bar R_{\sigma}^{(\sigma \eps)}
& N_{\mathds{1}}^{(\sigma\sigma)}+N_{\eps}^{(\sigma\sigma)}
& 0 & 0 & 0\cr
{\scriptstyle [\eps;{\mathds{1}}\eps]}&
0&0&0& L_{\mathds{1}}^{(\eps\eps)}& P_{\eps}^{(\eps {\mathds{1}})}
&\frac{1}{\sqrt{2}}Q_{\sigma}^{(\eps\sigma)}\cr
{\scriptstyle [\eps;\eps {\mathds{1}}]}&
0&0&0&\bar P_{\eps}^{({\mathds{1}}\eps)} &M_{\mathds{1}}^{({\mathds{1}}{\mathds{1}})}
&\sqrt{2}R_{\sigma}^{({\mathds{1}}\sigma)} \cr
{\scriptstyle [\eps;\sigma\sigma]}&
0&0&0&\frac{1}{\sqrt{2}}\bar Q_{\sigma}^{(\sigma \eps)}
&\sqrt{2}\bar R_{\sigma}^{(\sigma {\mathds{1}})}
&N_{\mathds{1}}^{(\sigma\sigma)}-N_{\eps}^{(\sigma\sigma)}
}.
\label{eq:(D_1+D_2)Psi}
\end{align}
This is clearly not equal to \eqref{eq:D3^2Psi}.
While the closed defect operators $D_d$ obey the defect algebra \eqref{eq:D3^2=D1+D2} which is strictly isomorphic to the Verlinde fusion algebra,
this is not the case for the open string defect operators $\dd_d$.
As discussed in section \ref{sec:Defects},
we need to take into account  the similarity
transformation \eqref{eq:U_explicit} and \eqref{eq:U_inv_explicit}
in order to connect \eqref{eq:(D_1+D_2)Psi} with \eqref{eq:D3^2Psi}.
In this case, the similarity
transformation is given by
\begin{align}
U_{\sigma\sigma}&=\bordermatrix{%
&{\scriptstyle[{\mathds{1}};{\mathds{1}},{\mathds{1}}]}
&{\scriptstyle[{\mathds{1}};\eps,\eps]}
&{\scriptstyle[{\mathds{1}};\sigma,\sigma]}
&{\scriptstyle[\eps;{\mathds{1}},\eps]}
&{\scriptstyle[\eps;\eps,{\mathds{1}}]}
&{\scriptstyle[\eps;\sigma,\sigma]}
\cr
{\scriptstyle\{{\mathds{1}}\sigma {\mathds{1}}\}}
&1 &0 &0 &0&0&0 \cr
{\scriptstyle\{{\mathds{1}}\sigma  \eps\}}
&0 & 0 & 0 & 1& 0& 0 \cr
{\scriptstyle \{\eps \sigma  {\mathds{1}}\}}
&0 & 0 & 0 &0&1&0   \cr
{\scriptstyle\{\eps\sigma \eps\}}&
0& 1& 0 & 0 & 0 & 0 \cr
{\scriptstyle\{\sigma {\mathds{1}}  \sigma\}}&
0& 0& \frac{1}{\sqrt{2}}& 0& 0& \frac{1}{\sqrt{2}}   \cr
{\scriptstyle\{\sigma \eps  \sigma\}}&
0&0&\frac{1}{\sqrt{2}}&0& 0& -\frac{1}{\sqrt{2}}  \cr
}
\\
&=
\bordermatrix{%
& {\scriptstyle[{\mathds{1}};{\mathds{1}}{\mathds{1}}]}
& {\scriptstyle[\eps;{\mathds{1}}\eps]}
& {\scriptstyle[\eps;\eps{\mathds{1}}]}
& {\scriptstyle[{\mathds{1}};\eps\eps]}
& {\scriptstyle[{\mathds{1}};\sigma\sigma]}
& {\scriptstyle[\eps;\sigma\sigma]}\cr
{\scriptstyle\{{\mathds{1}} \sigma  {\mathds{1}}\}}
&1 &  &  &  &  &  \cr
{\scriptstyle\{{\mathds{1}} \sigma  \eps\}}
& &1 &  &  &  &  \cr
{\scriptstyle\{\eps \sigma {\mathds{1}} \}}
&  &  & 1 & &  &  \cr
{\scriptstyle\{\eps \sigma  \eps\}}
&  &  &  & 1 &  &  \cr
{\scriptstyle\{\sigma  {\mathds{1}}\sigma\}}
&  &  &  &  & \frac{1}{\sqrt{2}} & \frac{1}{\sqrt{2}} \cr
{\scriptstyle\{\sigma \eps  \sigma\}}
&  &  &  &  & \frac{1}{\sqrt{2}} & -\frac{1}{\sqrt{2}} \cr
}=
\left(U_{\sigma\sigma}^{-1}\right)^{T}
.
\label{eq:U_SF}
\end{align}
The reader can easily check that
\be
\dd_\sigma^2\Psi=U_{\sigma\sigma}\left[\left(\dd_{\mathds 1}\oplus \dd_\eps\right)\Psi\right] U_{\sigma\sigma}^{-1}.
\ee
Similarly, to investigate the relation \eqref{eq:D3D2=D3}, we calculate $\dd_\sigma\dd_\eps \Psi$
\begin{align}
\dd_\sigma\dd_\eps\Psi&=\bordermatrix{%
&{\scriptstyle \{{\mathds{1}} \eps\sigma\}}
&{\scriptstyle \{\eps {\mathds{1}}\sigma\}}
&{\scriptstyle \{\sigma \sigma {\mathds{1}}\}}
&{\scriptstyle \{\sigma \sigma \eps\}}\cr
{\scriptstyle \{{\mathds{1}}\eps \sigma\}}
&L_{\mathds{1}}^{(\sigma\sigma)} & \frac{1}{\sqrt{2}}P_{\eps}^{(\sigma\sigma)}
& \frac{1}{2^{1/4}}Q_{\sigma}^{(\sigma {\mathds{1}})}
&- \frac{1}{2^{3/4}}Q_{\sigma}^{(\sigma \eps)} \cr
{\scriptstyle \{\eps {\mathds{1}}\sigma\}}
&\frac{1}{\sqrt{2}}\bar P_{\eps}^{(\sigma\sigma)} & M_{\mathds{1}}^{(\sigma\sigma)}
& 2^{1/4}R_{\sigma}^{(\sigma {\mathds{1}})}
& \frac{1}{2^{1/4}}R_{\sigma}^{(\sigma \eps)} \cr
{\scriptstyle \{\sigma \sigma {\mathds{1}}\}}
&\frac{1}{2^{1/4}}\bar Q_{\sigma}^{({\mathds{1}}\sigma)}
& 2^{1/4}\bar R_{\sigma}^{({\mathds{1}}\sigma)}
& N_{\mathds{1}}^{({\mathds{1}}{\mathds{1}})}
&- \sqrt{2}N_{\eps}^{({\mathds{1}}\eps)}\cr
{\scriptstyle \{\sigma \sigma \eps\}}
&-\frac{1}{2^{3/4}}\bar Q_{\sigma}^{(\eps\sigma)}
& \frac{1}{2^{1/4}}\bar R_{\sigma}^{(\eps\sigma)}
& -\sqrt{2}N_{\eps}^{(\eps {\mathds{1}})}
& N_{\mathds{1}}^{(\eps\eps)}
}
\end{align}
and $\dd_\eps \dd_\sigma \Psi$
\begin{align}
\dd_\eps\dd_\sigma\Psi&=\bordermatrix{%
&{\scriptstyle \{{\mathds{1}} \sigma\sigma\}}
&{\scriptstyle \{\eps \sigma\sigma\}}
&{\scriptstyle \{\sigma \eps {\mathds{1}}\}}
&{\scriptstyle \{\sigma {\mathds{1}} \eps\}}\cr
{\scriptstyle \{{\mathds{1}}\sigma \sigma\}}
&L_{\mathds{1}}^{(\sigma\sigma)} & -\frac{1}{\sqrt{2}}P_{\eps}^{(\sigma\sigma)}
& \frac{1}{2^{1/4}}Q_{\sigma}^{(\sigma {\mathds{1}})}
& \frac{1}{2^{3/4}}Q_{\sigma}^{(\sigma \eps)} \cr
{\scriptstyle \{\eps \sigma\sigma\}}
&-\frac{1}{\sqrt{2}}\bar P_{\eps}^{(\sigma\sigma)} & M_{\mathds{1}}^{(\sigma\sigma)}
& -2^{1/4}R_{\sigma}^{(\sigma {\mathds{1}})}
& \frac{1}{2^{1/4}}R_{\sigma}^{(\sigma \eps)} \cr
{\scriptstyle \{\sigma \eps {\mathds{1}}\}}
&\frac{1}{2^{1/4}}\bar Q_{\sigma}^{({\mathds{1}}\sigma)}
& -2^{1/4}\bar R_{\sigma}^{({\mathds{1}}\sigma)}
& N_{\mathds{1}}^{({\mathds{1}}{\mathds{1}})}
& \sqrt{2}N_{\eps}^{({\mathds{1}}\eps)}\cr
{\scriptstyle \{\sigma {\mathds{1}} \eps\}}
&\frac{1}{2^{3/4}}\bar Q_{\sigma}^{(\eps\sigma)}
& \frac{1}{2^{1/4}}\bar R_{\sigma}^{(\eps\sigma)}
& \sqrt{2}N_{\eps}^{(\eps {\mathds{1}})}
& N_{\mathds{1}}^{(\eps\eps)}
}.
\end{align}
Comparing these expressions with \eqref{eq:dd_sigmaPsi}, we find
\begin{align}
U_{\sigma\eps}&=\bordermatrix{%
&{\scriptstyle [\sigma; {\mathds{1}} \sigma]}
&{\scriptstyle [\sigma; \eps \sigma]}
&{\scriptstyle [\sigma; \sigma {\mathds{1}}]}
&{\scriptstyle [\sigma; \sigma \eps]}\cr
{\scriptstyle \{{\mathds{1}}\eps\sigma \}}
&1&
&
& \cr
{\scriptstyle \{\eps {\mathds{1}}\sigma\}}
& & 1
&
& \cr
{\scriptstyle \{\sigma\sigma {\mathds{1}}\}}
&
&
& 1
& \cr
{\scriptstyle \{\sigma \sigma \eps\}}
&
&
&
& -1
}
=
(U_{\sigma\eps})^{-1},
\end{align}

\begin{align}
U_{\eps\sigma}&=\bordermatrix{%
&{\scriptstyle [\sigma; {\mathds{1}} \sigma]}
&{\scriptstyle [\sigma; \eps \sigma]}
&{\scriptstyle [\sigma; \sigma {\mathds{1}}]}
&{\scriptstyle [\sigma; \sigma \eps]}\cr
{\scriptstyle \{{\mathds{1}}\eps \sigma\}}
&1&
&
&  \cr
{\scriptstyle \{\eps {\mathds{1}}\sigma\}}
& & -1
&
&  \cr
{\scriptstyle \{\sigma \eps {\mathds{1}}\}}
&
&
& 1
& \cr
{\scriptstyle \{\sigma \mathds 1 \eps\}}
&
&
&
&1
}
=
(U_{\eps\sigma})^{-1}.
\label{eq:U_epssigma}
\end{align}

\subsection{Defect action on Ising classical solutions}
Now let us discuss the action of the defect operators on the classical solutions of OSFT.
For illustration, let us focus on the $\sigma$-brane of the Ising model and let us consider the corresponding OSFT. At lowest nontrivial level $\half$ the string field takes the form $\Psi = t c_1 \ket{0} + a c_1 \ket{\eps}$ for which the potential is \cite{KRS}
\be
V(t,a)=-\frac{1}{2} t^2 -\frac{1}{4} a^2 + \frac{27\sqrt{3}}{64} t^3 + \frac{27}{16} t a^2.
\ee
From here one can already see the four critical points given by the perturbative as well as tachyon vacuum, and further two solutions related by a $Z_2$-symmetry describing the other two fundamental boundary conditions.\footnote{At higher levels one may find other solutions, see the discussion in \cite{KRS}.}

Going to higher levels, the string field will keep the form
\be\label{gen-Ising-sol}
\Psi_{\sigma \to {\mathds 1}}= \psi_{\mathds{1}}^{\sigma\sigma}
 + \psi_{\eps}^{\sigma\sigma},
\ee
where $\psi_{\mathds{1}}^{\sigma\sigma}$ and $\psi_{\eps}^{\sigma\sigma}$
will stand for the two components in the identity and $\eps$ Verma modules respectively. Correspondingly, the equations of motion split into two independent sets:
\bea
Q \psi_{\mathds{1}}^{\sigma\sigma}
+ \psi_{\mathds{1}}^{\sigma\sigma}*\psi_{\mathds{1}}^{\sigma\sigma}
+ \psi_{\eps}^{\sigma\sigma}*\psi_{\eps}^{\sigma\sigma} &=& 0\\
Q \psi_{\eps}^{\sigma\sigma}
+  \psi_{\mathds{1}}^{\sigma\sigma}* \psi_{\eps}^{\sigma\sigma}
+ \psi_{\eps}^{\sigma\sigma}*\psi_{\mathds{1}}^{\sigma\sigma} &=& 0.
\eea
Incidentally, these equations are enough to guarantee that three other string fields constructed from $\psi_{\mathds{1}}^{\sigma\sigma}$ and
$\psi_{\eps}^{\sigma\sigma}$ obey the equations of motion as well:
\be
\psi_{\mathds{1}}^{\sigma\sigma} - \psi_{\eps}^{\sigma\sigma},
\label{eq:psi_1-psi_eps}
\ee
and
\be
\begin{pmatrix} \psi_{\mathds{1}}^{\sigma\sigma}
& \pm \psi_{\eps}^{\sigma\sigma}\\
\pm \psi_{\eps}^{\sigma\sigma}
& \psi_{\mathds{1}}^{\sigma\sigma}\end{pmatrix}.
\label{eq:(4.12)}
\ee
Intuitively, following our discussion in section \ref{sec:Defects},  it is clear that first of these solutions should be the result of applying the $\dd_\eps$ defect to (\ref{gen-Ising-sol}).
Actually, we obtain $X^{\eps\sigma\sigma}_{{\mathds 1}\sigma\sigma}=-X^{\eps\sigma\sigma}_{\eps\sigma\sigma}=1$ from \eqref{X-sym}, then $\dd_\eps\Psi_{\sigma\to {\mathds 1}}$
results in \eqref{eq:psi_1-psi_eps}.

The remaining two solutions in \eqref{eq:(4.12)} also fit in with our discussion:
from the analysis in section {\ref{Sec:OSFT}},
we see that
the solution $\dd_\sigma\Psi_{\sigma\to {\mathds 1}}$ describes a $\sigma$-brane
in the theory around a system with a {$\mathds 1$}-brane and an $\eps$-brane.
We then denote it by $\Psi_{{\mathds 1}+\eps\to \sigma}$,
\be
\Psi_{{\mathds 1}+\eps\to\sigma}\equiv \dd_\sigma \Psi_{\sigma\to {\mathds 1}}.
\ee
From \eqref{X-sym}, we see that
\be
\dd_\sigma \psi_{\mathds 1}^{\sigma\sigma}=
\begin{pmatrix}
X^{\sigma\sigma\sigma}_{ {\mathds 1}{\mathds 1}{\mathds 1}}\psi_{\mathds 1}^{{\mathds 1}{\mathds 1}}&\\
&X^{\sigma\sigma\sigma}_{ {\mathds 1}\eps\eps}\psi_{\mathds 1}^{\eps\eps}
\end{pmatrix}
=
\begin{pmatrix}
\psi_{\mathds 1}^{{\mathds 1}{\mathds 1}}&\\
&\psi_{\mathds 1}^{\epsilon\epsilon}
\end{pmatrix},
\ee
\be
\dd_\sigma \psi_{\eps}^{\sigma\sigma}=
\begin{pmatrix}
&X^{\sigma\sigma\sigma}_{  \eps{\mathds 1}\eps}\psi_{\eps}^{{\mathds 1}\epsilon}\\
X^{\sigma\sigma\sigma}_{  \eps\eps{\mathds 1}}\psi_{\eps}^{\eps {\mathds 1}}&
\end{pmatrix}
=
\begin{pmatrix}
&\psi_{\eps}^{{\mathds 1}\eps}\\
\psi_{\eps}^{\eps {\mathds 1}}&
\end{pmatrix}.
\ee
and $\Psi_{{\mathds{1}}+\eps \to \sigma}$ is given by
\be
\Psi_{{\mathds 1}+\eps \to \sigma}=
\begin{pmatrix}
\psi_{\mathds 1}^{{\mathds 1}{\mathds 1}}&\psi_{\eps}^{{\mathds 1}\eps}\\
\psi_{\eps}^{\eps {\mathds 1}}&\psi_{\mathds 1}^{\eps\eps}
\end{pmatrix}.
\ee
Further acting $\dd_\sigma$ or $\dd_{\eps}\dd_\sigma$,
we obtain the two solutions in \eqref{eq:(4.12)}
\begin{equation}
\dd_\sigma \Psi_{{\mathds 1}+\eps \to \sigma}=
\begin{pmatrix} \psi_{\mathds{1}}^{\sigma\sigma}
& + \psi_{\eps}^{\sigma\sigma}\\
+ \psi_{\eps}^{\sigma\sigma}
& \psi_{\mathds{1}}^{\sigma\sigma}\end{pmatrix},
\label{eq:dd_sPsi}
\end{equation}
\begin{equation}
\dd_\eps \dd_\sigma
\Psi_{{\mathds 1}+\eps\to\sigma}=
\begin{pmatrix} \psi_{\mathds{1}}^{\sigma\sigma}
& - \psi_{\eps}^{\sigma\sigma}\\
- \psi_{\eps}^{\sigma\sigma}
& \psi_{\mathds{1}}^{\sigma\sigma}\end{pmatrix}.
\label{eq:dd_edd_sPsi}
\end{equation}
Note that these two solutions are related by a similarity transformation,
\be
\dd_\epsilon\dd_{\sigma}\Psi_{{\mathds 1}+\eps\to\sigma}=
U_{\eps\sigma}(\dd_\sigma\Psi_{{\mathds 1}+\eps\to\sigma})U_{\eps\sigma}^{-1},
\label{eq:dd_eps dd_sigma Psi_{1+eps to sigma}}
\ee
where
\be
U_{\eps\sigma}=
\begin{pmatrix}
U_{\eps\sigma}^{\{\mathds{1}\sigma\sigma\}[\sigma;\mathds{1}\sigma]}&\\
&U_{\eps\sigma}^{\{\eps\sigma\sigma\}[\sigma;\eps\sigma]}
\end{pmatrix}
=
\begin{pmatrix}
1&\\
&-1
\end{pmatrix}.
\ee
The matrix which is used here to connect these two solutions is nothing but the $U$ matrix discussed in section {\ref{sec:fusion of open string defects}} and which is a part of \eqref{eq:U_epssigma} we derived in the first half of this section.
Furthermore,we can consider the components of $U$ as proportional to
the identity string field, and then we can regard \eqref{eq:dd_eps dd_sigma Psi_{1+eps to sigma}} as a gauge transformation in OSFT $\Lambda(Q_B+\Psi)\Lambda^{-1}$.

The discussion here can be generalized
and we can prove the following statement: consider a Verlinde fusion algebra $a\times b=\sum_e {{\rm N}_{ab}}^ee$ and
let $\Psi$ be a general classical solution of OSFT.
Then two classical solutions $\Psi'=\dd_a\dd_b \Psi$ and $\Psi''=\oplus_{e} {{\rm N}_{ab}}^e \dd_e \Psi $ are
related by a gauge transformation with a constant gauge parameter, which is given by
the $U$ matrix,
\be
\Psi'=U_{ab}(Q_B+\Psi'')U_{ab}^{-1}.
\label{eq:6.33}
 \ee
This is true in general because the components of $U$ are
proportional to the identity string field and thus vanishes upon action of the BRST charge.
Note that there is no problem in  considering the components of $U$
to be proportional to the identity string field, for
if two final boundary conditions are not the same ($a''\neq \tilde{a}''$),
the corresponding components become zero,
as showed in \eqref{eq:U_general}.
This is consistent with the fact that
the identity Verma module cannot connect different boundary conditions.\ \\

Similarly, we can produce different classical solutions by acting defect operators on $\Psi_{\sigma\to{\mathds 1}}$.
We here summarize solutions obtained by acting combination of $\dd_\sigma$'s and $\dd_\eps$'s
with the number of $\dd_\sigma$'s less than three:
\begin{equation}
\xymatrix
{
\Psi_{\sigma\to{\mathds 1}}\ar@<1mm>[d]^{\dd_\eps}\ar[r]^{\dd_\sigma}&\Psi_{{\mathds 1}+\eps\to\sigma}\ar[r]^{\dd_\sigma}\ar@(ul,ur)[]^{\dd_\eps}&\Psi_{2\sigma\to {\mathds 1}+\eps}\ar[r]^{\dd_\sigma}\ar@<1mm>[d]^{\dd_\eps}&\dots\\
\Psi_{\sigma\to\eps}\ar@<1mm>[u]^{\dd_\eps}\ar[r]_{\dd_\sigma}&
\Psi'_{{\mathds 1}+\eps\to\sigma}\ar@(dl,dr)[]_{\dd_\eps}\ar[r]_{\dd_\sigma}&
\Psi_{2\sigma\to\eps+{\mathds 1}}\ar@<1mm>[u]^{\dd_\eps}\ar[r]_{\dd_\sigma}&\dots
}
\label{solution_arrow_diagram_1}
\end{equation}
where
\be
\Psi_{\sigma\to\eps}=\psi_{\mathds{1}}^{\sigma\sigma} - \psi_{\eps}^{\sigma\sigma},
\ee
\be
\Psi'_{{\mathds 1}+\eps\to\sigma}=
\begin{pmatrix} \psi_{\mathds{1}}^{{\mathds 1}{\mathds 1}}
&-\psi_{\eps}^{{\mathds 1}\eps}\\
-\psi_{\eps}^{\eps{\mathds 1}}
& \psi_{\mathds{1}}^{\eps\eps}\end{pmatrix},
\ee
besides $\Psi_{2\sigma\to{\mathds 1}+\eps}\equiv \dd_\sigma
\Psi_{{\mathds 1}+\eps\to\sigma}$ and $\Psi_{2\sigma\to\eps+{\mathds 1}}\equiv \dd_\eps \dd_\sigma
\Psi_{{\mathds 1}+\eps\to\sigma}$ in \eqref{eq:dd_sPsi} and \eqref{eq:dd_edd_sPsi}, respectively.
%
The relation between $\Psi_{{\mathds 1}+\eps\to\sigma}$ and
$\Psi'_{{\mathds 1}+\eps\to\sigma}$ is also given by a $U$ matrix,
\be
\dd_{\sigma}\dd_\eps\Psi_{\sigma\to\eps}
=U_{\sigma\eps}(\dd_{\sigma}\Psi_{\sigma\to\eps})U^{-1}_{\sigma\eps},
\ee
that is,
\be
\Psi_{{\mathds 1}+\eps\to\sigma}=
\begin{pmatrix}
1&\\
&-1
\end{pmatrix}
\Psi'_{{\mathds 1}+\eps\to\sigma}
\begin{pmatrix}
1&\\
&-1
\end{pmatrix}.
\label{eq:not_gauge_transf}
\ee
\newcommand{\s}{\sigma}
\newcommand{\e}{\eps}

Starting from $\Psi_{{\mathds 1}\to \sigma}$, another series of classical solutions can be obtained.
Since the state $\Psi_{{\mathds 1}\to \sigma}$ satisfy the boundary condition $\mathds 1$,
the whole solution will be solely composed by the identity Verma module
\be
\Psi_{{\mathds 1}\to\sigma}=\widetilde \psi_{\mathds 1}^{{\mathds 1}{\mathds 1}}.
\ee
Applying open string defect operators $\dd_\eps$ and $\dd_{\sigma}$ again and again,
we obtain the following sequence\footnote{Here we have distinguished $\Psi_{\sigma\to {\mathds 1}+\epsilon}$ and $\Psi'_{\sigma\to {\mathds 1}+\epsilon}$
because
$\dd_\sigma(\Psi_{\mathds 1\to \sigma }-\Psi_{\eps \to \sigma})\ne 0$ as
topological defects do not kill boundary fields \cite{GW}.
}:
\begin{equation}
\xymatrix
{
\Psi_{{\mathds 1}\to\s}\ar@<1mm>[d]^{\dd_\e}\ar[r]^{\dd_\s}&\Psi_{\s\to{\mathds 1}+\e}\ar[r]^{_\s}\ar@(ul,ur)[]^{\dd_\e}&\Psi_{{\mathds 1}+\e\to2\s}\ar[r]^{\dd_\s}\ar@(ul,ur)[]^{\dd_\e}&\Psi_{2\s\to2\cdot {\mathds 1}+2\e}\ar@(ul,ur)[]^{\dd_\e}\ar[r]^{\dd_\s}&\dots\\
\Psi_{{\eps}\to\s}\ar@<1mm>[u]^{\dd_\e}\ar[r]^{\dd_\s}&\Psi'_{\s\to{\mathds 1}+\e}\ar[r]^{_\s}\ar@(ul,ur)[]^{\dd_\e}&\Psi'_{{\mathds 1}+\e\to2\s}\ar[r]^{\dd_\s}\ar@(ul,ur)[]^{\dd_\e}&\Psi'_{2\s\to2\cdot {\mathds 1}+2\e}\ar@(ul,ur)[]^{\dd_\e}\ar[r]^{\dd_\s}&\dots
}
\end{equation}
where
\be
\Psi_{\eps\to\sigma}=\widetilde \psi_{\mathds 1}^{\eps\eps},
\ee
\be
\Psi_{\sigma\to {\mathds 1}+\eps}=
\Psi'_{\sigma\to {\mathds 1}+\eps}=
\widetilde \psi_{\mathds 1}^{\sigma\sigma},
\ee
\be
\Psi_{{\mathds 1}+\eps \to 2\sigma}=
\Psi'_{{\mathds 1}+\eps \to 2\sigma}=
\begin{pmatrix}
\widetilde \psi_{\mathds 1}^{{\mathds 1}{\mathds 1}}&\\
&\widetilde \psi_{\mathds 1}^{\eps\eps}
\end{pmatrix},
\ee
\be
\Psi_{2\sigma\to {\mathds 1}+\eps}=
\Psi'_{2\sigma\to {\mathds 1}+\eps}=
\begin{pmatrix}
\widetilde \psi_{\mathds 1}^{\sigma\sigma}&\\
&\widetilde \psi_{\mathds 1}^{\sigma\sigma}
\end{pmatrix}.
\ee
It is interesting that we constructed these solutions
without using $\eps$ Verma module,
{which is a possible excitation on a $\sigma$-brane, or on a $(\mathds 1+\eps)$-brane system.}

Following the same line of reasoning, we can generally prove that
for $a\times b=c$ with $a$, $b$ and $c$ general conformal boundary conditions,
$\Psi_{a\to c}$ can be constructed strictly within the identity Verma module,
for $\Psi_{a\to c}=\dd_a\Psi_{\mathds 1\to b}$.
This is a rather nontrivial fact derived by
considering defect action on the classical solutions.

\section{Conclusions}

In this paper, starting from CFT topological defects, we have build new operators acting on the open string star algebra, and we have used them to generate new solutions in  OSFT. To this end we have carefully studied the action of topological defects on boundaries and boundary fields, extending to the full boundary operator algebra the results of Graham and Watts \cite{GW}. We have also provided a clear geometric construction of open topological defects using defect networks. Our geometric picture is  two-dimensional and it doesn't  require the 3D topological description of \cite{TFT}. The action
of defects on boundary fields turns out to be much more involved than the action on bulk fields and the composition of open defect operators follows the fusion rules only up to a similarity transformation in Chan-Paton space. The pentagon identity is crucial for the consistency of open defect operators and their fusion.
 The concrete results we presented are valid for diagonal minimal models but the idea of using topological defects to generate new solutions  is clearly very general and applies  to any string background.

We have payed special attention to the issue of gauge freedom in the definition of the $F$-matrices,  which arise from the symmetries of the pentagon identity. We do not use the auxiliary concept of chiral vertex operators and we take conformal blocks to be canonically normalized as in \cite{francesco} which fixes the ``gauge'' of the $F$-matrices. However, in the computation we have performed, we did not have to specify any gauge choice for the $F$-matrices involved in defect network manipulations: our results concerning the action of topological defects on boundary fields and their composition are explicitly independent of the gauge chosen for defect networks. Combining defects and disorder operators may give further constraints relating the defect $F$-matrices with the $F$-matrices coming from the transformation of the conformal blocks.

A simple generalization of our results should be given by the study of open topological defects in RCFT's with charge conjugation modular invariant, the original setting of \cite{GW}.
In this case the representations won't be self-conjugate  and one will have to pay attention to the orientation inside defect networks. More work would be needed to address RCFT with non charge conjugation partition function (the simplest example being the Pott's model).

It would be interesting to explore open topological defects in non rational CFT's (although a complete classification of them is not available). For example, a typical properties of defects in non rational CFT's is that they can posses a moduli space as it happens for boundary conditions. It would be also useful to extend our work to conformal defects \cite{Bachas:2007td} to get more general solution generating techniques in OSFT.

A more interesting (and difficult) question, and actually one of the motivation for the present research, is whether defects can be used as part of the building blocks for constructing OSFT solutions.  This is certainly a new arena where interesting new structures may appear and which could overcome some limitations of the known solutions such as \cite{EM}.

Open topological defects are very natural objects inside the open string star-algebra which are relevant to the description of the open string landscape. We hope that our research is a useful step towards a better understanding of the space of solutions of OSFT.

\section*{Acknowledgments}

We would like to thank C. Bachas, Z. Bajnok, I. Brunner, T. Erler, S. Fredenhagen, J. Fuchs, D. Gaiotto,  M. Kudrna, M. Rap\v{c}\'{a}k, I. Runkel, Y. Satoh, C. Schweigert, A. Sen and R. Tateo for useful discussions at various stages of this work. TK, CM and TM thank the Academy of Science of Czech Republic for kind hospitality and support during various stages of this work, while MS thanks  Torino University for the same. Both CM and MS thank the organizers of the 2015 workshop ``Gauge theories, supergravity and superstrings'' in Benasque, for providing a stimulating environment during part of this research. The research of CM is funded by a {\it Rita Levi Montalcini} grant from the Italian MIUR. The research of MS is supported by the Czech funding agency grant P201/12/G028.


\appendix

\section{Comments on  Moore and Seiberg  ``gauge symmetry"}
\setcounter{equation}{0}
\label{app:Gauge_Symmetry}

An important property of the pentagon equation is its huge symmetry called a bit misleadingly a gauge symmetry
\be\label{Fgt}
\Fmat{p}{q}{a}{b}{c}{d} \to \frac{\Lambda(a,b,q)\Lambda(c,q,d)}{\Lambda(c,a,p)\Lambda(p,b,d)} \Fmat{p}{q}{a}{b}{c}{d},
\ee
where $\Lambda(i,j,k)$ is an arbitrary function of an admissible triplet. One can redefine $\Lambda(i,j,k)$ by multiplying it with $\phi_1(i)\phi_2(j)\phi_3(k)$ to get
\be
\Fmat{p}{q}{a}{b}{c}{d} \to \frac{\phi_1(p)}{\phi_1(a)} \frac{\phi_2(a)}{\phi_2(q)} \frac{\phi_3(p)}{\phi_3(q)} \frac{\Lambda(a,b,q)\Lambda(c,q,d)}{\Lambda(c,a,p)\Lambda(p,b,d)} \Fmat{p}{q}{a}{b}{c}{d},
\ee
so that the symmetry looks even bigger. Imposing that our normalization condition $\Fmat{c}{b}{1}{b}{c}{d}=1$ is preserved by the gauge transformation, implies
$\Lambda(1,b,b)=\Lambda(c,1,c), \forall b,c$, so that both expressions are label-independent, and hence we can freely normalize them to 1, i.e.
\be
\Lambda(1,b,b)=\Lambda(c,1,c)= \Lambda(1,1,1)= 1.
\ee
Under this condition $g_a' = \left(\Fmat{1}{1}{a}{a}{a}{a}\right)^{-1}$ are gauge invariant.

Imposing that gauge transformation preserve the 180$^\circ$ invariance condition  $\Fmat{p}{q}{a}{b}{c}{d} = \Fmat{p}{q}{d}{c}{b}{a}$ implies that $\Lambda$ must be cyclically invariant. The specular conditions $\Fmat{p}{q}{a}{b}{c}{d} = \Fmat{p}{q}{c}{d}{a}{b}$ and $\Fmat{p}{q}{a}{b}{c}{d} = \Fmat{p}{q}{b}{a}{d}{c}$ would similarly imply that $\Lambda$ is permutation invariant.

An important question however, is to what extent is this freedom physical. Here we wish to stress, that while different physical quantities in BCFT might be related by such a transformation, in general there is no gauge freedom when a concretely defined quantity is concerned.

In particular the structure constants of boundary CFT are given by one concrete gauge choice of $F$, which {\em cannot} be changed by a choice of normalization of the boundary operators.
Changing their normalization as
\be
\phi_i^{ab} \quad \to \quad \tilde\phi_i^{ab}=n_i^{ab} \phi_i^{ab},
\ee
the structure constants transform as well
\bea
C_{ij}^{(abc)\, p} \quad &\to& \quad \tilde C_{ij}^{(abc)\, p}= \frac{n_i^{ab}n_j^{bc}}{n_p^{ac}} C_{ij}^{(abc)\, p},
\\
C_{pp}^{(aca)\, 1} \quad &\to& \quad \tilde C_{pp}^{(aca)\, 1}= \frac{n_p^{ac}n_p^{ca}}{n_1^{aa}} C_{pp}^{(aca)\, 1}.
\eea
It is easy to check that the equation (\ref{bootstrap}) is obeyed also for the transformed structure constants, and that it is {\em not} required to transform $F$
\be
\frac{n_i^{ab} n_j^{bc}}{n_p^{ac}}
\frac{n_k^{cd} n_l^{da}}{n_p^{ca}} n_p^{ac} n_p^{ca} =
\frac{n_j^{bc} n_k^{cd}}{n_q^{bd}}
\frac{n_l^{da} n_i^{ab}}{n_q^{db}} n_q^{bd} n_q^{db}.
\ee
So the freedom in normalization of the boundary operators has no relation to the choice of gauge for $\Fblock{p}{q}{j}{k}{i}{l} $, which in fact is fixed in conformal field theory as we saw above.

In general there are {\em two special gauges} for the solutions of the pentagon identity (plus the identity condition and specular conditions): The {\em blocks gauge} and the {\em Racah gauge}. The blocks gauge has been discussed in section \ref{Runkel}. The Racah gauge is defined as
\be
\FRac{p}{q}{a}{b}{c}{d} = \Rac{b}{a}{q}{c}{d}{p}.
\ee
It has the special property
\bea
\FRac{1}{1}{a}{a}{a}{a} \FRac{1}{a}{b}{c}{b}{c} &=& \frac{1}{\sqrt{g_a'g_b'g_c'}}.
\eea
The Racah-gauge $F$-matrix is among other things important for expressing the boundary structure constants in terms of the bulk structure constants, see formula (\ref{RunkelRacahSolution}).


\section{Un-fusing defects from boundaries}

\setcounter{equation}{0}
\label{app:TFT}

\begin{figure}[!htbp]
\centering
\includegraphics[]{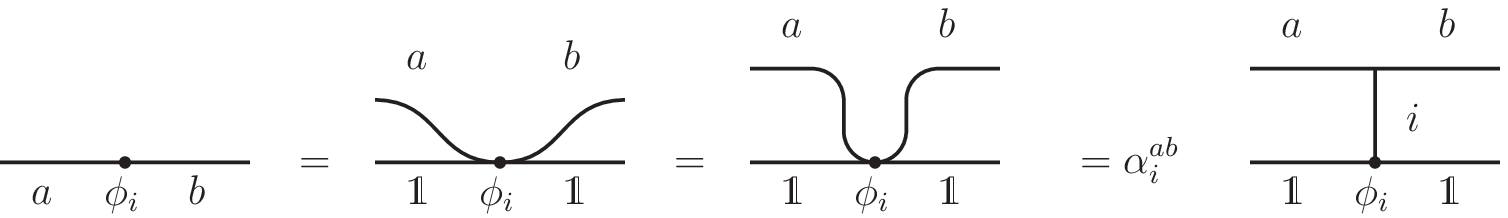}
\caption{A boundary field in the $i$ Virasoro representation can be traded for a defect network ending on a defect ending field carrying the same Virasoro representation but sitting on a boundary with the $\1$ boundary condition.}
\label{fig:TFT1}
\end{figure}

In this appendix we will derive a useful manipulation which we have used in section \ref{sec:geomX}
 to determine the defect coefficients $X_{ia'b'}^{dab}$ in the geometric approach.
We will assume that all fundamental boundary conditions in the game can be obtained by fusing a topological defect on a particular reference boundary condition, denoted as $\1$.
Consider a boundary field $\phi_i^{ab}$ changing an $a$ boundary to a $b$ boundary. By our assumption this is topologically equivalent to a network of defects with a leg ending on a defect ending field, carrying
the $i$ Virasoro representation and sitting at the $\1$ boundary, as in Figure \ref{fig:TFT1}.


\begin{figure}[!htbp]
\centering
\includegraphics[]{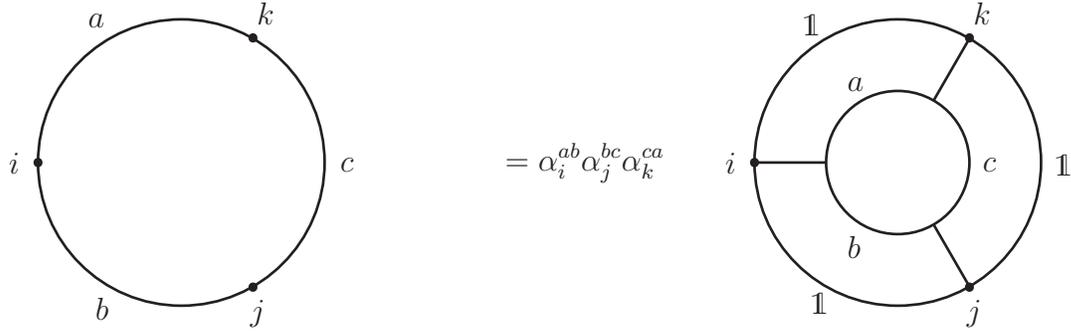}
\caption{Two equivalent ways of computing a boundary three-point function. The $\alpha$ coefficients are sensitive to both   the normalization choice for boundary fields and the $F$-matrix gauge used in defect manipulations.}
\label{fig:TFT2}
\end{figure}

The equality holds up to an unknown three-label coefficient $\alpha_i^{ab}=\alpha_i^{ba}$, symmetric in the boundary condition labels for parity reasons.
By the triviality of the identity representation we must have
\be
\alpha_\1^{aa}=1.
\ee
This is consistent with the fact that the normalized $g$ function of the defect is related to the disk partition function as
\be
\aver 1^{(a)}=\aver 1^{(1)}\frac1{\Fmat{1}{1}{a}{a}{a}{a}}=\aver 1^{(1)} g'_a=g_a.
\ee
We can determine the coefficients $\alpha_i^{ab}$ by computing a boundary three point function in two ways, either by using  the (given) boundary structure constants, or by
manipulating defects  after having performed the move in Figure \ref{fig:TFT1}, as represented in Figure \ref{fig:TFT2}.


\begin{figure}[!htbp]
\centering
\includegraphics[]{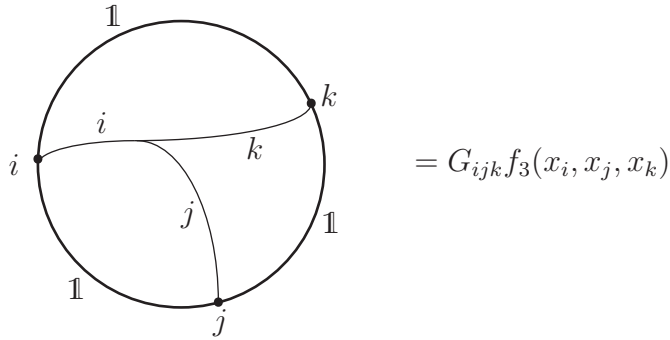}
\caption{A simple three-point function which depends on the gauge chosen for defect manipulation. The function $f_3(x_i,x_j,x_k)$ is the part of the three-point-function which is completely fixed by conformal invariance.}
\label{fig:TFT3}
\end{figure}

 After factoring out the universal dependence on the insertion points, we are left with the equality
\be\label{eq:TFT-main}
C_{ij}^{(abc)k}\, C_{kk}^{(aca)\1}\,g'_a=G_{ijk}\, \alpha_i^{ab}\alpha_j^{bc}\alpha_k^{ca}\,\frac{\Fmat{b}{k}{a}{c}{i}{j}}{\Fmat{\1}{k}{a}{c}{a}{c}}.
\ee
The new quantity $G_{ijk}$ is the non-trivial coefficient of the three point function involving the elementary defect network with boundary shown in Figure \ref{fig:TFT3}.


We will now see that consistency will relate $G$ and the $\alpha$'s to the choice of normalization
of the boundary fields and the chosen gauge for the defect networks.

Suppose we made a generic rescaling of the canonically normalized boundary fields with structure constants given by (\ref{eq:C-canonical}), $\phi_i^{ab}\to n_i^{ab}\phi_i^{ab}$  so that

\be
C_{ij}^{(abc)k}=\frac{n_i^{ab}n_j^{bc}}{n_k^{ac}}\sqrt{\frac{\sqrt{g'_ig'_jg'_k}}  {\theta(i,j,k)^{\rm blocks}}} \,\Rac{c}{a}{k}{i}{j}{b},
\ee

Then rewriting the rhs of (\ref{eq:TFT-main}) by expressing the generic $F$-matrices in terms of the Racah symbols, using  (\ref{Normalized6J}) and (\ref{Racah}), and isolating the unknowns $\alpha$ and $G$ to the right we get
\be\label{eq:alpha-def}
\frac{n_i^{ab}n_j^{bc}n_k^{ac}}{\sqrt{\gamma(i,j,k)^{\rm blocks}}}\sqrt{\frac{\gamma(i,j,k)}{\gamma(i,a,b)\gamma(j,b,c)\gamma(k,a,c)}}=G_{ijk}\, \alpha_i^{ab}\alpha_j^{bc}\alpha_k^{ca},
\ee
where we have defined for convenience
\be
\gamma(i,j,k)\equiv\frac  {\theta(i,j,k)}{\sqrt{g'_ig'_jg'_k}}.
\ee
Notice that specifying this quantity corresponds to pick a gauge for the $F$-matrices used  for defect manipulation. Racah gauge corresponds to $\gamma(i,j,k)=1$, for an admissible triplet.
The generic solution to equation (\ref{eq:alpha-def}), enforcing $\alpha_i^{ab}=\alpha_i^{ba}$, is then given by
\bea
\alpha_i^{ab}&=&\beta_i \frac{n_i^{ab}}{\sqrt{\gamma(i,a,b)}},\label{alphaTFT}\\
G_{ijk}&=&\frac1{\beta_i\beta_j\beta_k}\sqrt{\frac{\gamma(i,j,k)}{\gamma(i,j,k)^{\rm blocks}}}.
\eea
Notice the undetermined parameters $\beta_i$. Their presence is easily explained: they are just a normalization choice for the defect ending field appearing in Figure \ref{fig:TFT1}. This undetermined normalization always cancels in the defect manipulations we consider in this paper and it is  consistent to set it to 1.

\section{Ising data}
\setcounter{equation}{0}
\label{sec:X_Ising}

In this appendix, we present explicit data for the Ising model to facilitate
concrete calculations.

The standard $F$ matrices (the one entering in the transformation properties of the canonically normalized conformal blocks, and called $F^{\rm blocks}$ in the main text) are given by the expressions below for $\xi=\frac{1}{2}$. The Racah coefficients are given by the same expressions for $\xi=1$.
Note that the parameter $\xi$ corresponds to the gauge-freedom of Moore and Seiberg, but we  stress again  that there is no such freedom in the transformation property of the 4-pt conformal blocks, once the coefficient of their leading term is canonically set to~1. The general $\xi$-dependent $F$-matrices read
\begin{align}
\Fmat{1}{1}{\eps}{\eps}{\eps}{\eps}&=1,
\quad
\Fmat{1}{1}{\sigma}{\sigma}{\sigma}{\sigma}
=-\Fmat{\eps}{\eps}{\sigma}{\sigma}{\sigma}{\sigma}
=\frac{1}{\sqrt{2}}
\\
\Fmat{1}{\eps}{\sigma}{\sigma}{\sigma}{\sigma}&=\frac{\xi}{\sqrt{2}},
\quad
\Fmat{\eps}{1}{\sigma}{\sigma}{\sigma}{\sigma}=\frac{1}{\sqrt{2}\xi}
\\
\Fmat{1}{\sigma}{\eps}{\sigma}{\eps}{\sigma}
&=\Fmat{1}{\sigma}{\sigma}{\eps}{\sigma}{\eps}=\xi,
\quad
\Fmat{\sigma}{1}{\eps}{\eps}{\sigma}{\sigma}
=\Fmat{\sigma}{1}{\sigma}{\sigma}{\eps}{\eps}
=\frac{1}{\xi}
\\
\Fmat{\sigma}{\sigma}{\eps}{\sigma}{\sigma}{\eps}&=\Fmat{\sigma}{\sigma}{\sigma}{\eps}{\eps}{\sigma}=-1.
\end{align}

%
%
The modular S-matrix is (the order of rows and columns is 1, $\eps$, $\sigma$)
\be
S=
\begin{pmatrix}
\frac{1}{2}&\frac{1}{2}&\sqrt{\frac{1}{2}}\\
\frac{1}{2}&\frac{1}{2}&-\sqrt{\frac{1}{2}}\\
\sqrt{\frac{1}{2}}&-\sqrt{\frac{1}{2}}&0
\end{pmatrix}.
\ee
The normalized $g$-functions are therefore
\be
g'_1=1,\quad g'_\eps=1,\quad g'_\sigma=\sqrt{2}.
\ee
\be
{g'}_a=\frac{1}{\Fmat{1}{1}{a}{a}{a}{a}}
\ee

We can obtain the symmetric defect coefficients ($X^{dab}_{ia'b'}=X^{dba}_{ib'a'}$)
 \eqref{X-sym} by using the above $F$-matrices in the Racah gauge.
Defects action on $\mathds{1}$ boundary fields are given by
\begin{align}
X^{\eps \mathds{1} \mathds{1}}_{\mathds{1} \eps \eps}=
X^{\eps \eps \eps}_{\mathds{1} \mathds{1} \mathds{1}}=
X^{\eps \sigma \sigma}_{\mathds{1} \sigma \sigma}=
X^{\sigma \mathds{1} \mathds{1}}_{\mathds{1} \sigma \sigma}=
X^{\sigma \eps \eps}_{\mathds{1} \sigma \sigma}=
X^{\sigma \sigma \sigma}_{\mathds{1} \mathds{1} \mathds{1}}=
X^{\sigma \sigma \sigma}_{\mathds{1} \eps \eps}=1\nonumber,
\end{align}
on $\eps$ boundary fields
\begin{align}
X^{\eps \mathds{1} \eps}_{\eps \eps \mathds{1}}&=
-X^{\eps \sigma \sigma}_{\eps \sigma \sigma}=1,\quad
X^{\sigma \mathds{1} \eps}_{\eps \sigma \sigma}
=\frac{1}{\sqrt{2}},\quad
X^{\sigma \sigma \sigma}_{\eps \mathds{1} \eps}=\sqrt{2}\nonumber,
\end{align}
and on $\sigma$ boundary fields are obtained by
\begin{align}
X^{\eps \mathds{1} \sigma}_{\sigma \eps \sigma}&=\frac{1}{\sqrt{2}},\quad
X^{\eps \eps \sigma}_{\sigma \mathds{1} \sigma}=\sqrt{2},\quad
X^{\sigma \mathds{1} \sigma}_{\sigma \sigma \mathds{1}}
=-X^{\sigma \eps \sigma}_{\sigma \sigma \eps}=\frac{1}{2^{1/4}}\nonumber\\
X^{\sigma \mathds{1} \sigma}_{\sigma \sigma \eps}&=\frac{1}{2^{3/4}},\quad
X^{\sigma \eps \sigma}_{\sigma \sigma \mathds{1}}=2^{1/4}\nonumber.
\end{align}


\end{document}